\documentclass{jfm}
\usepackage{graphicx}
\usepackage{epstopdf, epsfig}
\usepackage{bm}
\usepackage{relsize}
\usepackage{amsmath}
\usepackage{xcolor}
\usepackage{rotating}
\usepackage[normalem]{ulem}

\shorttitle{JFM Perspectives}
\shortauthor{R. Pandit and N.B. Padhan}

\title{The Cahn-Hilliard-Navier-Stokes Framework for Multiphase Fluid Flows: Laminar, Turbulent, and Active}

\author{
    Nadia Bihari Padhan\aff{1,2}, 
    \corresp{\email{nadia\_bihari.padhan@tu-dresden.de}}
    \and Rahul Pandit\aff{1} 
    \corresp{\email{rahul@iisc.ac.in}}
}

\affiliation{\aff{1}Centre for Condensed Matter Theory, Department  of Physics, Indian Institute of Science, Bangalore, India \aff{2}Institute of Scientific Computing, TU Dresden, 01069 Dresden, Germany}

\begin{document}

\maketitle

\begin{abstract}
The Cahn-Hilliard-Navier-Stokes (CHNS) partial differential equations (PDEs) provide a powerful framework for the study of the statistical mechanics and fluid dynamics of multiphase fluids. We provide an introduction to the  equilibrium and nonequilibrium statistical mechanics of systems in which coexisting phases, distinguished from each other by scalar order parameters, are separated by an interface. We then introduce the coupled Cahn-Hilliard-Navier-Stokes (CHNS) PDEs for two immiscible fluids and generalisations for (a) coexisting phases with different viscosities, (b) CHNS with gravity, (c) the three-component fluids (CHNS3), and (d) the CHNS for active fluids. We discuss mathematical issues of the regularity of solutions of the CHNS PDEs. Finally we provide a survey of the rich variety of results that have been obtained by numerical studies of CHNS-type PDEs for diverse systems, including bubbles in turbulent flows, antibubbles, droplet and liquid-lens mergers, turbulence in the active-CHNS model, and its generalisation that can lead to a self-propelled droplet. 
\end{abstract}

\maketitle

\section{Introduction}

Multiphase flows  [see, e.g., ~\cite{brennen2005fundamentals,prosperetti2007introduction,balachandar2010turbulent}] abound in nature. They occur on astrophysical, atmospheric, industrial, laboratory, and cellular scales. Examples include the circumgalactic medium [\cite{faucher2023key,sharma2012thermal}], clouds [see, e.g., ~\cite{shaw2003particle,bodenschatz2010can,devenish2012droplet}], flows with droplets and bubbles [see, e.g., \cite{johnson1985fluid,stone1994dynamics,magnaudet2000motion,mercado2010bubble,anna2016droplets,mathai2020bubbly,magnaudet2020particles,pandey2020liquid,pal2022ephemeral,pandey2023kolmogorov}],
aerosols that transmit pathogens [see, e.g., ~\cite{bourouiba2021fluid}], and biomolecular condensates [see, e.g., ~\cite{banani2017biomolecular,gouveia2022capillary}]. An examination of these examples reveals that the term multiphase is used to mean different things in different settings. For example, the important overview by~\cite{balachandar2010turbulent} begins by noting, `` \ldots  Dispersed multiphase ﬂows are distinguished from other kinds of multiphase ﬂows, such as free-surface ﬂows. In dispersed multiphase ﬂows, the evolution of the interface between the phases is considered of secondary importance. Processes such as droplet or bubble break-up and agglomeration do indeed alter the interface between the phases. However, in the context of dispersed multiphase ﬂows, one accounts for the interface between the dispersed
and carrier phases in terms of particle-size spectra without considering the detailed evolution of the interface. '' In contrast, we restrict ourselves to systems in which the different \textit{phases coexist in thermodynamic equilibrium} and where it is necessary to account for the \textit{surface tension} and the break-up or coalescence of droplets of these phases. To understand such multiphase flows, we must combine theoretical methods from fluid dynamics and statistical mechanics. 

\textcolor{black}{The fundamental equations governing the dynamics of a viscous fluid, now known as the Navier-Stokes (NS) equations, were first introduced by Navier in 1822 ~\cite{navier1823memoire}. Over the following decades, these equations were refined and rigorously formulated by Stokes and others ~\cite{stokes1901mathematical,leray1934essai,batchelor1967introduction,doering1995applied,galdi2000introduction,foias2001navier,robinson2020navier,farwig2021jean}, leading to the modern form used today. For a detailed historical developments of this equation, see ~\cite{darrigol2005worlds} and ~\cite{bistafa2018development}.}
 The Cahn-Hilliard equation, which is 67 years old [see, e.g., ~\cite{cahn1958free,cahn1959free,cahn1961spinodal,lothe1962reconsiderations,lifshitz1961kinetics,hohenberg1977theory,gunton1983p,bray2002theory,chaikin1995principles,puri2009kinetics,onuki2002phase,badalassi2003computation,perlekar2014spinodal,berti2005turbulence}], plays a central role in the theory of two-phase mixtures, interfaces, and phase separation, domain growth, and coarsening in the wake of a quench from a high-temperature single-phase regime to a low-temperature two-phase region~\cite{puri2004kinetics,bray2002theory}. \textcolor{black}{The Cahn-Hilliard equation, initially formulated to
describe phase separation in binary alloys [see, e.g., ~\cite{lebowitz1976computer,binder1979computer,puri2009kinetics}], has been applied, with suitable adaptations, to a wide range of phenomena, including self-assembly in diblock copolymers~\cite{hill2017numerical} and tumour-growth modelling~\cite{hilhorst2015formal,ebenbeck2020cahn,garcke2021phase}. For a comprehensive review of the applications of the Cahn-Hilliard equation, see~\cite{kim2016basic}.} If the two phases under consideration are immiscible fluids, it is natural to combine the above two equations to obtain the Cahn-Hilliard-Navier-Stokes (CHNS) equations that provide a powerful mathematical framework for the study of two-fluid flows, be they laminar or turbulent. The CHNS approach and its generalisations have (a) found extensive applications, across diverse length and time scales, e.g., in droplet formation in the atmosphere and the manipulation of microscale droplets in microfluidic devices, (b) led to new insights into a variety of multiphase flows, and (c) provided a convenient and efficient numerical scheme for direct numerical simulations (DNSs) of such flows. We provide on overview of this rich and rapidly developing field.

The CHNS framework, also known as the diffuse-interface or the phase-field method, is related to Model H that is used in dynamic critical phenomena [see, e.g., ~\cite{hohenberg1977theory,anderson1998diffuse,gurtin1996two,puri2009kinetics}]. It has a rich history that dates back to the pioneering studies of Fixman and of Kawasaki [see, e.g.,  ~\cite{fixman1967transport} and ~\cite{kawasaki1970kinetic}], who developed coupled hydrodynamical equations for studying the behaviour of binary-fluid mixtures. Since then, the CHNS framework has evolved into a powerful tool for modelling a wide range of complex multiphase flows. It uses diffuse interfaces with a smooth transition region instead of a sharp boundary, so the CHNS model provides a convenient representation of the dynamics of fluid-fluid interfaces[see, e.g., ~\cite{lowengrub2009phase,kim2012phase}]. This approach enables us to study a variety of phenomena, including droplet formation, coalescence, and breakup, as well as phase separation and mixing. In Fig.~\ref{fig:flow-chart}, we present a schematic overview of multiphase flows, their wide-ranging applications, and the mathematical and numerical models employed for studying these systems. This diagram includes the CHNS or phase-field model, various numerical methods that are used to solve the CHNS partial differential equations (PDEs), and the broad spectrum of applications of the CHNS model. \textcolor{black}{Below, we briefly outline the key differences and advantages of the CHNS method in comparison with other approaches: 
\begin{figure}
  \centerline{\includegraphics[width=14.5cm]{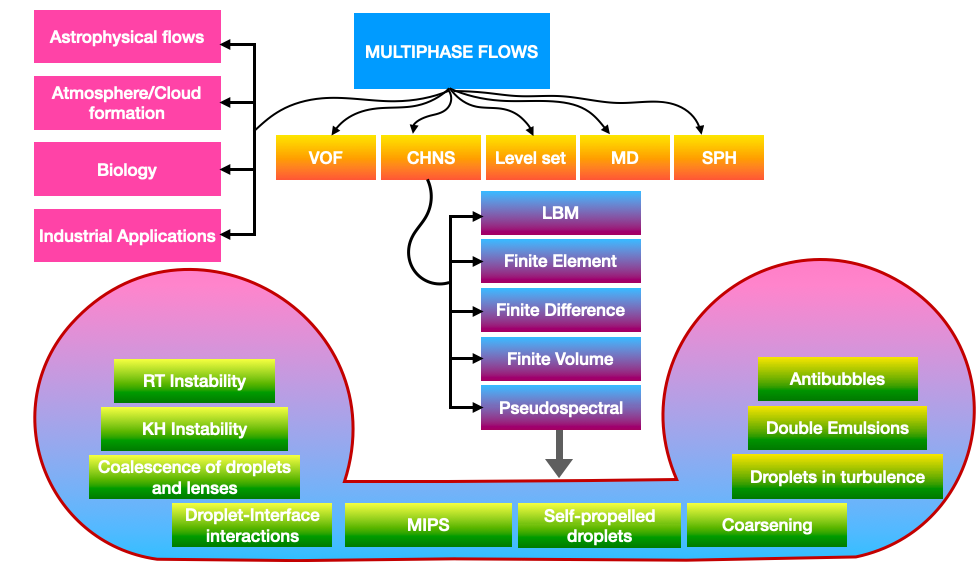}}
  \caption{A schematic overview of multiphase flows, their wide-ranging applications, the mathematical and numerical models employed for studying them; we include the CHNS and phase-field models, various numerical methods that are used to solve the CHNS model, and the broad spectrum of applications of the CHNS model. [VOF: Volume of fluids; MD: Molecular dynamics; SPH: Smooth particles hydrodynamics; LBM: Lattice-Boltzmann method; RT: Rayleigh-Taylor; KH: Kelvin Helmholtz; MIPS: Motility-induced phase separation].}
\label{fig:flow-chart}
\end{figure}
\begin{enumerate}
    \item Molecular Dynamics (MD): Molecular dynamics (MD) provides a microscopic perspective on  and treatment of phase separation and interfacial dynamics [see, e.g., ~\cite{singh2015phase, li2023molecular}]; MD becomes computationally costly and challenging when dealing with large interfacial fluctuations and turbulence in macroscopic multiphase systems. Although MD has been used successfully for studying problems like droplet coalescence, it lacks the ability to model large-scale hydrodynamic effects efficiently. See ~\cite{heinen2022droplet} for a comparative study of MD and phase-field simulations for problems related to the coalescence of droplets.
    \item Volume of Fluid (VOF): VOF is a widely used numerical technique for tracking and locating free surfaces in multiphase flows [see, e.g., ~\cite{lafaurie1994modelling, popinet1999front, yokoi2007efficient, tryggvason2011direct, mohan2024volume}]. It is particularly effective for simulating fluid flows with interfaces between different fluids. VOF models the interface as a sharp boundary, which can introduce numerical challenges when handling complex topological changes. However, in reality, interfaces are not infinitely sharp, but they have a finite thickness, especially in systems undergoing phase separation. In contrast, the phase-field (CHNS) framework represents the interface as a diffuse region rather than a sharp boundary. This approach naturally accommodates topological changes, such as droplet coalescence or breakup, without requiring complex interface-tracking algorithms. Both methods have been successfully applied in the study of antibubble dynamics by ~\cite{pal2022ephemeral}. A comprehensive comparison of these methods for various multiphase flows can be found in ~\cite{mirjalili2019comparison}.
    \item Smoothed Particle Hydrodynamics (SPH): SPH is a mesh-free, particle-based method that is well-suited for simulating free-surface flows and multiphase interactions [see, e.g., ~\cite{wang2016overview,diaz2023smoothed,colagrossi2003numerical,violeau2016smoothed}]. SPH is well-suited for studies of certain types of free-surface flows, but it can be computationally intensive because it requires the tracking of individual particles and, in some cases, stabilization techniques. SPH typically uses additional formulations to capture, accurately, interfacial dynamics, such as surface tension [see ~\cite{jeske2023implicit}].
    \item Level Set Method: The Level-Set Method represents interfaces as sharp boundaries using a level-set function. This approach is effective for tracking evolving interfaces, particularly in cases where the topology changes, such as in droplet merging and splitting [see, e.g., ~\cite{sethian2003level,valle2020energy,yuan2018simple}]. A detailed comparative study via numerical simulations of both Level-Set and phase-field methods is given in ~\cite{zhu2021comparison}, where the authors write, ``\ldots In the rising process of the bubble, the tracking efficiency of the phase field method is higher than that of the level set method. The phase field method is easier to capture the subtle changes of the bubble, and has higher calculation efficiency and accuracy, and has better simulated calculation results."
\end{enumerate}
}
The remaining part of this paper is organised as follows. In Section~\ref{sec:StatMech} we begin with a discussion of the statistical mechanics of systems in which two coexisting phases, distinguished from each other by a scalar order parameter $\phi$, are separated by an interface. Our discussion leads naturally to the time-dependent Ginzburg-Landau (TDGL) partial differential equation (PDE), when $\phi$ is \textit{not conserved}, and the Cahn-Hilliard (CH) PDE, if $\phi$ is \textit{conserved}. In Section~\ref{sec:Models} we define the models that we use when the coexisting phases are fluids; in the simple case of two immiscible fluids we have the Cahn-Hilliard-Navier-Stokes (CHNS) equations; we give its generalisations for (a) coexisting phases with different viscosities, (b) CHNS with gravity, (c) three-component fluids (CHNS3), and (d) CHNS for active fluids. Section~\ref{sec:Numerics} gives an overview of the numerical schemes that are used to study these models; we include details of pseudospectral direct numerical simulations (DNSs) and the volume-penalisation method that we use in our work; furthermore, we contrast the CHNS diffuse-interface approach with schemes that track the spatiotemporal evolution of sharp interfaces. Section~\ref{sec:Maths} discusses mathematical issues of the regularity of solutions of the CHNS PDEs. Section~\ref{sec:validations} contains a survey of the types of results that have been obtained by numerical studies of CHNS-type PDEs. In Section~\ref{sec:beyond} we present results for a variety of challenging problems for which we have to go beyond the binary-fluid CHNS. The concluding Section~\ref{sec:Conclusions} summarises the CHNS framework for multiphase flows and discusses and new challenges in this area.      


\section{Overview: Statistical Mechanics of Interfaces}
\label{sec:StatMech}

This Section contains a short outline of the statistical mechanics of systems with interfaces. We include material that is required for the development of Cahn-Hilliard models and their generalisations. Subsection~\ref{subsec:EqStatMech} is devoted to the equilibrium statistical mechanics of systems with interfaces; in particular, it uses the Ising and lattice-gas models to explore interfacial statistical mechanics. In Subsection~\ref{subsec:Kin_Ising} we turn to a discussion of time-dependent phenomena, especially in the context of the Kinetic Ising model. This leads naturally to the discussion, in Subsection~\ref{subsec:TDGL}, of time-dependent Ginzburg-Landau theory, for a non-conserved order parameters, and, in Subsection~\ref{subsec:CH_TD}, the Cahn-Hilliard equation for a conserved order parameter.

\subsection{Equilibrium Statistical Mechanics}
\label{subsec:EqStatMech}

Interfaces separate coexisting phases. In \textit{equilibrium}, all thermodynamic properties of interfaces follow from the \textit{intensive interfacial free energy} $f_I$, which is precisely the interfacial tension $\sigma_{\alpha,\beta}$ between coexisting bulk phases $\alpha$ and $\beta$. For pedagogical reasons, we illustrate this by considering the ferromagnetic, spin-half, Ising model, which is used to model magnets that are anisotropic (in spin space). This Ising model, which can be mapped onto a simple lattice-gas model for a fluid that can have two coexisting bulk phases (see below), has the Hamiltonian
\begin{equation}
\mathcal{H} = - J \sum_{\langle i,j\rangle} S_iS_j - h \sum_i S_i \,.
\label{eq:Ising_H}
\end{equation}
The spins $S_i = \pm 1$ occupy the $N \sim L^d$ sites $i$ of a $d-dimensional$ hypercubic lattice, with side $L$, in a region $\Omega$; the exchange interaction $J$ couples spins on nearest-neighbour pairs of sites $\langle i,j\rangle$; a ferromagnetic interaction $(J > 0)$  favours spin alignment, i.e., spins of the same sign on nearest-neighbour sites; the external magnetic field $h$ favours $S_i = sgn (h)$ and it distinguishes between the two phases: up-spin $\uparrow$ (with $S_i = +1\,\, , \forall i$, at zero temperature $T=0$) and down-spin $\downarrow$ (with $S_i = -1\,\, , \forall i$, at $T=0$). If we are interested only in bulk statistical physics, it is convenient to use periodic boundary conditions in all $d$ directions. 

The transformation $n_i = (1+S_i)/2$, yields the lattice gas variables $n_i = 0 \, \rm{or} \, 1$ and defines the lattice-gas model, whose low- and high-density phases are the analogues of the $\downarrow$ and $\uparrow$ phases in the ferromagnetic Ising model and in which the chemical potential is the counterpart of $h$ in the Ising model [see, e.g., ~\cite{griffiths1972rigorous,goldenfeld2018lectures}].\footnote{A similar transformation relates the Ising-model Hamiltonian to a lattice model for a binary alloy.} \textcolor{black}{Just as a positive (negative) external field $h$ favours the up-spin $\uparrow$ (down-spin $\downarrow$) phase, a large and positive (negative) values of the chemical potential favour the high-density (low-density) phase  in the lattice-gas model [see Table~\ref{tab:my_label}].}

\begin{table}
\textcolor{black}{
    \centering
    \begin{tabular}{cccc}
        System \hspace{0.1cm} & $\uparrow$ Phase & \hspace{0.1cm} $\downarrow$ Phase & \hspace{0.1cm} Phase Transition\\
        \hline
        Ising Ferromagnet \hspace{0.1cm} & ferromagnetic up  & \hspace{0.1cm} ferromagnetic down & \hspace{0.1cm} spin flip\\
        \hline
        Liquid-Gas \hspace{0.1cm} & high-density (liquid) & \hspace{0.1cm} low-density (gas) & \hspace{0.1cm} condensation\\
        \hline
        Binary (A-B) Mixture \hspace{0.1cm} & A-rich & \hspace{0.1cm} B-rich & \hspace{0.1cm} phase-separation\\
        \hline
        Active Fluid \hspace{0.1cm} & high-density & \hspace{0.1cm} low-density & \hspace{0.1cm} active phase-separation\\
        \hline
    \end{tabular}
    \caption{Correspondences between Ising ferromagnetic systems, with lattice-gas or binary-mixture counterparts, and their phases and phase transitions; here, $\uparrow$ and $\downarrow$ represent the up-spin and down-spin phases of an Ising ferromagnet (see text). Rows $2-4$ mention equilibrium phases and transitions; the last row mentions nonequilibrium active fluids which can exhibit active phase-separation [see Section~\ref{subsec:ActiveH}].}
    \label{tab:my_label}
}
\end{table}

The intensive bulk free energy $f_B$ of the Ising model, which is a function of the temperature $T$, $h$, and $J$, is defined as follows:
\begin{eqnarray}
f_B(T,h,J) &=& \lim_{N\to \infty} \frac{1}{N}\bigg[F(T,h,J,\Omega)\bigg]\,; \nonumber \\
F(T,h,J,\Omega)] &=& -k_BT\ln[\mathcal{Z}] \,; \nonumber \\
{\mathcal{Z}} &=& \sum_{\{S_i = \pm 1\}} \exp [-\mathcal{H}/(k_BT)]\,;
\label{eq:Ising_fB}
\end{eqnarray}
here, $k_B,\,\, F$, and ${\mathcal Z}$ are, respectively, the Boltzmann constant, the total free energy and the partition function, and the sum is over all the spin states. Note that $F$ depends on $\Omega$ and, therefore, on boundary conditions;
by contrast, $f_B$ does not depend on $\Omega$, the boundary conditions, and boundary couplings and fields because we have taken the \textit{thermodynamic limit} $N \to \infty$. If the thermodynamic limit exists (as it does for the Ising model we consider), then: (a) $f_B$ is a \textit{convex-up} function of its arguments, which guarantees continuity of $f_B$ with respect to $T$, $h$, and $J$; (b) $\partial f_B/\partial T\,, \partial f_B/\partial h,$ and $\partial f/\partial J$ exist almost everywhere ($f_B$ has slope discontinuities at first-order phase boundaries); and (c) $\partial f_B/\partial T,\, \partial f_B/\partial h,$ and $\partial f/\partial J$ are, respectively, monotone, non-increasing functions of $T,\, h,$ and $J$. Bulk thermodynamic functions follow from derivatives of $f_B$; e.g., the magnetization per site
\begin{equation}
 M = \sum_i \langle S_i \rangle /N = -\partial f_B/ \partial h \,,
 \label{eq:mag}
\end{equation}
where the angular brackets denote the thermal average. When the second derivatives of $f_B$ exist, they lead to stability conditions such as the positivity of the specific heat at constant $h$, to wit, $C_h \equiv -k_BT\partial^2f_B/\partial T^2 \geq 0$. 

\begin{figure}
  \centerline{\includegraphics[width=12cm]{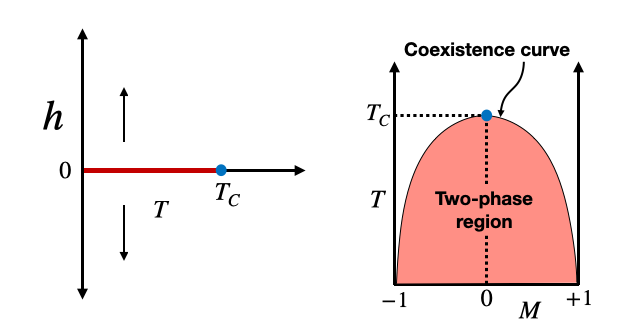}}
  \caption{Schematic phase diagrams for the $d$-dimensional Ising model~\ref{eq:Ising_H} in the magnetic-field $h$ and temperature $T$ plane (left panel) and the $T$ and magnetization $M$ plane (right panel). There is a first-order phase boundary (brown line), $h=0$ for $T < T_c$, that ends in a critical point at $h=0,\, T=T_c$ (blue point); along the first-order boundary, the $\uparrow$ and $\downarrow$ phases coexist. In the $T-M$ phase diagram, two-phase coexistence occurs in the peach-shaded region everywhere below the coexistence curve (black), atop which is the critical point.}
\label{fig:IsingPD}
\end{figure}

This magnetization $M \in [-1,1]$ is the \textit{order parameter}, which is positive (negative) in the up-spin (down-spin) phase; its lattice-gas counterpart is the density $\rho = \sum_i \langle n_i \rangle /N \in [0,1]$. For dimension $d >1$, the Ising-model phase diagram in the $h/J-k_BT/J$ plane (henceforth, we set $k_B=1$ and $J=1$) shows a first-order phase boundary, $h=0$ for $T < T_c$, where $T_c$ is the critical or Curie temperature at which this model exhibits a continuous phase transition from the ferromagnetic to the paramagnetic phase [left panel of Fig.~\ref{fig:IsingPD}]. Along the first-order boundary, the $\uparrow$ and $\downarrow$ phases coexist. In the $T-M$ counterpart [right panel of Fig.~\ref{fig:IsingPD}] of the $h-T$ phase diagram, two-phase coexistence occurs everywhere below the \textit{coexistence curve} in the peach-shaded region: \textcolor{black}{the system undergoes macroscopic phase separation, forming distinct $\uparrow$ and $\downarrow$ phases, rather than a fully mixed state; in equilibrium, these coexisting phases are separated by an interface (see below). If we use the Ising-model-lattice-gas correspondence [Table~\ref{tab:my_label}], we see that its binary-mixture
counterpart is the coexistence of immiscible fluids, below the critical temperature above which the two fluids mix.}

We do not concentrate on the critical properties of the Ising model here; but we do mention critical-point behaviours occasionally. Therefore, we note that, in the vicinity of the critical point at $h=0$ and $T=T_c$, thermodynamic functions display power-law forms, with universal exponents and scaling functions [see, e.g., ~\cite{goldenfeld2018lectures,kardar2007statistical,chaikin1995principles}]. For example, at $h=0$ and $(T-T_c)/T_C \ll 1$, $M \sim [(T_c-T)/T_c]^\beta$, for $T \lesssim T_c$; and the specific heat $C_h \sim [(T_c-T)/T_c]^{-\alpha}$ diverges; $\beta$ and $\alpha$ are universal critical exponents that depend on $d$ and the number of components $n$ of the order parameter (for the Ising model we have a scalar order parameter and $n=1$). If we were to obtain order-parameter correlation functions and, therefrom, the correlation length $\xi$, then we would find the divergence $\xi \sim [(T_c-T)/T_c]^{-\nu}$. [At such a critical point, only two exponents are independent; the others can be related by scaling laws that are linear relations between these exponents (see, e.g.,~\cite{goldenfeld2018lectures,kardar2007statistical,chaikin1995principles}). For the $d=2$ Ising model, $\alpha = 0$ because $C_h$ has a logarithmic divergence at $h=0$ and $T=T_c$, $\beta = 1/8$, and $\nu = 1$.]

\begin{figure}
  \centerline{\includegraphics[width=15cm]{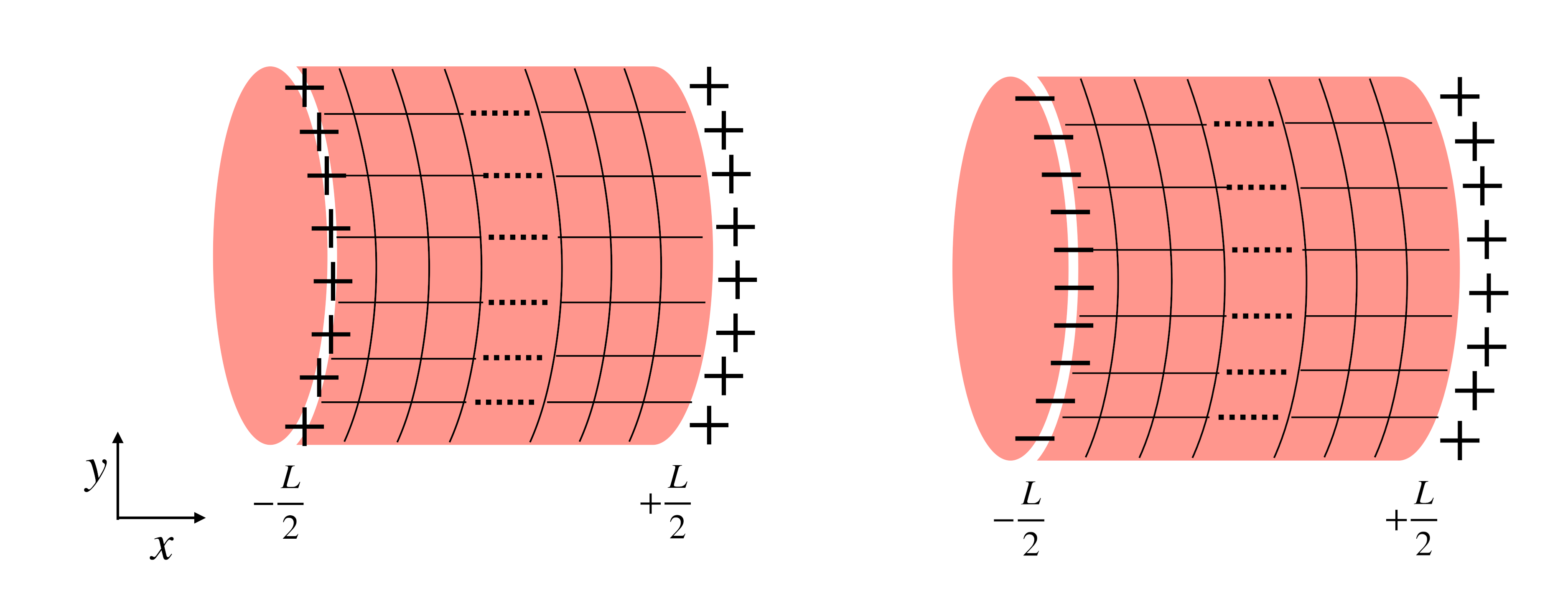}
  \put(-390,150){\rm {\bf(a)}}
  \put(-190,150){\rm {\bf(b)}}
  }
  \caption{Schematic diagrams for (a) $++$ and (b) $-+$ boundary conditions (BCs) for the $d$-dimensional Ising model~\eqref{eq:Ising_H}; $d=2$ in this illustration. In $d-1$ directions ($y$ here) we use periodic boundary conditions.}
\label{fig:IsingBC}
\end{figure}

To define the intensive interfacial free energy $f_I$ (per unit area of the interface), we must distinguish between boundary conditions (BCs), for the Ising Hamiltonian~\ref{eq:Ising_H}, that yield a $(d-1)-$dimensional interface between coexisting bulk phases and those that do not. We illustrate two such BCs, $++$ and $-+$, via the schematic diagrams for $d=2$ in Fig.~\ref{fig:IsingBC}. We use periodic boundary conditions in $(d-1)$ directions; in the remaining direction, denoted by $x$, we have two $(d-1)$-dimensional surfaces at $x=-L/2$ and $x=+L/2$; the $++$ BC does not yield an interface; by contrast, the $-+$ BC yields an interface with $N_I \sim L^{(d-1)}$ the number of sites in the $T=0$ interface.\footnote{The precise value of $x$ at the interface location depends on whether we work in the fixed-$h$ or fixed-$M$ ensemble; in the former, the interface can lie anywhere between $x=-L/2$ and $x=+L/2$, whereas, in the latter, the position of the interface is such that the proportion of the $\uparrow$ and $\downarrow$ regions 
yields the fixed magnetization $M$.}      
\begin{equation}
f_I(T,h,J) = \lim_{N_I \to \infty; N\to \infty}\frac{1}{N_I}\bigg[F^{-+} - F^{++}\bigg]\,,
\label{eq:Ising_fI}
\end{equation}
where $F^{++}$ and $F^{-+}$ are, respectively, the total free energies of the Ising model with $++$ and $+-$ boundary conditions [Fig.~\ref{fig:IsingBC}]. [Note that the free-energy contributions from the two surfaces cancel when we subtract $F^{++}$ from $F^{-+}$ in Eq.~(\ref{eq:Ising_fI}).] We have mentioned above that bulk statistical mechanics and phase diagrams follows from $f_B$ and its derivatives; similarly, interfacial statistical mechanics and interfacial phase diagrams follow from $f_I$ and its derivatives.~\footnote{If the order parameter has more than one component, e.g., if $n=2$ as in the XY model or a superfluid, then $f_I$ has to be replaced by the helicity modulus $\Upsilon$, because the interface is not sharp [see, e.g., ~\cite{fisher1973helicity}]; in such cases, $-+$ boundary conditions lead to a correction to the total free energy that scales as $L^{(d-2)}$. We restrict ourselves to scalar order parameters that have $n=1$.} Furthermore, $f_I$ gives us the strict definition of the interfacial (or surface) tension [see, e.g., ~\cite{rottman1984statistical}]\footnote{\textcolor{black}{Both $f_I$ and surface tension have the same physical units: $f_I$ is typically given in energy per unit area, whereas the surface tension is specified as a force per unit length, both of which are dimensionally equivalent.}}. In particular, if $T \to T_c$ from below, the coexisting phases come together at the critical point (the consolute point in a binary-fluid mixture) and the interfacial free energy vanishes as $f_I \sim [(T_c-T)/T_c]^{\mu}$, with the exponent $\mu = 2 \nu$.

It is important to distinguish between surface and interfacial free energies. The former is associated with a geometrically imposed surface [say on the (1,0,0, \ldots) face of the hypercubic lattice we consider]. The intensive surface free energy is [see, e.g., ~\cite{rottman1984statistical,fisher1977wall,caginalp1979wall}]:
\begin{equation}
f_s(T,h,J) = \lim_{N_s \to \infty; N\to \infty} \frac{1}{2N_s}\bigg[F^{++} - Nf_B\bigg]\,,
\label{eq:Ising_fS}
\end{equation}
where $N_s \sim L^{(d-1)}$ is the number of sites in the geometrically imposed surface (or surfaces); for the $++$ BC in Fig.~\ref{fig:IsingBC}, there are two surfaces at $x=-L/2$ and $x=+L/2$, so we have a factor of $2$ in Eq.~(\ref{eq:Ising_fS}). Geometrically imposed surfaces do not fluctuate, but interfaces do, because of thermal or other (e.g., turbulent) fluctuations.

It is well known that the $d-$dimensional Ising model can be solved exactly (i.e., $f_B$ can be obtained analytically) only if (a) $d=1$ or (b) $d=2$ and $h=0$ [see, e.g., ~\cite{onsager1944crystal,huang2008statistical,thompson2015mathematical,baxter2016exactly}]. In other dimensions, we must either use approximations, such as mean-field theory, or numerical simulations, such as the Monte Carlo method [see, e.g., ~\cite{plischke1994equilibrium,goldenfeld2018lectures,kardar2007statistical,chaikin1995principles,gould1996introduction}]. For $d=2$, a variety of elegant results can be obtained for $f_I$; these are of relevance to equilibrium crystal shapes [see~\cite{rottman1984statistical}].

We give a brief introduction to mean-field theory because it leads directly to our discussion of the Cahn-Hilliard system. We consider a generalization of the Hamiltonian~\eqref{eq:Ising_H} in which the magnetic field $h$ is replaced by a site-dependent magnetic field $h_i$, so the second term becomes $\sum_i h_i S_i$. In its most rudimentary form, mean-field theory (known as Curie-Weiss theory for our Ising-model example) uses:
\begin{eqnarray}
S_iS_j &=& (S_i - \langle S_i \rangle) (S_j - \langle S_j \rangle) + S_i \langle S_j \rangle + S_i \langle S_j\rangle -  \langle S_i \rangle  \langle S_j \rangle \,;\nonumber \\
S_iS_j&\simeq& + S_i \langle S_j \rangle + S_i \langle S_j\rangle -  \langle S_i \rangle  \langle S_j \rangle \,,
\label{eq:CW_decoup}
\end{eqnarray}
The second equation follows from the first one if we neglect $(S_i - \langle S_i \rangle) (S_j - \langle S_j \rangle)$ because it is quadratic in deviations from the equilibrium values of the site magnetizations $M_i \equiv \langle S_i \rangle$. We now substitute the second row of Eq.~(\ref{eq:CW_decoup}) into the Hamiltonian~(\ref{eq:Ising_H}); this yields a mean-field Hamiltonian $H_{MF}$ in which a spin at site $i$ experiences an effective magnetic field $h^{eff}_i = h_i + J\sum_{j\in[nni]} M_j$, where $[nni]$ indicates all the nearest-neighbour sites of $i$. The single-site magnetization for $\mathcal{H}_{MF}$ follows simply and we get the Curie-Weiss self-consistency equations
\begin{equation}
M_i = \tanh\bigg[\frac{(h_i + J\sum_{j\in[nni]} M_j)}{(k_BT)}\bigg]\,,
\label{eq:CW_Mi}
\end{equation}
which can have many solutions. The variational formulation of this mean-field theory [see, e.g., ~\cite{plischke1994equilibrium,falk1970inequalities,girardeau1973variational}] implies that we must select the solution that leads to the \textit{global minimum} of the \textit{variational free energy}
\begin{eqnarray}
\mathcal{F}_{CW}(\{\mathcal{M}_i\}) &=& -\sum_ih_i\mathcal{M}_i - J\sum_{<i,j>} (\mathcal{M}_i\mathcal{M}_j) \nonumber \\
&+& \frac{k_BT}{2}\sum_i\bigg[(1+\mathcal{M}_i)\ln\frac{(1+\mathcal{M}_i)}{2} + (1-\mathcal{M}_i)\ln\frac{(1-\mathcal{M}_i)}{2}\bigg]\,,
\label{eq:CW_varF}
\end{eqnarray}
where $\mathcal{M}_i$ are the variational parameters and the subscript $CW$ stands for Curie-Weiss. Note that $\mathcal{F}(\{\mathcal{M}_i\})$ is \textit{not the equilibrium free energy}, so it might not be a convex function of its arguments. If we minimise $\mathcal{F}(\{\mathcal{M}_i\})$ by setting $\partial \mathcal{F}(\{\mathcal{M}_i\})/\partial \mathcal{M}_i = 0$, we recover the self-consistency equations~\eqref{eq:CW_Mi}. At the global minimum of $\mathcal{F}(\{\mathcal{M}_i\})$, the equilibrium value of $\mathcal{M}_i$ is $\mathcal{M}_{i,eq}$; when we substitute this value of $\mathcal{M}_i$ in $\mathcal{F}(\{\mathcal{M}_i\})$, we obtain the mean-field expression for the equilibrium free energy, which satisfies all the convexity properties mentioned earlier.

If the variational parameters $\mathcal{M}_i$ vary slowly in space, say over a length scale $\ell$, with $a/\ell \ll 1$, where $a$ is the lattice spacing of the hypercubic lattice, then we can make the following continuum approximation:
\begin{eqnarray}
\mathcal{M}_i &\to& \phi(\mathbf{r}) \,; \;\; h_i\to h(\mathbf{r})\,; \;\; \sum_i \to \frac{1}{a^d}\int d\mathbf{r} \,; \nonumber \\  
&-& J\sum_{<i,j>}(\mathcal{M}_i\mathcal{M}_j) \to \bigg[-\frac{qJ}{2a^d}\int d\mathbf{r}[\phi(\mathbf{r})]^2+\frac{J}{2a^{(d-2)}}\int d\mathbf{r}[\nabla \phi(\mathbf{r})]^2 \bigg] \nonumber \\ 
&+& \mathcal{O}((a/\ell)^4)\,,
\label{eq:Cont_app}
\end{eqnarray}
where $q = 2d$ is the nearest-neighbour coordination number for the $d-$dimensional hypercubic lattice and $\phi$ is a scalar order parameter. We now use Eq.~(\ref{eq:Cont_app}) to obtain the continuum limit of the variational free energy functional~(\ref{eq:CW_varF}):
\begin{eqnarray}
a^d\mathcal{F}_{CW}(\phi,\nabla \phi) &=& \int d\mathbf{r} \bigg[-h(\mathbf{r}) \phi(\mathbf{r})  - \frac{qJ}{2} [\phi(\mathbf{r})]^2 + \frac{Ja^2}{2} [\nabla \phi(\mathbf{r})]^2 
+ \frac{k_BT}{2} V_{CW}(\phi)  \bigg] \,, \nonumber \\  
V_{CW}(\phi) &\equiv& \left[(1+\phi)\ln\frac{(1+\phi)}{2} + (1-\phi)\ln\frac{(1-\phi)}{2}\right]\,,
\label{eq:CW_varFunc}
\end{eqnarray}
where, in the CW approximation, the first three terms in Eq.~(\ref{eq:CW_varFunc}) yield the energy contribution and the fourth term the entropy contribution [see, e.g., ~\cite{puri2004kinetics}]; furthermore, $\phi \in [-1,1]$. 

\begin{figure}
  \centerline{\includegraphics[width=10cm, height=10cm]{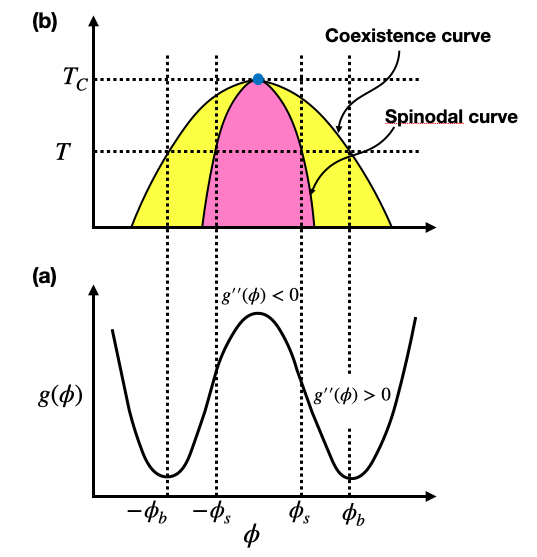}}
  \caption{Schematic plots of:  (a) the variational free energy $g(\phi)$ [see Eq.~(\ref{eq:LG_varFunc})] versus the order parameter $\phi$, for magnetic field $h=0$ and temperature $T < T_c$, with two equally deep minima at $\phi = \pm \phi_b$, the LG equilibrium values;
  (b) the Landau-Ginzburg (LG) phase diagram in the $T-\phi$ plane showing the coexistence curve (also called the \textit{binodal}), below which the phases coexist, and the \textit{spinodal} curve, which is the locus of points at which $d^2g(\phi)/d\phi^2 = 0$.}
\label{fig:gphi_coex}
\end{figure}

If we expand $V_{CW}(\phi)$ to $\mathcal{O}(\phi^4)$ and set $a=1$, we obtain the Landau-Ginzburg (LG) variational free-energy functional [see, e.g., ~\cite{landau2013statistical,kadanoff1967static,goldenfeld2018lectures,kardar2007statistical,chaikin1995principles}]. 
\begin{eqnarray}
\mathcal{F}_{LG}(\phi,\nabla \phi) &=& \int d\mathbf{r} \bigg[ g(\phi) + \frac{\sigma}{2}[\nabla \phi]^2 \bigg] \,, \nonumber \\  
g(\phi) &\equiv& \bigg[-h \phi  + \frac{k_B}{2} (T - T_c^{CW}) \phi^2 + \frac{k_BT}{12} \phi^4 +\mathcal{O}(\phi^6)\bigg] \,,
\label{eq:LG_varFunc}
\end{eqnarray}
where we suppress the $\mathbf{r}$ argument of $h$ and $\phi$, the Curie-Weiss critical temperature $T_c^{CW} \equiv qJ$, and $\sigma \equiv J$ is the bare surface-tension cost for large gradients in the scalar-order-parameter field $\phi$. If we consider uniform ordering, then minimisation of $g$ via $\partial g(\phi)/\partial \phi = 0$ yields the mean-field-theory phases, transitions, and exponents for this Ising model. We present a schematic plot of $g(\phi)$ in Fig.~\ref{fig:gphi_coex}(a) for $T < T_c^{CW}$ and $h=0$. Note that $g$ displays two, equally deep quadratic minima at 
\begin{equation}
\phi_{b} \equiv \sqrt{\frac{3(T_c^{CW} - T)}{T}} \;\; {\rm{and}} \;\; -\phi_{b}\,,     
\end{equation}
so the $\uparrow$ and $\downarrow$ phases coexist along the first-order phase boundary $h=0,\, 0\leq T < T_c^{CW}$, whose counterpart in the $T-\phi$ phase is the coexistence curve shown in Fig.~\ref{fig:gphi_coex}(b). If $T \to T_c^{CW}$ from below, with $h=0$, these minima merge at $T = T_c^{CW}$, the coefficient of the quadratic term vanishes, and elementary steps yield the mean-field order-parameter exponent $\beta_{MF}=1/2$ and interfacial-free-energy exponent $\mu_{MF} = 2\nu_{MF} = 1$. The interfacial free energy vanishes for $T > T_c$, for there are no coexisting phases and, therefore, no interface; and $g(\phi)$ has only one quadratic minimum if $T >  T_c^{CW}$ and $h=0$.

To study non-uniform order-parameter configurations [see, e.g.,~\cite{pandit1982surfaces,bray2002theory,puri2004kinetics,puri2009kinetics}], we must minimise $\mathcal{F}_{CW}$ or $\mathcal{F}_{LG}$. We illustrate this for $\mathcal{F}_{LG}$ below by considering BCs (cf., Fig.~\ref{fig:IsingBC}) that yield an interface for $T < T_c^{CW}$ and $h=0$:
\begin{eqnarray}
\delta\mathcal{F}_{LG}/\delta \phi& =& 0 \,; \nonumber\\
&\Rightarrow&  \bigg[ -\sigma \nabla^2\phi  + k_B (T - T_c^{CW}) \phi + \frac{k_BT}{3} \phi^3  \bigg] = h(\mathbf{r})\,.
\label{eq:Inhom_LGPDE}
\end{eqnarray}
If we assume that $\phi$ varies only in the $x$ direction and that $h(\mathbf{r}) = 0$, we obtain the following ordinary differential equation (ODE), which must be solved with the boundary conditions (BCs) given in the second row:
\begin{eqnarray}
\sigma d^2\phi/dx^2 &=& k_B (T - T_c^{CW}) \phi + \frac{k_BT}{3} \phi^3 \,; \nonumber\\
{\rm{BCs:}}\;\; \phi(x) &=& \pm \phi_b \;\; {\rm{at}} \;\; x = \pm \infty \,.
\label{eq:LG_Interface_PDE}
\end{eqnarray}
This has the following well-known solution [see, e.g., ~\cite{pandit1982surfaces,puri2004kinetics,puri2009kinetics}]:
\begin{eqnarray}
\phi(x) &=& \phi_b \tanh\bigg[ (x-x_0)/\xi_{MF} \bigg]\,; \nonumber\\
\xi_{MF} &=& \sqrt{\frac{\sigma}{2k_B(T_c^{CW} - T)}}\,;
\label{eq:LG_kink}
\end{eqnarray}
here, $x_0$ is the point at which this interfacial (or kink) profile goes through zero and $\xi$ is the width of the interface. If we portray the solutions of Eq.~(\ref{eq:LG_Interface_PDE}) in the phase space $[d\phi/dx,\,\phi]$, the bulk phases correspond to hyperbolic fixed points at $d\phi/dx=0,\,\phi=\pm \phi_{b}$ and the interfacial profile is a heteroclinic trajectory that goes from one of these fixed points to the other [see~\cite{pandit1982surfaces}]. The mean-field free energy $f_{I,MF}$ (surface tension) for the kink solution follows by subtracting $\mathcal{F}_{LG}$ without an interface from its value with an interface:
\begin{eqnarray}
L^{(d-1)}f_{I,MF}(\phi,\nabla \phi) &=& \int_{-\infty}^{\infty} dx \bigg[ \Delta g(\phi) + \frac{\sigma}{2}[d\phi/dx]^2 \bigg] \,, \nonumber \\  
\Delta g(\phi) &\equiv& \bigg[\frac{k_B}{2} (T - T_c^{CW}) [\phi^2 - \phi_b^2] + 
\frac{k_BT}{12} [\phi^4 - \phi_b^4]\bigg] \,,
\label{eq:fIMF1}
\end{eqnarray}
If we multiply the first equation in ~\eqref{eq:Inhom_LGPDE} by $d\phi/dx$ and integrate, we obtain
[see, e.g., ~\cite{puri2009kinetics} for details]
\begin{eqnarray}
\frac{k_B}{2}\phi^2 + \frac{k_B}{12}\phi^4 + \frac{\sigma}{2}[d\phi/dx]^2 &=&
\frac{k_B}{2}(T_c^{CW} - T)\phi_b^2 + \frac{k_B}{12}\phi_b^4 \,,\nonumber \\
{\rm{whence}} \;\; f_{I,MF} = \int_{-\infty}^{\infty} dx [d\phi/dx]^2.
\label{eq:fIMF2}
\end{eqnarray}
Note that as $T$ approaches $T_c^{CW}$ from below, $\phi_b \to 0$, so $f_{I,MF}$ vanishes as $ \sim (T_c^{CW} - T)^{\mu_{MF}}$, with $\mu_{MF} = 2\nu_{MF} = 1$, where $\nu_{MF}$ is the exponent that characterises the divergence of $\xi_{MF}$ [see Eq.~(\ref{eq:LG_kink})]. To go beyond this mean-field treatment, we must include the thermal fluctuations that are neglected by mean-field theory and which modify the mean-field values of critical exponents for $d \leq 4$ [see, e.g., ~\cite{ma2018modern,kardar2007statistical,goldenfeld2018lectures,chaikin1995principles,plischke1994equilibrium,amit2005field,zinn2021quantum}]. 

In most of this paper, we concentrate on physical parameter ranges in which systems are far from critical points. Therefore, we continue to use the Landau-Ginzburg framework outlined above and PDEs such as Eq.~(\ref{eq:Inhom_LGPDE}) with no noise terms that account for thermal fluctuations. If turbulence is present in the flows we consider, then turbulent fluctuations overpower thermal fluctuations; the latter are normally not required in studies of turbulent flows.~\footnote{For studies that suggest the inclusion of  thermal noise in dissipation-range fluid turbulence see~\cite{betchov1977transition,bandak2022dissipation}.}  

\begin{figure}
  \centerline{\includegraphics[width=15cm]{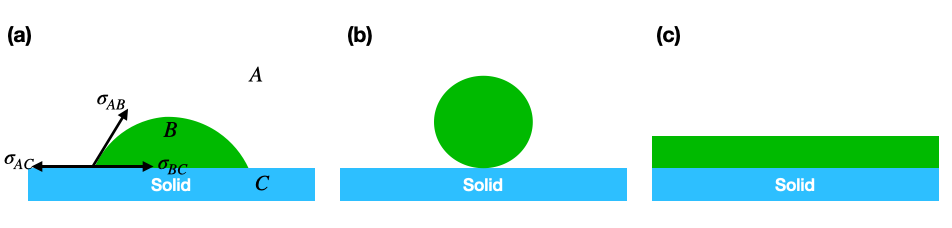}
  \put(-375, 30){\rm {$\theta$}}
  }
  \caption{Schematic diagram of a droplet on a solid substrate for (a) partial wetting, (b) complete drying, and (c) complete wetting.
  The white, green, and blue regions indicate the three phases A, B, and C (the last of which is a solid substrate); the surface tensions between the coexisting phases are $\sigma_{AB}$, $\sigma_{AC}$, and $\sigma_{BC}$; the contact angle $\theta$ is given by the 
  Young-Dupr\'e equation $\sigma_{AC}=\sigma_{BC}+\sigma_{AB} \cos(\theta)$; in (a) $0 < \theta < \pi$, (b) $\theta = \pi$, and (c) $\theta = 0$.}
\label{fig:wet}
\end{figure}

We can also use the above inhomogeneous mean-field theory for the following:
\begin{itemize}
    \item To study the wetting transition on an attractive substrate, which we illustrate by the schematic diagram in Fig.~\ref{fig:wet} [see, e.g., ~\cite{cahn1977critical,pandit1982surfaces,pandit1982systematics} for a detailed treatment of this problem].
    \item To study curvature corrections to the surface tension [see~\cite{fisher1984curvature} for details] at a curved interface; the dominant contribution at a curved interface yields Laplace's result [see, e.g., ~\cite{fisher1984curvature,rowlinson2013molecular}] for the pressure difference $\Delta P$ between the interior (say $A$) and exterior (say $B$) phases that are separated by a curved interface with surface tension $\sigma_{AB}$ and radius of curvature $R_{AB}$:
    \begin{equation}
        \Delta P = \frac{2 \sigma_{AB}}{R_{AB}}\,.
    \end{equation}
\end{itemize}

\subsection{Kinetic Ising Models}
\label{subsec:Kin_Ising}

We turn now to a brief description of some time-dependent phenomena in multiphase systems. For pedagogical reasons, we have used the Ising model to illustrate various aspects of the equilibrium statistical mechanics of systems with interfaces. We adopt a similar strategy here by beginning with the kinetic Ising model [see, e.g., ~\cite{kawasaki1970kinetic,puri2004kinetics,puri2009kinetics}], employing a mean-field approximation, taking a continuum limit, and obtaining, therefrom, time-dependent Ginzburg-Landau (TDGL) descriptions for the spatiotemporal evolution of the order parameter in two cases, viz., when it is (a) \textit{not conserved} and (b) \textit{conserved}. Case (b) yields the Cahn-Hilliard (CH) equation. Our discussion follows that of ~\cite{puri2009kinetics}, which the reader should consult for details. 

The Ising-model Hamiltonian~(\ref{eq:Ising_H}) contains only one component (say the $z$ component in spin space), so it does not have intrinsic dynamics. We assume instead that the spins in this model are in contact with a heat bath (provided, e.g., by coupling to phonons in a crystalline solid) that leads to stochastic flipping of these spins. To study time-dependent phenomena we use, therefore, master equations for the conditional probability $P(\{S_i^0\},0;\{S_i\},t)$
that the spin at site $i$ is in the state $S_i$ at time $t$, given that its $t=0$ state was 
$S_i^0$; henceforth, for notational convenience, we suppress the arguments $\{S_i^0\},0$. [The braces indicate that we consider all the $N$ spins in the $d-$dimensional hypercubic lattice.] We give below two such master equations:
\begin{itemize}
    \item for the Glauber model [see ~\cite{glauber1963time}], indicated by the subscript $G$ in Eq.~(\ref{eq:PG}]), in which the total magnetization $\sum_i S_i$ is \textit{not conserved} because we consider single-spin flips;
    \item for the Kawasaki model [see ~\cite{kawasaki1966diffusion,kawasaki1970kinetic}], indicated by the subscript $K$ in Eq.~(\ref{eq:PK}), in which the total magnetization is \textit{conserved} because we consider exchanges of nearest-neighbour spins. 
\end{itemize}

\subsubsection{Non-conserved order parameter: Time-dependent Ginzburg-Landau Theory}
\label{subsec:TDGL}

The master equation for the Glauber model is 
\begin{eqnarray}
    \frac{dP_G(\{S_i\},t)}{dt} &=& - \sum_{j=1}^N w_G(S_1, \ldots S_j, \ldots S_N|S_1, \ldots -S_j, \ldots S_N)P_G(\{S_i\},t)\nonumber \\
    &+& \sum_{j=1}^N w_G(S_1, \ldots -S_j, \ldots S_N|S_1, \ldots S_j, \ldots S_N)P_G(\{S'_i\},t) \,;
    \label{eq:PG}
\end{eqnarray}
on the right-hand side (RHS) the first and second terms arise, respectively, from the loss and gain of probability. The loss is associated with the spin flip $S_j \to -S_j$; the gain is because of the flip $S'_j \to -S'_j$, in a state $\{S'_i\}$ with $S'_i = S_i\;\; \forall \; i\neq j$ and $S'_j = -S_j$. The master equation~\eqref{eq:PG} has the long-time ($t \to \infty$) equilibrium solution
\begin{equation}
    P_{eq}(\{S_i\}) = \frac{1}{\mathcal{Z}}\exp\bigg[\frac{-\mathcal{H}}{k_BT}\bigg]
     \label{eq:PGeq}
\end{equation}
if the transition matrix $w_G(\{S_i\}|\{S'_i\})$ satisfies the condition of detailed balance [see, e.g., ~\cite{van1981stochastic,martinelli1999lectures,puri2009kinetics}]
\begin{equation}
    w_G(\{S_i\}|\{S'_i\}) P_{eq}(\{S_i\}) = w_G(\{S'_i\}|\{S_i\}) P_{eq}(\{S'_i\})\,.
    \label{eq:detbal}
\end{equation}
This condition does not specify $w_G(\{S_i\}|\{S'_i\})$ completely; one convenient choice is the Suzuki-Kubo form [see, e.g., ~\cite{suzuki1968dynamics,puri2009kinetics}]
\begin{equation}
    w_G(\{S_i\}|\{S'_i\}) =  \frac{1}{2\tau_G}\bigg[1-\tanh\bigg(\frac{\Delta\mathcal{H}}{2k_BT}\bigg)\bigg]\,,
    \label{eq:SK}
\end{equation}
where $\tau_G$ is the relaxation time scale and $\Delta\mathcal{H}$ is the change in the energy because of the spin flip. If we substitute Eq.~(\ref{eq:SK}) in the master equation~(\ref{eq:PG}) and then calculate the mean value of the magnetization at site $i$
\begin{equation}
    \langle S_i \rangle = \sum_{\{S_j\}} S_i P_G(\{S_j\},t)\,,
    \label{eq:GmeanSi}
\end{equation}
standard calculations yield [see, e.g., ~\cite{puri2009kinetics}]
\begin{equation}
    \tau_G\frac{d}{dt}\langle S_i \rangle = -\langle S_i \rangle +\bigg\langle \tanh\bigg[\frac{(h_i + J\sum_{j\in[nni]} S_j)}{(k_BT)}\bigg]\bigg\rangle \,.
    \label{eq:GdmeanSidt}
\end{equation}
This equation can be solved analytically only if $d=1$ and $h_i=0\;\; \forall i$ [see, e.g., ~\cite{glauber1963time,siggia1977pseudospin,pandit1981finite}]. For $d > 1$, we must either use approximations, such as mean-field theory, or finite-size calculations, numerical simulations, or renormalization groups [see, e.g., ~\cite{pandit1981finite,munkel1993large,goldenfeld2018lectures,kardar2007statistical,chaikin1995principles,ma2018modern,tauber2013critical}]. We follow ~\cite{puri2009kinetics} and use following mean-field approach, which builds on  the Curie-Weiss approximation~(\ref{eq:CW_Mi}) for the equilibrium magnetization $M_i$. A direct expansion of the second term on the RHS of Eq.~(\ref{eq:GdmeanSidt}) yields moments, of all orders, of the spins $S_i$. In the mean-field dynamical model we use the approximation
\begin{equation}
    \tau_G\frac{d}{dt}\langle S_i \rangle \simeq -\langle S_i \rangle + \tanh\bigg[\frac{(h_i + J\sum_{j\in[nni]} \langle S_j\rangle)}{(k_BT)}\bigg] \,,
    \label{eq:GdmeanSidtCW}
\end{equation}
whose steady-state solution yields the CW self-consistency equation for the mean-field magnetisation given in Eq.~(\ref{eq:CW_Mi}). If we now take the continuum limit~(\ref{eq:Cont_app}), we get
\begin{equation}
    \tau_G\frac{\partial \phi}{\partial t} = -\frac{\delta\mathcal{F}_{CW}}{\delta \phi} \,,
    \label{eq:dphidtCW}
\end{equation}
and, if we make a small-$\phi$ expansion (valid, strictly speaking, near the CW critical point), we obtain the \textit{time-dependent Ginzburg-Landau} (TDGL) equation
\begin{equation}
     \frac{\partial\phi}{\partial t} = -D\frac{\delta\mathcal{F}_{LG}}{\delta \phi} \,,
    \label{eq:TDGL}
\end{equation}
where we measure time in units of $\tau_G$ (so we set $\tau_G =1 $). The TDGL equation~\eqref{eq:TDGL} is often postulated phenomenologically and its RHS includes a Gaussian white-noise term $\theta(\mathbf{r},t)$ with zero-mean and a variance $2Dk_BT$ that satisfies the \textit{fluctuation-dissipation theorem}; this is referred to as Model A in the critical-dynamics literature [see, e.g., ~\cite{ma2018modern,tauber2013critical,hohenberg1977theory,puri2009kinetics}]. This noise term plays an important role in the calculation of critical exponents, in general, and the dynamic critical exponent $z$, in particular, which characterises the critical-point divergence of the correlation time $\tau_c \sim \xi^z$. At the simplest mean-field level, Model A yields $z^A_{MF} = 2$.

\subsubsection{Conserved order parameter: Cahn-Hilliard Theory}
\label{subsec:CH_TD}

The master equation for the Kawasaki model is [see~\cite{kawasaki1966diffusion,kawasaki1970kinetic,puri2009kinetics}] 
\begin{eqnarray}
   \frac{dP_K(\{S_i\},t)}{dt} &=& - \sum_{j=1}^N\sum_{k\in[nnj]} w_K(S_1, \ldots S_j,S_k \ldots S_N|S_1, \ldots S_k,S_j, \ldots S_N)P_K(\{S_i\},t)\nonumber \\
    + \sum_{j=1}^N\sum_{k\in[nnj]} &w_K&(S_1, \ldots S_k,S_j, \ldots S_N|S_1, \ldots S_j,S_k \ldots S_N)P_K(\{S'_i\},t) \,,
    \label{eq:PK} 
\end{eqnarray}
where the subscript $K$ stands for Kawasaki, we set $h=0$ in the Hamiltonian~(\ref{eq:Ising_H}), and we work in the ensemble with fixed total magnetization $M_{tot}=\sum_i S_i$. Note that Eq.~(\ref{eq:PK}) accounts for nearest-neighbour spin exchanges $S_j \leftrightarrow S_k$ that conserve $M_{tot}$. If we proceed as we did in the Glauber case, with the Suzuki-Kubo form~(\ref{eq:SK}) for the transition matrix $w_K$, we obtain [see, e.g., ~\cite{puri2009kinetics} for details]
\begin{eqnarray}
    \tau_K\frac{d}{dt}\langle S_i \rangle &=& -2d\langle S_i \rangle + \sum_{j\in[nni]} \langle S_j\rangle \nonumber \\
    +\sum_{k\in[nni]} \bigg\langle (1&-&S_iS_k)\tanh\bigg[\frac{J\big(\sum_{j\in[nni;\neq k]}S_j-\sum_{j\in[nni;\neq i]}S_j\big)}{(k_BT)}\bigg]\bigg\rangle \,.
    \label{eq:GdmeanSidt}
\end{eqnarray}
The dynamical analogue of the CW approximation, followed by a continuum approximation, and an expansion in powers of $\phi$, yields the Cahn-Hilliard (CH) equation [see ~\cite{cahn1958free,cahn1959free,cahn1961spinodal,puri2009kinetics}]
\begin{equation}
    \frac{\partial\phi}{\partial t} = D \nabla^2 \bigg[\frac{\delta\mathcal{F}_{LG}}{\delta \phi} \bigg] \,,
    \label{eq:CH}
\end{equation}
where we set $\tau_K =1 $; note that $\int d\mathbf{r} \phi$ is conserved by this CH equation~(\ref{eq:CH}), which is often postulated phenomenologically [see, e.g., ~\cite{ma2018modern,tauber2013critical,hohenberg1977theory,puri2009kinetics}] and which has been used extensively in studies of phase separation, nucleation, and spinodal decomposition. If the RHS of the CH equation~(\ref{eq:CH}) includes a $\phi$-conserving Gaussian white-noise term $\theta(\mathbf{r},t)$ with zero-mean and with a variance that satisfies the fluctuation-dissipation theorem, it is known as Model B or the Cahn-Hilliard-Cook equation. This noise term plays an important role in the calculation of critical exponents. For Model B, at the simplest level, Eq.~(\ref{eq:CH}) yields $z^B_{MF} = 4$, which should be contrasted with the Model-A result $z^A_{MF} = 2$.

Before we proceed to our discussion of the Cahn-Hilliard-Navier-Stokes (CHNS) PDEs, we summarise below the types of time-dependent phenomena that can be studied by using the TDGL and CH equations:
\begin{itemize}
    \item Space and time dependent correlation functions, e.g., 
    \begin{equation}
    C(\mathbf{r},t) \equiv \langle \phi(\mathbf{R},t_0) \phi(\mathbf{R}+\mathbf{r},t_0+t)\rangle_{t_0,\mathbf{R}}
    - \langle \phi(\mathbf{R},t_0)\rangle_{t_0,\mathbf{R}}\rangle^2_{t_0,\mathbf{R}}\,,
    \label{eq:Crt}
    \end{equation}
    which can be used to study critical dynamics [see, e.g., ~\cite{hohenberg1977theory,tauber2013critical}] and the kinetics of phase separation [see, e.g., ~\cite{puri2004kinetics,puri2009kinetics,bray2002theory}].
    \item Consider the schematic plot of $g(\phi)$, which appears in Eq.~(\ref{eq:LG_varFunc}) for $\mathcal{F}_{LG}$, in Fig.~\ref{fig:gphi_coex} (b) [we set $h=0$] and the associated schematic phase diagram shown in Fig.~\ref{fig:gphi_coex} (a). The early stages of phase separation can occur via \textit{nucleation}, or droplet-type fluctuations, and \textit{spinodal decomposition}, or small-amplitude long-wavelength fluctuations [see, e.g., ~\cite{oki1977separation,gunton1983p} and 
    Fig. 3 in ~\cite{gunton1983p}, which has been reproduced from ~\cite{oki1977separation}]. At the level of Landau-Ginzburg theory, the loci of points along which $g''(\phi) = 0$ is the \textit{spinodal curve} that provides a clear boundary between the regions with nucleation-dominated ($g''(\phi) > 0$) growth and spinodal decomposition ($g''(\phi) < 0$) in Fig.~\ref{fig:gphi_coex} (a). If we go beyond mean-field or Landau-Ginzburg theory and consider the early stages of phase separation, the sharp spinodal curve is replaced by a crossover regime across which the phase-separating system moves from nucleation-dominated to spinodal-decomposition-initiated growth [see, e.g.,~\cite{oki1977separation,gunton1983p}].
    \item Late stages of phase separation: Here, we must distinguish Lifshitz-Slyozov and Lifshitz-Allen-Cahn scaling laws,
    for conserved and non-conserved $\phi$, respectively; the former [latter] leads to domain growth in time $t$ that is characterised by the power-law growth of the length $\mathcal{L}(t) \sim t^{1/3}$ [$\mathcal{L}(t) \sim t^{1/2}$] as discussed, e.g., in ~\cite{puri2004kinetics,puri2009kinetics,bray2002theory,lifshitz1962kinetics,lifshitz1961kinetics,fogedby1988lifshitz}; hydrodynamical effects can modify the domain-growth power-law exponent [see, e.g.,~\cite{puri2004kinetics,puri2009kinetics,bray2002theory}].
\end{itemize}


\section{Cahn-Hilliard-Navier-Stokes Models}
\label{sec:Models}

Henceforth, we will not consider critical phenomena in the Cahn-Hilliard-Navier-Stokes (CHNS) system and its generalisations. We will concentrate on turbulence in these systems, well below the transition temperature $T_c$, and the effects it has on the suppression of phase separation (also called coarsening arrest) or droplet motion in turbulent binary- or ternary-fluid flows, the coalescence of droplets or lenses, and turbulence and self-propelled droplets in active-CHNS models. Therefore, it will be convenient to work at a fixed temperature below $T_c$ and use a Landau-Ginzburg free-energy functional in which parameters are scaled in such a way that the minima of $F_{LG}$ are at $\phi_b = \pm 1$ in the two-fluid case. For related derivations of the CHNS model and reviews of diffuse-interface models see, e.g., ~\cite{gurtin1996two} and ~\cite{anderson1998diffuse}. We define below the CHNS models that we consider. Subsections~\ref{subsec:chns2} and \ref{subsec:chns3} cover, respectively, binary- and ternary-fluid systems. In Subsection~\ref{subsubsec:chns3boussinesq} describes the Boussinesq approximation for the ternary-fluid case. We then turn to active models: In Subsection~\ref{subsec:ActiveH} we introduce the active Model H and in Subsection~\ref{subsec:genactivechns} a generalisation that allows us to study the self propulsion of an active droplet.

\subsection{Binary-fluid CHNS}
\label{subsec:chns2}

\begin{enumerate}
    \item 
For the binary-fluid case [see, e.g., ~\cite{guo2021diffuse,borcia2022phase}] we use the following free-energy functional for our CHNS equations:
    \begin{eqnarray}
    \mathcal F[\phi, \nabla \phi] = \int_{\Omega} d\Omega \left[\mathcal V(\phi) + \frac{3}{4} \sigma \epsilon |\nabla \phi|^2\right]\,,\label{eq:functional}
    \end{eqnarray}
    where $\mathcal V(\phi)$ is the potential, $\sigma$ is the bare interfacial tension, and $\epsilon$ is the width of the interface. The following two forms of $\mathcal V(\phi)$ are used in the CHNS literature: 
\begin{enumerate}
    \item Curie-Weiss-type potential:
    \begin{eqnarray}
        \mathcal V_{CW}(\phi) = \frac{a}{2} \left[(1+\phi)\ln(1+\phi) + (1-\phi)\ln(1-\phi)\right] - \frac{b}{2} \phi^2\,;
    \end{eqnarray}
    this potential has the virtue that $\phi \in [-1,1]$. [In some studies $\mathcal{V}_{CW}$ is called the Flory-Huggins logarithmic potential [see, e.g., ~\cite{giorgini2020weak}], as in the theory of polymer solutions [see, e.g., ~\cite{rubinstein2003polymer}].]
    \item Landau-Ginzburg-type $\phi^4$ potential:
    \begin{eqnarray}
        \mathcal V_{LG}(\phi) =  \frac{3}{16} \frac{\sigma}{\epsilon}(\phi^2-1)^2\,;
    \end{eqnarray}
    this parametrization has the virtue that the global minima of $\mathcal V_{LG}(\phi)$ are at $\phi_b = \pm 1$. However, now $\phi \in [-\infty,\infty]$. We use $\mathcal V_{LG}(\phi)$ in most of our CHNS studies.
    \end{enumerate}   
    \item If we allow the shear viscosity $\eta$ and the density $\rho$ to depend on $\phi$, so that the two coexisting phases have different viscosities ($\eta_1$ and $\eta_2$) and densities ($\rho_1$ and $\rho_2$), far away from interfaces, the incompressible CHNS equations can be written as follows:
    \begin{eqnarray}
\rho(\phi) (\partial_t {\bm u} + (\bm{u} \cdot \nabla) \bm{u}) &=& -\nabla P + \nabla \cdot \left[\eta(\phi) (\nabla \bm u + \nabla \bm u^{T})\right] \nonumber \\
&-& \phi \nabla \mu + \rho(\phi) \bm{g} - \alpha \bm u + \bm{f}^{ext}\,; \nonumber\\
\nabla \cdot \bm u &=& 0\,; \nonumber\\
\partial_t \phi + (\bm{u} \cdot \nabla) \phi &=& M \nabla^2\mu\,; \nonumber \\
\mu  &=& \frac {\delta\mathcal{F}}{\delta \phi}\,; \nonumber\\
\eta(\phi) &=& \eta_1\left(\frac{1+\phi}{2}\right) + \eta_2\left(\frac{1-\phi}{2}\right)\,; \nonumber\\
\rho(\phi) &=& \rho_1\left(\frac{1+\phi}{2}\right) + \rho_2\left(\frac{1-\phi}{2}\right)\,.
\label{eq:chnsphi}
\end{eqnarray}
Here, $\bm{g}$, $\bm{f}^{ext}$, and $\alpha$ are, respectively, the acceleration because of gravity,
an external forcing, and the coefficient of friction (that is required typically in two-dimensional
settings, e.g., to account for bottom friction).
\begin{table}
\centering
\begin{tabular}{c c}
 \hspace{0.5cm} {Dimensionless numbers}& \hspace{7cm}Formulae\\
\hline
Reynolds number& \hspace{6cm}$\textrm{Re} = U_0 L_0 / \nu$\\
\hline
Peclet number& \hspace{6cm}$\textrm{Pe} = U_0 L_0/\nu$\\
\hline
Bond number& \hspace{6cm}$\textrm{Bo} = \mathcal{A}\rho L_0^2 g/\sigma$\\
\hline
Ohnesorge number&\hspace{6cm} $\textrm{Oh} = \nu \sqrt{\frac{\rho}{\sigma L_0}}$\\
\hline
Capillary number& \hspace{6cm}$\textrm{Ca} = \rho \nu U_0/\sigma$\\
\hline
Weber number& \hspace{6cm}$\textrm{We} = \rho U_0^2 L_0/\sigma$\\
\hline
Cahn number&\hspace{6cm} $\textrm{Cn} = \epsilon/L_0 $\\
\hline
\end{tabular}
\caption{\label{tab:Dimensionless}
Important dimensionless numbers for the binary-fluid CHNS system}
\end{table}
\item Boussinesq approximation: If the density difference between the two phases is not too large, in the Navier-Stokes part of the above equations we can use the Boussinesq  approximation [see, e.g., ~\cite{lee2013numerical,celani2009phase,boffetta2010rayleigh,boffetta2010statistics,zanella2020two,khan2019simulation,shah2017artificial,forbes2022extended,huang2022linear}]. \textcolor{black}{Furthermore, this approximation holds only when the vorticity is not too strong; otherwise, non-Boussinesq effects, such as interfacial instabilities, can emerge even at low Atwood numbers [see, e.g., \cite{ramadugu2022surface}]. In this Boussinesq approximation the Navier-Stokes equations can be written as follows:}
\begin{eqnarray}
\partial_t {\bm u} + (\bm u \cdot \nabla) \bm u &=& - \frac{1}{\rho_0}\nabla P + \nu \nabla^2 \bm u- \frac{1}{\rho_0}(\phi \nabla \mu) + \frac{[\rho(\phi) - {\rho}_0]}{{\rho}_0} \bm{g} - \alpha \bm u\,,\nonumber\\
\nabla \cdot \bm u &=& 0\,,
\label{eq:chns_Boussinesq}
\end{eqnarray}
where 
the mean density is $\rho_0 = (\rho_1+\rho_2)/2$. This approximation neglects density differences except in the term with \textit{gravity}. We can write 
\begin{equation}
    \frac{(\rho(\phi) - \rho_0)}{\rho_0} = -\frac{\Delta \rho}{\rho_0} \phi \equiv -\mathcal A \phi\,,
    \label{eq:Atwood}
\end{equation}
with the Atwood number $\mathcal A \ll 1$.\\

In Table~\ref{tab:Dimensionless} we give the important dimensionless parameters that govern the states of the binary-fluid CHNS system. The larger the Reynolds number $\textrm{Re}$, the more the inertia-induced turbulence, which can lead, e.g., to coarsening arrest [Subsection~\ref{subsec:PhaseSepBinary}]. The Bond number $\textrm{Bo}$ plays a crucial role in gravity-driven flows that include the evolution of antibubbles [Subsection~\ref{subsec:antibubble}] or a bubble passing through a liquid-liquid interface [Subsection~\ref{subsec:bubblepass}]; the Ohnesorge number 
$\textrm{Oh}$ is important in the coalescence of droplets or liquid lenses [Subsection~\ref{subsec:coalescence}]; and the Weber ($\textrm{We}$) and Capillary ($\textrm{Ca}$) numbers affect interface-induced  low-$\textrm{Re}\,$turbulence [Subsection~\ref{subsec:interfaceturb}]. The Cahn number governs the thickness of the interface. Other dimensionless numbers must be introduced as we increase the complexity of the multi-phase system; e.g., in three-phase flows, density, surface-tension, and viscosity ratios are relevant [Subsection~\ref{subsec:PhaseSepTernary}], and the activity parameter is important in active CHNS systems [Subsection~\ref{subsec:activechnsturb} and \ref{subsec:dropprop}]. 
    
\end{enumerate}    
\section{Numerical Methods}
\label{sec:Numerics}
\textcolor{black}{Various numerical methods have been employed successfully to simulate multiphase flows using phase-field methods, in general, and the CHNS equations, in particular. To capture the underlying physics, it is crucial to resolve interfaces accurately and to track their evolution. Common numerical discretization techniques include the finite element method (FEM) [see, e.g., ~\cite{vey2007amdis,lowengrub2009phase,barrett1999finite,barrett2001fully,elliott1987numerical}], finite difference methods [see, e.g., ~\cite{teigen2011diffuse,furihata2001stable,kim2009numerical}], and the Lattice-Boltzmann method (LBM) [see, e.g., ~\cite{timm2016lattice,chen1998lattice,benzi1990two,benzi1992lattice,shan1993lattice,shan1994simulation,scarbolo2013turbulence}], each one of which offers distinct advantages that depend on both the complexity of the problem and the computational demands
of the numerical simulation. In this article, we adopt direct numerical simulations (DNSs) that are based on the pseudospectral approach [see, e.g., \cite{canuto1988ta}].} \\

Direct numerical simulations (DNSs) play an important role in obtaining solutions of the CHNS equations. Interfaces are diffuse in CHNS systems, so we do not have to impose boundary conditions on complicated interfaces that evolve in time. A DNS must, of course, accurately resolve all relevant scales of motion, given initial and boundary conditions. Such DNSs can achieve a high level of accuracy by retaining the governing equations in their complete (rather than reduced) forms. However, DNSs can be computationally expensive, particularly at high Reynolds numbers and for coupled hydrodynamical equations, such as the CHNS equations, which contain many nonlinear terms. In Subsections~\ref{subsec:pseudospec} and \ref{subsec:volpenal} we give, respectively, overviews of the pseudsoectral and volume-penalisation methods that we use in our direct numerical simulations of CHNS models.

\subsection{Pseudospectral Method}
\label{subsec:pseudospec}
The pseudospectral method, a widely used numerical technique for solving hydrodynamical partial differential equations (PDEs), was pioneered over $50$ years ago by~\cite{patterson1971spectral} and has been used, \textit{inter alia}, to study fluid turbulence [see, e.g., ~\cite{mcwilliams1984emergence,canuto1988ta,pandit2009statistical,celani2010turbulence,san2012high,buaria2022intermittency,buaria2023forecasting}], magnetohydrodynamic (MHD) turbulence [see, e.g., ~\cite{verma2004statistical,sahoo2011systematics,sahoo2020direct,basu1998multiscaling,dritschel2012two,gomez2005parallel,muller2000scaling,yadav2022statistical}], CHNS turbulence [see, e.g.,~\cite{pandit2017overview,pal2016binary,vela2021deformation,scarbolo2013turbulence,vela2022memoryless,perlekar2014spinodal,pal2022ephemeral}], the coalescence of droplets and liquid lenses [see, e.g., ~\cite{padhan2023unveiling,soligo2019coalescence}], elastic turbulence in polymer solutions [see, e.g., ~\cite{berti2008two,plan2017lyapunov,gupta2017melting,singh2024intermittency}], and active turbulence and active droplets [see, e.g., \cite{mukherjee2023intermittency,mukherjee2021anomalous,kiran2023irreversibility,puggioni2023flocking,rana2020coarsening,gibbon2023analytical,padhan2023activity,li2022dynamics,gao2017self,backofen2024nonequilibrium}].

We employ the Fourier pseudospectral method, which is known for its accuracy and efficiency in comparison to other numerical methods [see, e.g., ~\cite{orszag1975numerical,canuto1988ta}]; in its most common form, this method uses plane waves as basis functions. We solve the coupled CHNS equations in a periodic domain $\mathcal{D}\equiv l^d$, with $l$ the length of the side of (typically) $d-$dimensional hypercubic domain. To illustrate this method, we simplify Eq.~(\ref{eq:chnsphi}) by making the following two key assumptions: all fluids have identical densities (i.e., $\rho_1 = \rho_2 = \rho_0 = 1$) and identical viscosities (i.e., $\eta_1 = \eta_2 = \eta$). The equations are given below for 2D and 3D.

\subsubsection{Equations in 2D}
In 2D, we solve the CHNS equations in its vorticity-streamfunction form [see,e.g., ~\cite{boffetta2012two,padhan2023activity,pal2016binary}]:
\begin{eqnarray}
    \partial_t \omega + (\bm u \cdot \nabla) \omega &=& \nu \nabla^2 \omega -\alpha \omega - [\nabla \times (\phi \nabla \mu)]\cdot\hat{e_z} + f^\omega\,;\nonumber\\
    \nabla \cdot \bm u &=& 0\,; \quad \omega = (\nabla \times \bm u)\cdot \hat e_z\,;\nonumber\\
    \partial_t \phi + (\bm u \cdot \nabla) \phi &=& M \nabla^2 \mu\,;\;\;\;\; \mu  = -\frac{3}{2}\sigma \epsilon \nabla^2 {\phi} + \frac{3}{4}\frac{\sigma}{\epsilon} ({\phi^3} - {\phi})\,.\label{eq:chns_2d}
\end{eqnarray}

\subsubsection{Equations in 3D}
In 3D, we use
\begin{eqnarray}
    \partial_t {\bm u} + (\bm{u} \cdot \nabla) \bm{u} &=& -\nabla P + \nu \nabla^2\bm u - \phi \nabla \mu - \alpha \bm u + \bm{f}^{ext}\,; \nonumber\\
\nabla \cdot \bm u &=& 0\,; \nonumber\\
\partial_t \phi + (\bm{u} \cdot \nabla) \phi &=& M \nabla^2\mu\,; \;\;\;\; \mu  = -\frac{3}{2}\sigma \epsilon \nabla^2 {\phi} + \frac{3}{4}\frac{\sigma}{\epsilon} ({\phi^3} - {\phi})\,.
\label{eq:chns_3d}
\end{eqnarray}
Here $\nu = \eta / \rho_0$ is the kinematic viscosity. In most 3D applications, $\alpha = 0$.

Our representation of the phase fields, velocity fields, vorticity fields, and pressure fields utilizes the (truncated) Fourier projection onto a grid of $N^d$ points, which are expressed as follows (carets denote spatial Fourier transform):
\begin{eqnarray}
    \phi(\bm x, t) &=& \displaystyle\sum_{|k|<N} \hat{\phi}(\bm k, t) \exp(\dot{\iota} \bm k \cdot \bm x)\,;\\
    \bm{u}(\bm x, t) &=& \displaystyle\sum_{|k|<N} \hat{\bm u}(\bm k, t) \exp(\dot{\iota} \bm k \cdot \bm x)\,;\\
    \bm{\omega}(\bm x, t) &=& \displaystyle\sum_{|k|<N} \hat{\bm \omega}(\bm k, t) \exp(\dot{\iota} \bm k \cdot \bm x)\,;\\
    P(\bm x, t) &=& \displaystyle\sum_{|k|<N} \hat{P}(\bm k, t) \exp(\dot{\iota} \bm k \cdot \bm x)\,.
\end{eqnarray}
There are $N^d$ wavenumbers: $\bm{k} \equiv k_0 \bm{n} = k_0\sum_{i=1}^{d} n_i \hat{e}_i$, where $k_0 = \frac{2\pi}{l}$ is the lowest wavenumber and ${n_i}$ are integers with values ranging from $(-\frac{N}{2}+1)$ to $\frac{N}{2}$. In each direction, the largest wavenumber is $k_{max} = \frac{k_0 N}{2}$. [The spectral representation mentioned above aligns with representing the fields in physical space on an $N^d$ grid with uniform spacing $\Delta x = \frac{l}{N} = {\pi}/{k_{{max}}}$.] We write, below, the CHNS equations in terms of the above Fourier modes. 
\subsubsection{2D CHNS equations in Fourier space}
We write Eq.~(\ref{eq:chns_2d}) in Fourier space as follows:
\begin{eqnarray}
    \partial_t \hat{\omega}(\bm k, t) &=& -\widehat{(\bm u \cdot \nabla \omega)} (\bm k, t) - \nu k^2 \hat{\omega}(\bm k, t) - \dot{\iota} \bm{k}\times \widehat{(\phi \nabla \mu)}(\bm k, t) -\alpha \hat{\omega}(\bm k, t)\,;\label{eq:FFT_u}\\
    \dot{\iota}\bm k \cdot \hat{\bm u}(\bm k, t) &=& 0\,; \;\;\; \hat{\omega}(\bm k, t) = \dot{\iota}\bm k\times \hat{\bm u}(\bm k, t)\,;\label{eq:FFT_incom}\\
    \hat{\psi}(\bm k, t) &=& \frac{\hat{\omega}(\bm k, t)}{k^2}\,;\label{eq:FFT_stream}\\
    \partial_t \hat{\phi}(\bm k, t) &=& -\widehat{(\bm u \cdot \nabla \phi)}(\bm k, t) - M k^2 \hat{\mu}(\bm k, t)\,;\label{eq:FFT_phi}\\
    \mu (\bm k, t) &=& \frac{3}{2}\sigma \epsilon k^2 \hat{\phi}(\bm k, t) + \frac{3}{4}\frac{\sigma}{\epsilon} [\hat{\phi^3}(\bm k, t) - \hat{\phi}(\bm k, t)]\label{eq:FFT_mu}.
\end{eqnarray}

\subsubsection{3D CHNS equations in Fourier space}
\label{subsec:CHNSFourier}

We write Eq.~\eqref{eq:chns_3d} in Fourier space as follows:
\begin{eqnarray}
    \partial_t \hat{\bm u}(\bm k, t) &=& \mathbb{P}(\bm k)\cdot \left[-\widehat{(\bm u \cdot \nabla \bm u)} (\bm k, t) - \widehat{(\phi \nabla \mu)}(\bm k, t)\right] - \nu k^2 \hat{\bm u}(\bm k, t) -\alpha \hat{\bm u}(\bm k, t)\,;\label{eq:3D_FFT_u}\\
    \dot{\iota}\bm k \cdot \hat{\bm u}(\bm k, t) &=& 0\,;\label{eq:3D_FFT_incom}\\
    \partial_t \hat{\phi}(\bm k, t) &=& -\widehat{(\bm u \cdot \nabla \phi)}(\bm k, t) - M k^2 \hat{\mu}(\bm k, t)\,;\label{eq:3D_FFT_phi}\\
    \mu (\bm k, t) &=& \frac{3}{2}\sigma \epsilon k^2 \hat{\phi}(\bm k, t) + \frac{3}{4}\frac{\sigma}{\epsilon} [\hat{\phi^3}(\bm k, t) - \hat{\phi}(\bm k, t)]\,\label{eq:3D_FFT_mu};
\end{eqnarray}
$\mathbb{P}$ is the transverse projection operator with components
\begin{eqnarray}
    \mathbb{P}_{ij}(\bm k) = \left(\delta_{ij} - \frac{k_i k_j}{k^2}\right).
\end{eqnarray}
To perform Fourier transforms, we utilize the FFTW library [\cite{frigo1998fftw}], which employs the Cooley-Tukey fast-Fourier-transform algorithm [see ~\cite{cooley1965algorithm}]. The nonlinear terms such as $\bm u \cdot \nabla \bm u\,,\;$ $\bm u \cdot \nabla \phi\;$, and $\mu \nabla \phi$ lead to convolution sums that require $N^{2d}$ operations; and the cubic nonlinear term $\phi^3$ requires $N^{3d}$ operations in Fourier space. To bypass the high computational cost of these operations, we use the pseudospectral method in which we evaluate these nonlinear terms as follows: First, we transform the fields back to real space; and then we perform the multiplications in real space and transform them back to Fourier space again. This procedure improves greatly the computational efficiency of the pseudospectral method compared to the ordinary spectral methods. However, the Fourier pseudospectral method leads to aliasing errors while evaluating the nonlinear terms, which we remove by using the $1/2$-dealiasing scheme [see, e.g., \cite{orszag1971elimination,hou2007computing,patterson1971spectral,padhan2023activity}], because of the cubic nonlinearity. 

The computational cost of such a pseudospectral DNS [see, e.g., ~\cite{moin1998direct}] increases with the number of grid points, which must increase in turn to resolve small-scale structures. For statistically homogeneous and isotropic turbulence in 3D, the standard estimate (based on Kolmogorov 1941 phenomenology) is $N \sim Re^{9/4}$, where $Re$ is the Reynolds number. A DNS of turbulence in the CHNS PDEs the grid spacing $\Delta x$ must be small enough to capture both the (Kolmogorv) dissipation scale and the widths of interfaces.

For the time integration of Eqs.~(\ref{eq:FFT_u})-(\ref{eq:3D_FFT_mu}) we use a semi-implicit exponential time-differentiating Runge-Kutta2 (ETDRK2) approach, which treats the linear terms implicitly with their exact solutions as described in ~\cite{cox2002exponential}. This method, which combines the ETD method with the RK2 method, offers several advantages over other numerical techniques, including enhanced accuracy, efficiency, stability, and ease of implementation. We write the CHNS equations~\eqref{eq:FFT_u}), \eqref{eq:FFT_phi}), and ~\eqref{eq:FFT_mu} in the following general form:
\begin{eqnarray}
    \frac{dq_1(t)}{dt} &=& \lambda_1 q_1(t) + \mathcal G(q_1, q_2)\,;\label{eq:chns_gen1}\\ 
    \frac{dq_2(t)}{dt} &=& \lambda_2 q_2(t) + \mathcal H(q_1, q_2)\,;\label{eq:chns_gen2}
\end{eqnarray}

here,
\begin{eqnarray}
q_1(t) &\equiv& \hat{\omega}(\bm k, t)\,;\;\;\;\; \lambda_1 \equiv (-\nu k^2 - \alpha)\,; \nonumber \\
q_2(t) &\equiv& \hat{\phi}(\bm k, t)\,; \;\;\;\; \lambda_2 \equiv (-\frac{3}{2} M \sigma \epsilon k^4 + \frac{3}{4} \frac{\sigma}{\epsilon} M k^2)\,; \nonumber \\ 
\mathcal G(q_1, q_2) &\equiv& -\widehat{(\bm u \cdot \nabla \bm \omega)} (\bm k, t) + \dot{\iota} \bm{k}\times \widehat{(\mu \nabla \phi)}(\bm k, t)\,; \nonumber \\
\mathcal{H}(q_1, q_2) &\equiv& -\widehat{(\bm u \cdot \nabla \phi)}(\bm k, t) - \frac{3}{4} \frac{\sigma}{\epsilon}M k^2 \hat{\phi}(\bm k, t)\,.
\end{eqnarray}

In the ETDRK2 algorithm, the Eqs.~(\ref{eq:chns_gen1})-(\ref{eq:chns_gen2}) have the solution
\begin{eqnarray}
    a_1^n &=& q_1^n \exp(\lambda_1 \Delta t) + \mathcal{G}(q_1^n, q_2^n) \left(\frac{\exp(\lambda_1 \Delta t) - 1}{\lambda_1}\right)\,;\\
    a_2^n &=& q_2^n \exp(\lambda_2 \Delta t) + \mathcal{H}(q_1^n, q_2^n) \left(\frac{\exp(\lambda_2 \Delta t) - 1}{\lambda_2}\right)\,;\\
    q_1^{n+1} &=& a_1^{n} + \left[\mathcal{G}(a_1^n, a_2^n) - \mathcal{G}(q_1^n, q_2^n)\right] \left(\frac{\exp(\lambda_1 \Delta t) - 1 - \lambda_1 \Delta t}{\lambda_1^2 \Delta t}\right)\,;\\
    q_2^{n+1} &=& a_2^{n} + \left[\mathcal{H}(a_1^n, a_2^n) - \mathcal{H}(q_1^n, q_2^n)\right] \left(\frac{\exp(\lambda_2 \Delta t) - 1 - \lambda_2 \Delta t}{\lambda_2^2 \Delta t}\right)\,;
\end{eqnarray}
for simplicity, we use $q_1^{n+1} \equiv q_1(t + \Delta t)$ and $q_1^n \equiv q_1(t)$.

\subsection{Volume-penalized CHNS (VPCHNS)}
\label{subsec:volpenal}
\textcolor{black}{Fluid-structure interactions play a crucial role in multiphase flows, where solid boundaries or immersed objects influence phase separation, interfacial dynamics, and transport properties [see, e.g., ~\cite{hester2023fluid,huat2015fluid,hou2024advances,zheng2021fluid,pavuluri2023interplay,treeratanaphitak2023diffuse}].}
These interactions are challenging in computational fluid dynamics (CFD) because of the enforcement of boundary conditions at the fluid-solid interface [see, e.g., ~\cite{mokbel2018phase,pramanik2024computational}]. The numerical implementation of these boundary conditions is especially difficult when the immersed solid is in motion. The immersed boundary method (IBM) has been used widely in CFD to handle such interactions [see, e.g., \cite{mittal2005immersed, mittal2023origin}]. The volume penalization method (VPM) is a type of IBM that is simple and versatile in modelling objects in fluid flows; it is gaining in popularity, given its ease of implementation [see, e.g., ~\cite{mittal2023origin, morales2014simulation,kolomenskiy2009fourier,sinhababu2022pseudo,hester2021improving,kadoch2012volume,engels2016flusi,engels2022computational,puggioni2023flocking}]. Without imposing explicit boundary conditions, the VPM uses an extra force as a penalization term in the classical Navier-Stokes equations; the modified equation is known as the Brinkman model [see, e.g., \cite{angot1999penalization}]. The VPM approximates solid boundaries in a fluid by applying rapid linear damping within a fictitious solid region. This approach allows for the simulation of complex, moving objects within general numerical solvers without the need for specialized algorithms or boundary-conforming grids. The volume penalization method treats solids as porous media that are characterized by negligible permeability, resulting in the velocity of the adjacent fluid becoming zero at the fluid-solid interface. This method can be easily incorporated into a DNS solver that employs periodic boundary conditions. The VPM has been successfully implemented in the Navier-Stokes equations, passive scalar equations, magnetohydrodynamic (MHD) equations, Cahn-Hilliard equations, and the Toner-Tu-Swift-Hohenberg (TTSH) equations [see, e.g., ~\cite{kadoch2012volume,engels2016flusi,engels2022computational,morales2014simulation,sinhababu2021efficient,puggioni2023flocking}]. We extend the VPM method to the CHNS equations as follows (for illustration we use the 2D CHNS system):
\begin{eqnarray}
    \partial_t \omega + (\bm u \cdot \nabla) \omega &=& \nu \nabla^2 \omega -\alpha \omega - [\nabla \times (\phi \nabla \mu)]\cdot\hat{e_z} + f^\omega - \nabla \times \left[\frac{\chi}{\eta_p} \bm u\right]\,;\nonumber\\
    \nabla \cdot \bm u &=& 0\,; \quad \omega = (\nabla \times \bm u)\cdot \hat e_z\,;\nonumber\\
    \partial_t \phi + [(1-\chi)\bm u] \cdot \nabla \phi &=& \nabla \cdot ([M(1 - \chi) + M_p \chi] \nabla \mu)\,.\label{eq:vpchns}
\end{eqnarray}
Here, $\eta_p$ and $M_p$ are the penalization parameters (or permeabilities) associated with the velocity and $\phi$ fields, respectively. $\chi$ is the mask function to distinguish solid and fluid regions, which is defined as follows:
\begin{eqnarray}
    \chi(\bm x) &=& \begin{cases}
    1 &\text{for}\quad \bm x \in \Omega_s \,;\\
    0 &\text{for}\quad \bm x \in \Omega_f \,;
    \label{eq:mask}
    \end{cases}
\end{eqnarray}
$\Omega_s$ and $\Omega_f$ are the volumes representing solid and fluid regions. The term $[M(1 - \chi) + M_p \chi]$ takes into account the no-flux boundary conditions at the solid-fluid interface $\nabla \phi \cdot \hat{n}|_{\Omega_f} = 0$; $\hat{n}$ is the unit vector that is normal to the solid wall. 

\section{Mathematical Challenges}
\label{sec:Maths}

The Euler PDEs for an inviscid fluid [see, e.g., ~\cite{euler1761principia,gibbon2008three,darrigol2008newton}] predate their viscous version by 61 years. Given analytic initial data, solutions of the incompressible 2D Euler equation, do not exhibit a finite-time singularity (FTS); however, it is still not known if any solutions of the 3D Euler equations develop a
singularity in a finite time, if we start with analytic initial data [see, e.g., ~\cite{majda2002vorticity,bardos2007euler,gibbon2008three,drivas2023singularity}]; there has been some recent progress on a possible FTS for the 3D axisymmetric Euler equation [see \cite{luo2014potentially,kolluru2022insights,hertel2022cauchy}]. For the NS PDEs global regularity of solutions, with analytic initial data, can be proved in 2D but not in 3D [see, e.g., ~\cite{constantin1988navier,doering1995applied,galdi2000introduction,doering20093d,robinson2016three,lemarie2018navier,robinson2020navier}]. The proof of regularity (or lack thereof), for all time, of solutions of the 3D NS PDEs is one of the Clay Millennium Prize Problems (for domains without boundary)
[see, e.g., ~\cite{robinson2020navier}].

Since its introduction by ~\cite{cahn1958free}, the Cahn-Hilliard (CH) PDE has been used extensively in multi-phase statistical mechanics, nucleation, spinodal decomposition, and the late stages of phase separation [see, e.g., \cite{lothe1962reconsiderations,lifshitz1961kinetics,hohenberg1977theory,gunton1983p,bray2002theory,chaikin1995principles,puri2009kinetics,onuki2002phase,badalassi2003computation,anderson1998diffuse,perlekar2014spinodal,cahn1961spinodal,berti2005turbulence}]. The well-posedness of the CH PDE has been considered in several works, such as ~\cite{elliott1986cahn,elliott1991existence,liu2006regularity,dlotko1994global,miranville2019cahn,wu2021review}, to which we refer the reader.

The important questions for the CHNS PDEs concern the smoothness, or lack thereof, of the contours of $\phi$ within fluid interfaces and their interplay with the regularity of the solutions of the NS part of this system. We turn now to a brief overview of mathematical results for the regularity of 
solutions of the CHNS PDEs. 

Several results are available in two dimensions (2D); for the 2D CHNS system see, e.g., ~\cite{abels2009existence,abels2009longtime} and \cite{gal2010asymptotic}. In particular, ~\cite{gal2010asymptotic} have shown that, in a bounded domain or with periodic boundary conditions, and with suitably smooth initial data, this system possesses a global attractor $\mathfrak{A}$; the existence of an exponential attractor $\mathcal{E}$ has also been established, whence it is concluded [see ~\cite{gal2010asymptotic}] that the fractal dimension of $\mathfrak{A}$  is finite; this dimension can be estimated. Furthermore, it has been demonstrated that each trajectory converges to a single equilibrium.

As we have mentioned above, the regularity of solutions of the 3D NS equations continues to be a major open challenge  [see, e.g., ~\cite{constantin1988navier,doering1995applied,galdi2000introduction,doering20093d,robinson2016three,lemarie2018navier,robinson2020navier}]. The 
coupling to the 3D CH equations brings with it the difficulties associated with the smoothness of contours of $\phi$, which could, e.g., lead to finite-time singularities in arbitrarily large spatial derivatives of $\phi$. Such issues have been addressed by ~\cite{gibbon2016regularity,gibbon2018role}, who adopt an approach related to the 
Beale-Kato-Majda (BKM) strategy [see ~\cite{beale1984remarks}] for the  incompressible 3D Euler equations. In particular, BKM showed that,
if 
\begin{equation}
\mathcal{I}_\omega(T^*) = \int_{0}^{T^*}\|\bm{\omega}\|_{\infty}\,d\tau  \,,
\label{eq:BKM}
\end{equation}
where the subscript $\infty$ indicates the $L^{\infty}$-norm [also called the sup- or maximum norm],
is finite, then the solution is regular up until time $T^{*}$; by contrast, if $\mathcal{I}_\omega(T^*)$ becomes infinite at $T^{*}$, then the solution loses 
regularity or, in common parlance, it \textit{blows up}. To monitor such blow up, it suffices to consider  Eq.~(\ref{eq:BKM}) \textit{alone}; specifically, if it is finite, high-order spatial derivatives of the velocity cannot develop singularities. Similarly, for the 3D CHNS system it has been shown that 
[see ~\cite{gibbon2016regularity,gibbon2018role}] it is possible to obtain a BKM-type regularity criterion by using the 
the energy of the CHNS system
\begin{equation}
E(t) = \int_V [ \frac{\Lambda}{2} |\nabla \phi|^{2} + \frac{\Lambda}{4\epsilon^{2}} (\phi^{2} - 1)^{2} + \frac{1}{2} |\bm u|^{2} ]\,dV\,,
\label{eq:CHNSEt}
\end{equation}
which comprises a combination of squares of $L^{2}$-norms, whose
$L^{\infty}$ version, the \textit{maximal energy}, is 
\begin{equation}
E_{\infty}(t) = \shalf\Lambda\|\nabla\phi\|_{\infty}^{2} + \frac{\Lambda}{4\epsilon^{2}} \left(\|\phi\|_{\infty}^{2} -1\right)^{2} 
+ \shalf \|\bm u\|_{\infty}^{2}\,;
\label{eq:Einf}
\end{equation}
the coefficient $\Lambda = \frac{3}{4} \sigma \epsilon$.
~\cite{gibbon2016regularity,gibbon2018role} have proved that 
\begin{equation}
\mathcal{I}_E(T^*) = \int_{0}^{T^{*}} E_{\infty}(\tau)\,d\tau
\label{eq:EinfBKM}
\end{equation}
controls the regularity of solutions of the 3D CHNS PDEs exactly as $\mathcal{I}(T^*)_\omega$
does for the 3D Euler equations [see ~\cite{beale1984remarks}]. Although we have used the $LG$ form of Eq.~(\ref{eq:LG_varFunc}) here,
these results also follow for the $CW$ form of Eq.~(\ref{eq:CW_varFunc}).
Furthermore, ~\cite{gibbon2016regularity,gibbon2018role} have examined 
the time dependence of scaled $L^{2m}$-norms of the vorticity and gradients of $\phi$ and compared them with their 
counterparts for the NS and related equations [see ~\cite{donzis2013vorticity,gibbon2014regimes,gibbon2016depletion}].

Several other groups [see, e.g., ~\cite{giorgini2019uniqueness,giorgini2020weak,giorgini2021well}] have considered weak and strong solutions of the 
CHNS system (they refer to it as the Navier-Stokes-Cahn-Hilliard system), in bounded smooth domains in 2D and 3D, with a concentration-dependent viscosity and both $LG$ and $CW$ forms of Eq.~(\ref{eq:LG_varFunc}) and Eq.~(\ref{eq:CW_varFunc}) for the variational free energy.
In 2D and 3D, they have proved the existence of global weak solutions and the existence of strong solutions with bounded and strictly positive
density. Furthermore [see ~\cite{giorgini2019uniqueness,giorgini2020weak,giorgini2021well}], the strong solutions are local in time [in 3D] and global in time [in 2D].
\begin{figure}
  \centerline{\includegraphics[width=0.9\textwidth]{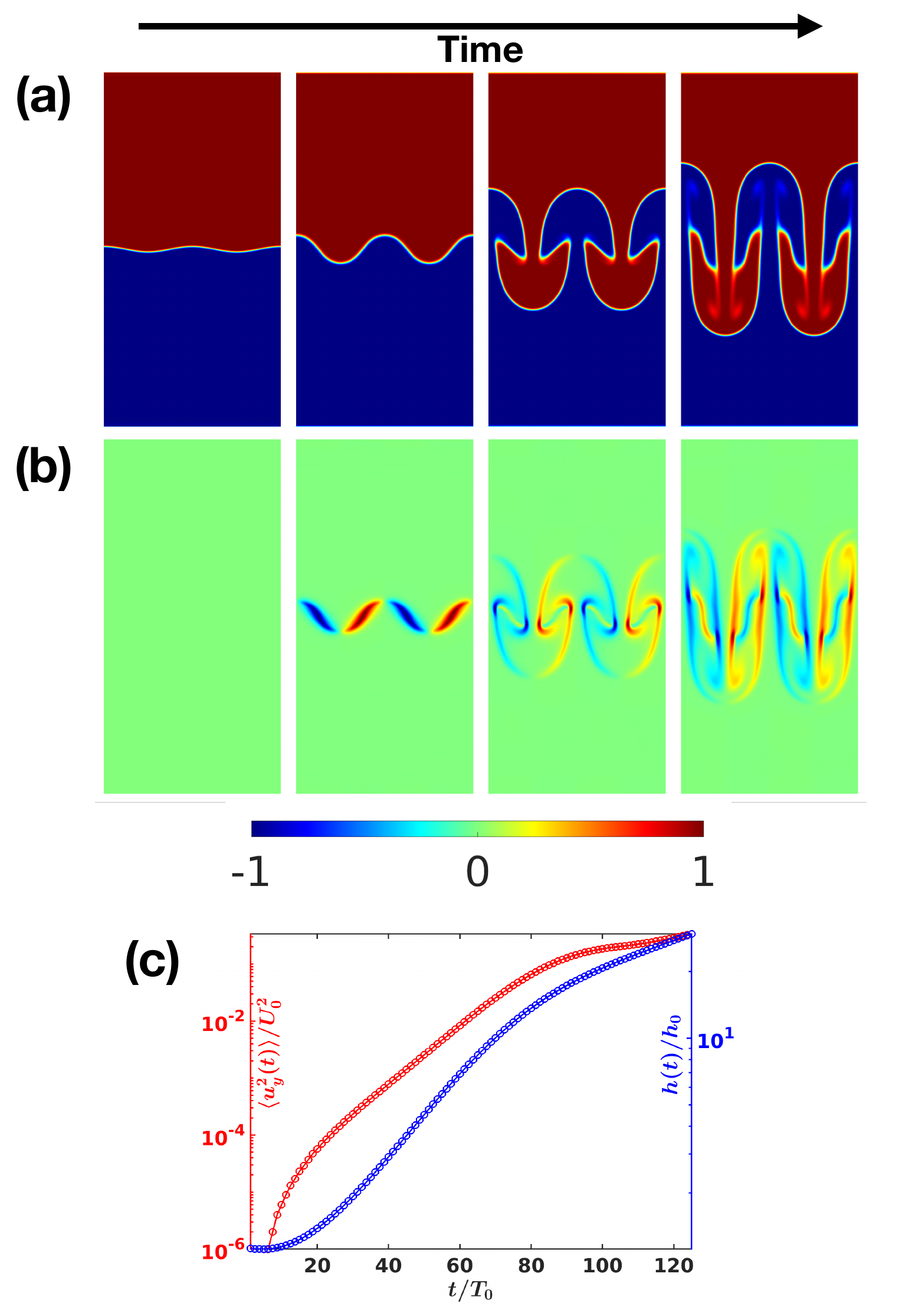}}
  \caption{Rayleigh-Taylor instability in the 2D CHNS system: Pseudocolour plots of (a) the $\phi$ field and (b) the corresponding vorticity field at different times (increasing from left to right). (c) Semi-log plots versus the scaled 
  time $T/T_0$ of $[\langle u_y^2(t)\rangle]/[U_0^2]$ (red) and the maximum scaled height $h(t)/h_0$ (blue) [see text]. The simulation box size is $(L_x, L_y) = (2\pi, 4\pi)$ with $512\times 1024$ grid points; $\nu = 0.01,\; \alpha = 0,\; \sigma = 0.1,\; g = 1,\; \mathcal A = 0.6,\; \rho_0 = 1,\; h_0 = 0.1$, and $m = 2$. The characteristic velocity and time scales are $U_0 = g h_0^2/\nu$ and $T_0 \equiv h_0/U_0 = \frac{\nu}{g h_0}$. The simulation box is periodic in the $x$-direction; boundary conditions in the $y$-direction are implemented by the volume-penalization method (see text).}
\label{fig:RT}
\end{figure}
\section{Illustrative CHNS-based DNS studies of Multiphase Flows}
\label{sec:validations}

We now present illustrative examples of the use of the CHNS framework for studies of a variety of multiphase flows. 
We begin with the Rayleigh-Taylor and Kelvin-Helmholtz instabilities in the binary-fluid CHNS system in Subsections~\ref{subsec:RTI}
and \ref{subsec:KHI}, respectively. Then, in Subsection~\ref{subsec:PhaseSepBinary} 
we describe how the CHNS
framework can be used to study phase separation, and its turbulence-induced suppression, in binary-fluid mixtures. Subsection~\ref{subsec:antibubble}
discusses the spatiotemporal evolution of antibubbles.
In Subsection~\ref{subsec:interfaceturb} we show how interfacial fluctuations 
in a binary mixture of immiscible fluids can lead to turbulence at low Reynolds numbers. 

\subsection{Rayleigh-Taylor instability: CHNS (2D)}
\label{subsec:RTI}

We illustrate how the CHNS framework can be used to simulate the gravity-induced Rayleigh-Taylor (RT) instability [see, e.g., ~\cite{drazin2002introduction,charru2011hydrodynamic,livescu2011direct}], which occurs when a dense fluid is positioned, initially, above a less dense one [see, e.g., \cite{tryggvason1988numerical,garoosi2022numerical,lee2013numerical,talat2018phase,khan2019simulation,khomenko2014rayleigh,youngs1992rayleigh,pandya2023interface,livescu2011direct,celani2009phase,gonzalez2019numerical,lherm2021rayleigh,kadau2004nanohydrodynamics,lherm2022rayleigh,young2001miscible,gibbon2016regularity}]. The CHNS framework has been shown to be useful in studying RT instabilities in immiscible fluids [see, e.g., \cite{lee2013numerical,celani2009phase,gibbon2016regularity}]. We demonstrate this by an illustrative pseudospectral DNS of the 2D CHNS equations, with the Boussinesq approximation [Eqs.~(\ref{eq:chns_Boussinesq})-(\ref{eq:Atwood})], in a 2D rectangular box $(L_x, L_y) = (2\pi, 4\pi)$; we incorporate impenetrable boundaries in the $y$-direction by using the volume-penalization method, with $6$ grid points on both the top and bottom boundaries for penalization. The initial conditions we use are as follows:
\begin{eqnarray}
    \phi(x, y, t = 0) &=& \tanh\left[\frac{y - L_y/2 - h_0 \cos(m x)}{\epsilon/2}\right]\,;\\ 
    \omega(x, y, t = 0) &=& 0\,;\nonumber
    \label{eq:init_RT}
\end{eqnarray}
here, $h_0$ is the amplitude of the perturbation we impose on top of a flat interface at $L_y/2$. In Figs.~\ref{fig:RT}(a)-(b), we present pseudocolor plots of $\phi$ and $\omega$; these show how the RT instability develops in time (which increases from left to right); the top phase (in red) has a higher density than the bottom phase (in blue); and the Atwood number is $\mathcal A = 0.6$. We can quantify the temporal growth of the RT instability by computing the normalised square of the vertical velocity
\begin{eqnarray}
    \frac{\langle u_y^2(t)\rangle}{U_0^2} \equiv \frac{2}{L_xL_y}\int_{0}^{L_x} dx \int_{\frac{L_y}{4}}^{\frac{3L_y}{4}} u_y^2(t) dy\,,
\end{eqnarray}
with $U_0 = g h_0^2/\nu$ the natural velocity scale for this problem, which is shown in the red semi-log
plot in Fig.~\ref{fig:RT}(c). The blue semi-log plot in Fig.~\ref{fig:RT}(c) shows the maximum scaled height $h(t)/h_0$ of the interface (the maximal deviation of the $\phi = 0$ contour line from the $t=0$ interface position $L_y/2$ ). The linear regions in both semi-log plots in Fig.~\ref{fig:RT}(c) are consistent with the exponential growth of the RT instability at early times.

For a CHNS-based study of the Rayleigh-Taylor (RT) instability in 3D, we refer the reader to \cite{gibbon2016regularity}. This DNS study was designed to obtain $E_{\infty}(t)$ [Eq.~\eqref{eq:Einf}] by obtaining large-$m$ estimates for the $L^m$ norm of the energy $E(t)$ [Eq.~\eqref{eq:CHNSEt}]; and the results, based on the BKM-type criterion~\eqref{eq:EinfBKM}, were consistent with no finite-time singularity, given the resolution of the DNS runs.

\subsection{Kelvin-Helmholtz instability: CHNS (2D)}
\label{subsec:KHI}
\begin{figure}
  \centerline{\includegraphics[width=\textwidth]{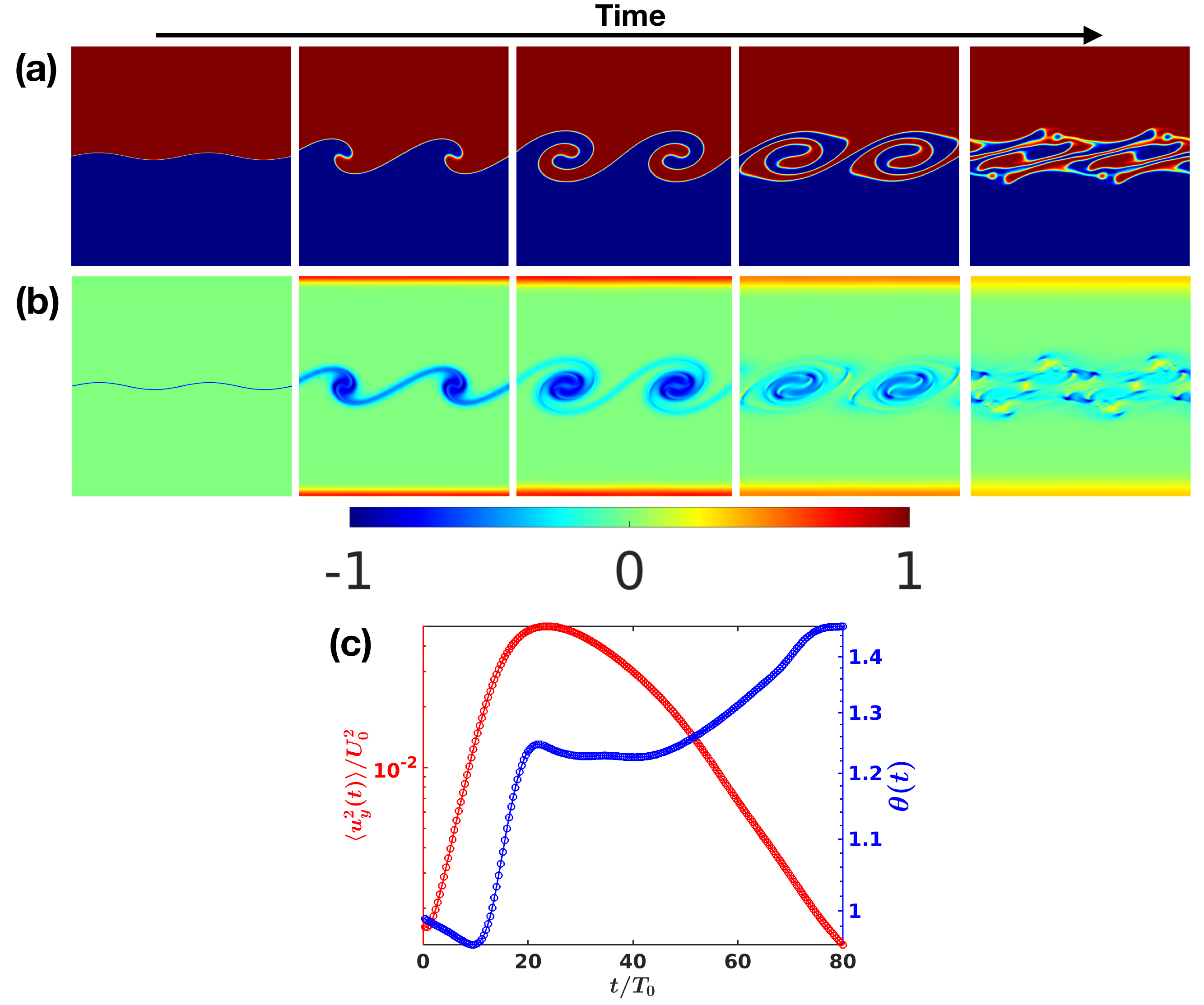}}
  \caption{Our DNS of the Kelvin-Helmholtz instability in the 2D CHNS system: Pseudocolour plots of (a) the $\phi$ field and (b) the corresponding vorticity field at different simulation times (increasing from left to right); the vorticity field is normalized by its absolute maximum value. (c) Plots versus the scaled time $t/T_0$ 
  of   $[\langle u_y^2(t) \rangle]/[U_0^2]$ (red semi-log plot), where $T_0 \equiv h_0/U_0$, and 
  $\theta(t)$ (blue linear plot) [see Eqs.~(\ref{eq:vertvel}) and (\ref{eq:momtheta})].  The simulation box size is $(L_x, L_y) = (2\pi, 2\pi)$ with $1024 \times 1024$ grid points; and $U_0 = 2, h_0 = 0.1, \nu = 0.01; \alpha = 0.01, \sigma = 0.05, g = 1, \mathcal A = 0.01, \rho_0 = 1$. The simulation box is periodic in the $x$-direction and we use volume penalization in the $y$-direction to incorporate solid boundaries; we incorporate impenetrable boundaries in the $y$-direction by using the volume-penalization method, with $6$ grid points on both the top and bottom boundaries for penalization.}
\label{fig:KH}
\end{figure}
If there is a significant difference in the velocities of two fluid layers, which are separated by an interface, then this interface becomes unstable because of the well-known Kelvin-Helmholtz (KH) instability [see, e.g., \cite{drazin2002introduction,charru2011hydrodynamic,budiana2020meshless,tian2016numerical,lee2015two,yilmaz2011numerical,delamere2021kelvin,cushman2005kelvin,hoshoudy2018kelvin,kumar2023kelvin,jia2023experimental,zhou2020phase,aref1981evolution}] that plays an important role in various marine, geophysical, solar, and astrophysical processes [see, e.g., ~\cite{gregg2018mixing,mishin2016kelvin}]. The CHNS framework has been used to study KH instabilities in binary and ternary fluids [see ~\cite{lee2015two}].  To illustrate how this is done, we perform a pseudospectral DNS of the CHNS equations within the Boussinesq approximation [Eqs.~(\ref{eq:chns_Boussinesq})-(\ref{eq:Atwood})] in a 2D box $(L_x, L_y) = (2\pi, 2\pi)$. Our simulation box is periodic in the $x$-direction and we incorporate solid boundaries in the $y$-direction by using the volume-penalization method, where we consider $6$ grid points on both the top and bottom sides for penalization. We use the following stably stratified initial conditions, so that there is no Rayleigh-Taylor (RT) instability: 
\begin{eqnarray}
    \phi(x, y, 0) &=& \tanh \left[\frac{y - L_y/2 - h_0 \sin(2x)}{\epsilon/2}\right]\,;\nonumber\\
    u_x(x, y, 0) &=& U_0  \phi(x, y, 0)\,;\\
    u_y(x, y, 0) &=& 0\,. \nonumber   
\end{eqnarray}
In Figs.~\ref{fig:KH}(a) and (b), we portray the $\phi$ and vorticity fields, respectively, via pseudocolor plots. 
We also quantify the temporal growth of the KH instability by computing the normalised square of the vertical velocity
\begin{eqnarray}
    \frac{\langle u_y^2(t) \rangle}{U_0^2} \equiv \frac{2}{L_xL_y} \int_{0}^{L_x}dx\int_{\frac{L_y}{4}}^{\frac{3L_y}{4}} u_y^2(t) dy\,,
    \label{eq:vertvel}
\end{eqnarray}
which we plot versus the scaled time $t/T_0$ in Fig.~\ref{fig:KH}(c) (red semi-log plot), where $T_0 \equiv h_0/U_0$; the limits on the integral over $y$ are chosen to exclude the effects of the boundaries. The initial increase in $\frac{\langle u_y^2(t) \rangle}{U_0^2}$ signals the KH instability; this ratio decreases eventually because, in our DNS, the shear is present only in the initial condition. We also calculate the momentum thickness~\cite{aref1981evolution}
\begin{eqnarray}
    \theta(t) = \int_{\frac{L_y}{4}}^{\frac{3L_y}{4}} dy \frac{\sqrt{\langle u_x^2(y, t)\rangle_x}}{\sqrt{\langle u_x^2(L_y/2, t)\rangle_x}}\,; 
    \label{eq:momtheta}
\end{eqnarray}
a linear plot of $\theta(t)$ versus the scaled time $t/T_0$ is shown in Fig.~\ref{fig:KH}(c)(blue line).

\subsection{Phase separation in the binary-fluid CHNS}
\label{subsec:PhaseSepBinary}

A homogeneous binary-fluid mixture spontaneously phase separates into domains of pure phases when the system is quenched from a high temperature to a low temperature, which lies below the critical temperature $T_c$ [see, e.g., ~\cite{bray2002theory,puri2009kinetics,chaikin1995principles}]; in equilibrium, pure phases are separated by an interface. If the order parameter is conserved, the early stages of phase separation proceed via nucleation or spinodal decomposition; thereafter, the system approaches the state with complete phase separation via the coarsening of domains of the pure phases [see Subsection~\ref{subsec:CH_TD}]. 

The CHNS framework has been used for studying the coarsening of phase-separating binary-fluid mixtures in both two and three dimensions [see, e.g., \cite{berti2005turbulence,perlekar2017two,wang2019progress}].  To study the coarsening process in a symmetric binary fluid mixture, we define the domain length scale ${L}(t)$ and integral length scale ${L}_I(t)$ in terms of the phase-field spectrum $S(k, t)$ and energy spectrum $E(k, t)$, respectively, as follows:
\begin{eqnarray}
    {L}(t) &=& 2\pi \frac{\displaystyle \sum_{k} k^{-1}S(k, t)}{\displaystyle \sum_{k} S(k, t)}\,;\;\;\;\;
    {L}_I(t) = 2\pi \frac{\displaystyle \sum_{k} k^{-1}E(k, t )}{\displaystyle \sum_{k} E(k, t)}\,; \nonumber \\
    S(k, t) &=& \displaystyle \sum_{k-1/2<k'<k+1/2} \left[\hat{\phi}(\mathbf{k}',t)\cdot \hat{\phi}(-\mathbf{k}',t)\right]\,; \nonumber \\
    E(k,t) &=& \frac{1}{2}\displaystyle \sum_{k-1/2<k'<k+1/2} \left[\hat{ \bm{u}}(\mathbf{k}',t)\cdot \hat{ \bm{u}}(-\mathbf{k}',t)\right].
    \label{eq:LTST}
\end{eqnarray}
The length scale $L(t)$ displays self-similar power-law growth with $L(t) \sim t^{\beta}$: (a) in the diffusion-dominated regime (governed by the Cahn-Hilliard equation) $\beta \simeq 1/3$, the well known Lifshitz-Slyozov exponent [see, e.g., ~\cite{lifshitz1961kinetics,puri2009kinetics,bray2002theory}]; (b) in the viscous-hydrodynamic regime, $\beta \simeq 1$ [see \cite{siggia1979late} and \cite{perlekar2014spinodal}]; and (c) in the inertia-dominated regime, $\beta \simeq 2/3$ [\cite{furukawa2000spinodal}]. 

\begin{figure}
  \centerline{
  \includegraphics[width=\textwidth]{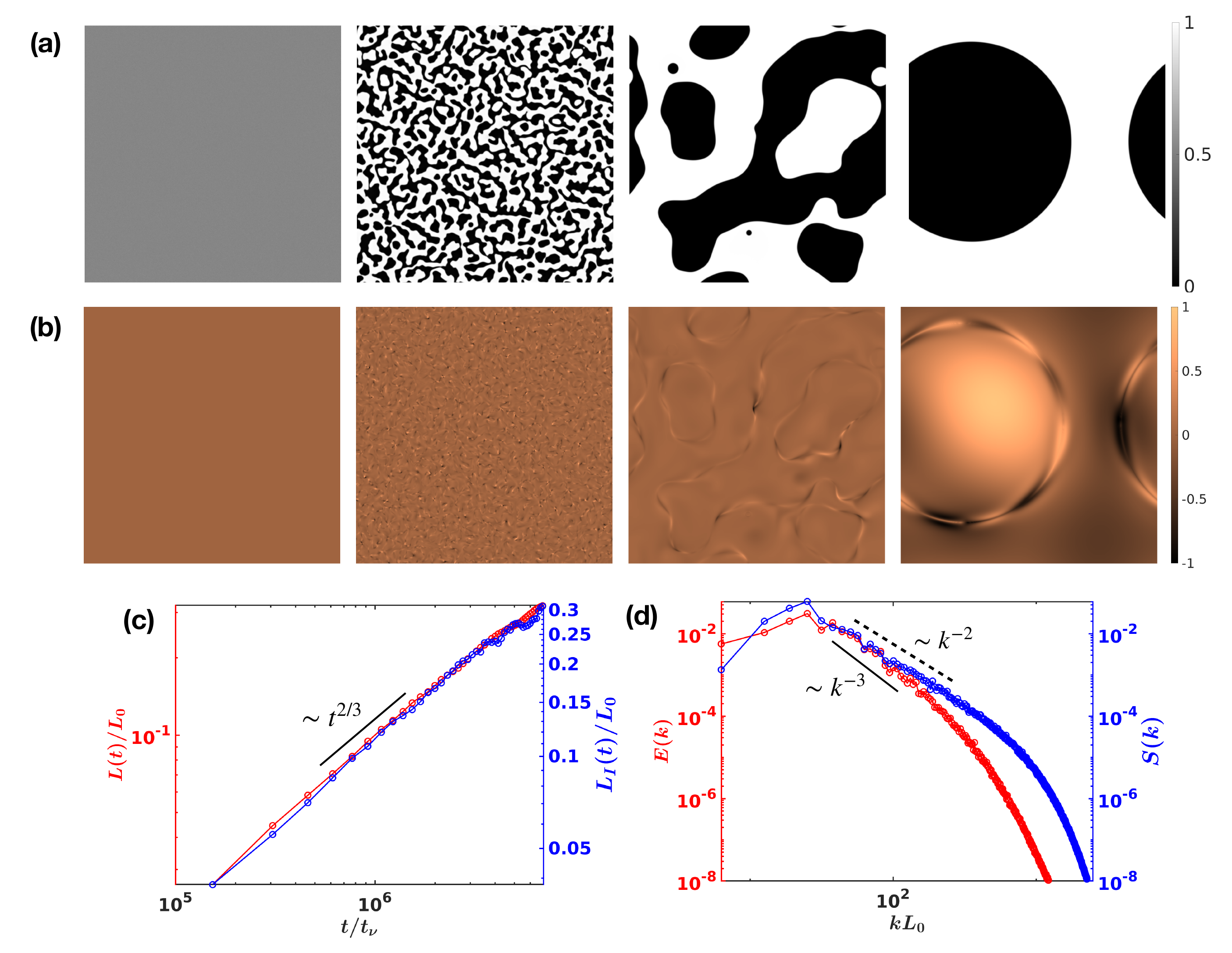}
  \put(-350,300){\rm {$\xrightarrow[\textit{\bf{time}}]{\hspace*{8cm}}$}}
  }
  \caption{Coarsening in a binary-fluid mixture from our DNS of the 2D CHNS equations. (a) Pseudo-gray-scale plots of the phase-field $\phi(\bm x, t)$ at simulation times $t = 0, 0.24, 5.52 (\equiv t_{*}), 8.88$, increasing from left to right; (b) pseudo-colour plots of the corresponding vorticity fields $\omega(\bm x, t)$ (normalized by their absolute maximum for ease of visualization). (c) Log-log plots versus $t/t_\nu$, with $t_{\nu} = \nu^3/\sigma^2$, of the scaled lengths $L(t)/L_0$ (red) and ${L}_I(t)/L_0$ (blue), with $L_0 = 2\pi$ the side of the simulation domain. (d) The energy spectrum (red line) $E(k,t=t_*)$ and the phase-field spectrum (blue line) $S(k, t=t_*)$ show power-law behaviour with scaling exponents $\sim k^{-3}$ and $\sim k^{-2}$, respectively.}
\label{fig:binary_emulsion}
\end{figure}

We carry out an illustrative pseudospectral DNS of the CHNS(2D) equations to obtain the scaling exponent $\beta \simeq 2/3$ in the inertial-hydrodynamic regime. We illustrate domain coarsening by the pseudocolor plots of $\phi$ and $\omega$ in Figs.~\ref{fig:binary_emulsion}(a)-(b). The power-law growth of $L(t)$ and ${L}_I(t)$, in the low-viscosity inertial regime, is shown in Fig.~\ref{fig:binary_emulsion}(c); the growth of both these lengths is
consistent with the same power-law exponent $\beta \simeq 2/3$. We show the energy spectrum and phase-field spectrum in Fig.~\ref{fig:binary_emulsion}(d); $E(k) \sim k^{-3}$ (consistent with the scaling exponent obtained in 2D turbulence \cite{boffetta2012two})and $S(k) \sim k^{-2}$ (consistent with the results obtained by ~\cite{furukawa2000spinodal}). We solve the 2D CHNS equations via a pseudospectral DNS in a doubly periodic box of size $2\pi \times 2\pi$ with grid points $1024 \times 1024$. The relevant dimensionless numbers are the Reynolds number $Re = u_{rms} L_I/\nu$ and the Ohnesorge number $Oh = \nu (\frac{\rho}{\sigma L_I})^{1/2}$. The simulation parameters are $\nu = 5\times10^{-3}, \sigma = 0.4, \rho = 1, \epsilon = 0.018$, and $M = 10^{-4}$. The dimensionless numbers, calculated at $t = t^*$ (see Fig.~\ref{fig:binary_emulsion}), are $Re = 24$ and $Oh = 0.01$.

Fluid turbulence can mix immiscible fluids and lead to the arrest of phase separation, which is also referred to as coarsening arrest. The CHNS PDEs have been used to study such turbulence-induced coarsening arrest in both 2D and 3D [see, e.g., \cite{perlekar2017two} and \cite{perlekar2014spinodal}]. If parameters are chosen such that the Cahn-Hilliard equations lead to complete phase separation of the two fluids in the binary mixture, then the inclusion of turbulence, obtained by forcing the coupled CHNS equations, suppresses this phase separation. This can be characterised by using the lengths and spectra that we have defined in Eq.~(\ref{eq:LTST}). In the absence of turbulence, the spectrum $S(k,t)$ displays an inverse cascade to small wavenumbers $k$ as time $t$ increases; this leads to the power-law growth of ${L}_I(t)$ as $t\to\infty$, with the Lifshitz-Slyozov, viscous-hydrodynamic, and inertial-hydrodynamic exponents mentioned above. This inverse cascade and the associated power-law growth of ${L}_I(t)$ are arrested by turbulence. In particular, it has been shown [see the pseudospectral DNS of \cite{perlekar2017two} and the Lattice-Boltzmann study of \cite{perlekar2014spinodal} in 2D and 3D, respectively] that ${L}_I(t) \sim L_H$ as $t \to \infty$,
where the Hinze length scale $L_H \sim \varepsilon_{inj}^{-2/5}\sigma^{3/5}$, with $\varepsilon_{inj}$ the energy injection into the NS part of the CHNS system. We will show in Subsection~\ref{subsec:activechnsturb} that active turbulence, which occurs in the active-CHNS system, also leads to qualitatively similar coarsening arrest.

\subsection{Spatiotemporal evolution of antibubbles}
\label{subsec:antibubble}

A shell of a low-density fluid inside a high-density fluid is known as an antibubble. It seems to have been described first by Hughes and Hughes [see, e.g., ~\cite{hughes1932liquid,dorbolo2006vita,kim2006antibubbles,kalelkar2017inveterate,vitry2019controlling,zia2022advances,pal2022ephemeral}]. An antibubble has two surfaces, and a certain volume of fluid is trapped between these two surfaces. Clearly, an antibubble is unstable under gravity: if the fluid in its inner core
is denser than that in the outer core, the antibubble rises; and the fluid in the shell forms a bulb at the top while its 
bottom thins until the shell collapses completely.  For experimental investigations of the dynamics of an
antibubble see, e.g., ~\cite{dorbolo2006vita,kim2006antibubbles,vitry2019controlling,zia2022advances}; and for theoretical work 
consult, e.g., ~\cite{scheid2012antibubble,zou2013collapse,sob2015theory,pal2022ephemeral}. It is important to note that the inherent instability of antibubbles
makes experimental studies challenging; in some cases surfactant molecules have to be introduced to obtain some stabilization 
of an antibubble. Furthermore, antibubbles have several applications that include sonoporation [see, e.g., ~\cite{kotopoulis2014sonoporation}],
drug delivery [see, e.g., ~\cite{johansen2015nonlinear,kotopoulis2015acoustically}], and active leakage
detection [see, e.g., ~\cite{johansen2015ultrasonically}]. 

It was recognised by \cite{pal2022ephemeral} that the CHNS system provides an ideal theoretical framework for the elucidation of the spatiotemporal evoluition of 
antibubbles. The initial condition for $\phi$ is an annulus in 2D or a shell in 3D of the lighter fluid (shown in red) with the heavier fluid (shown in blue) both inside and outside the shell. The initial outer and inner radii of the shell are, respectively, $R_0$ and $R_1$, and $h_0$ is the initial thickness of the antibubble shell, so it is natural to define the Bond number Bo with $L_0$ replaced by $h_0$
 [see Table~\ref{tab:Dimensionless}]. The spatiotemporal evolution of such an 
antibubble in the 2D CHNS system is shown via pseudocolor plots of $\phi$ in Figs.~\ref{fig:antibubble_small} (a) and (c) at representative times;
the associated evolution of the vorticity $\omega$ is given,
respectively, in Figs.~\ref{fig:antibubble_small} (b) and (d). [These figures have been provided very kindly by Nairita Pal.] From Figs.~\ref{fig:antibubble_small} (a) and (c) we see that gravity induces a thinning of the bottom of the antibubble while forming a slight dome on the top of it; eventually this makes the antibubble rupture. A similar rupture also occurs for antibubbles in the 3D CHNS system [see \cite{pal2022ephemeral} for details]. These DNSs yield the scaled rupture time $\tau_1/\tau_g$, with $\tau_g \equiv \sqrt{R_0/\mathcal{A}g}$, the velocity $v_{rim}$ of the retracting rim of the collapsing antibubble, and they find that $v_{rim} \sim \sqrt{\sigma}$, in agreement with the theoretical estimate of ~\cite{sob2015theory} and experiments of \cite{scheid2012antibubble} and \cite{zou2013collapse}. Furthermore, this CHNS study [\cite{pal2022ephemeral}] obtains the dependence of $\tau_1/\tau_g$ on the Bond number Bo and shows, by obtaining the energy [$E(k)$] and concentration [$S(k)$] spectra, that the rupture of the antibubble leads to some turbulence. Finally  \cite{pal2022ephemeral} provide a comparison of the spatiotemporal evolution of antibubbles by using both 
the CHNS framework and a volume-of-fluid DNS.

\begin{figure}
  \centerline{
  \includegraphics[width=\textwidth]{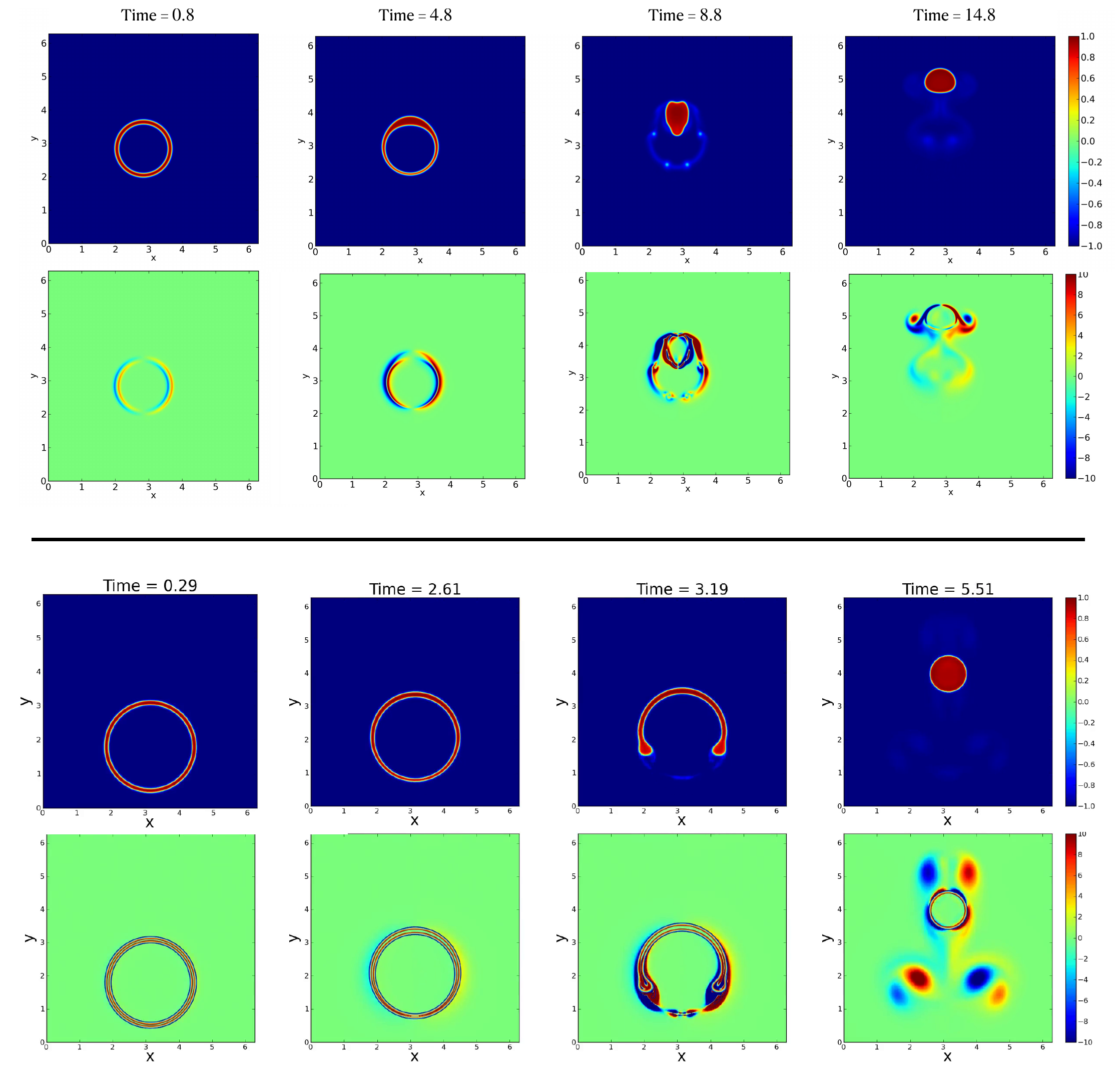}
  \put(-390,350){\rm {\bf(a)}}
  \put(-390,270){\rm {\bf(b)}}
  \put(-390,160){\rm {\bf(c)}}
  \put(-390,75){\rm {\bf(d)}}
  }
  \caption{Illustrative pseudocolor plots of (a) and (c) the field $\phi$ and the associated plots of the vorticity $\omega$ showing the spatiotemporal evolution of antibubbles for low [(a) and (b)] and high [(c) and (d)] values of $R_0$ and $R_1$, the initial outer and inner radii of the shell
  of the antibubble [see text and \cite{pal2022ephemeral} for details]. We thank Nairita Pal for these figures.}
\label{fig:antibubble_small}
\end{figure}
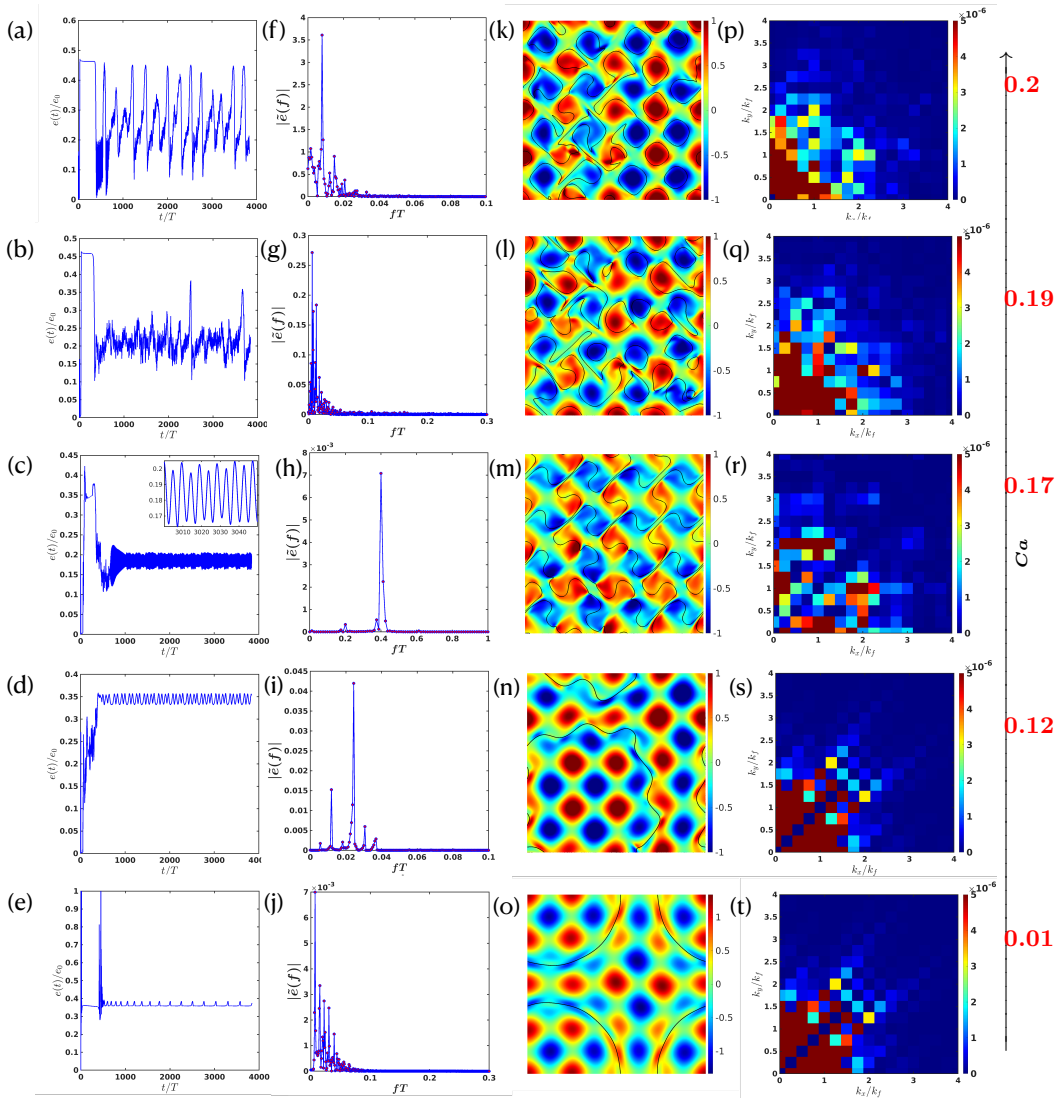
\begin{figure}
\hspace{-0.75cm}
{\includegraphics[width=0.99\textwidth]{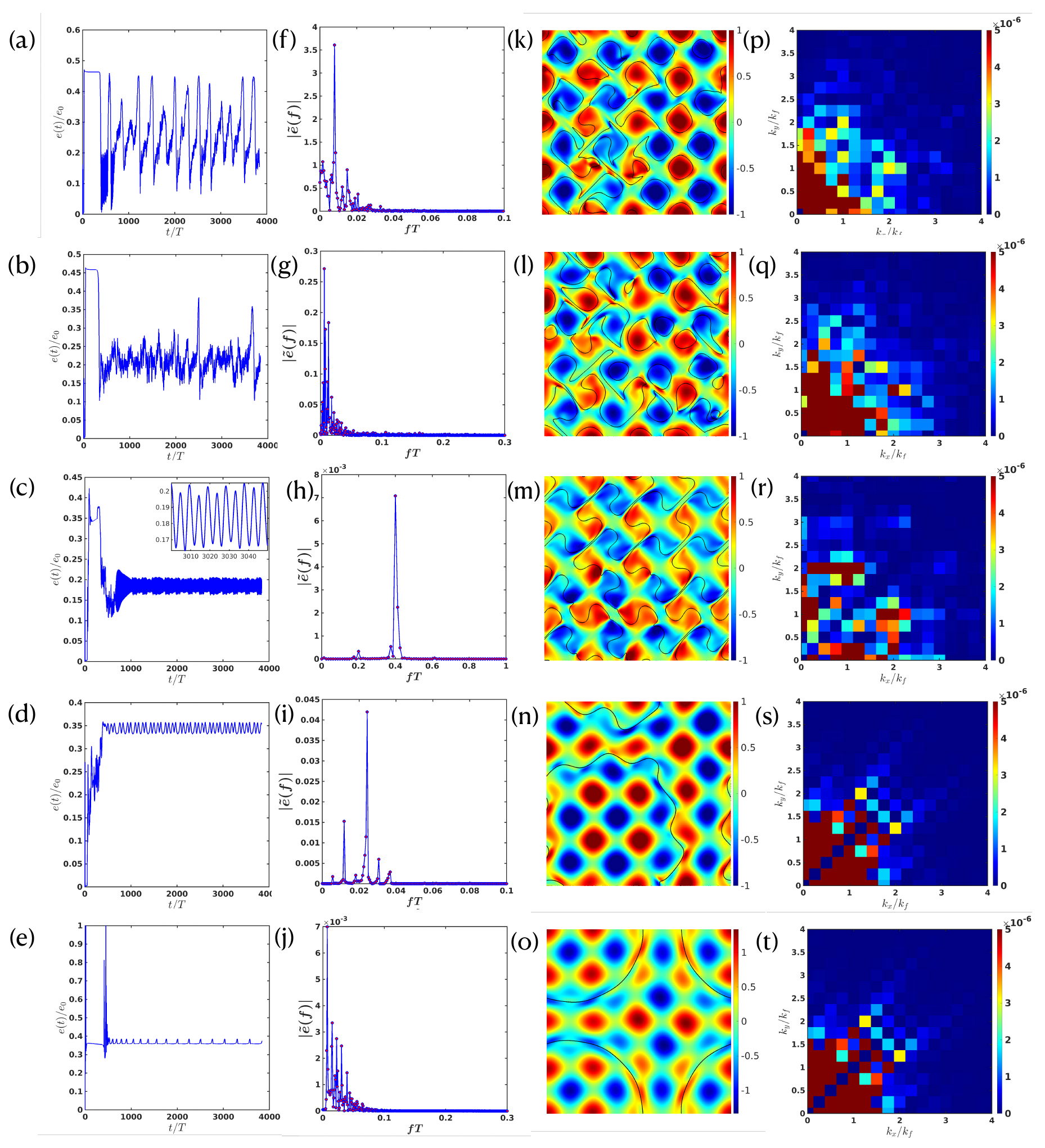}
\put(-5, 20){\rm {\begin{turn}{90}$\xrightarrow[{\bm{Ca}}]{\hspace*{13cm}}$\end{turn}}}
\put(-2, 60){\rm {\bf \textcolor{red}{0.01}}}
\put(-2, 140){\rm {\bf \textcolor{red}{0.12}}}
\put(-2, 230){\rm {\bf \textcolor{red}{0.17}}}
\put(-2, 300){\rm {\bf \textcolor{red}{0.19}}}
\put(-2, 380){\rm {\bf \textcolor{red}{0.2}}}
}
\caption{\label{fig:chns_melting} (a)-(e) The scaled total kinetic energy $e(t)/e_0$ versus time $t$ [here, $e_0 = U^2$ and $U = f_0/(\nu k_f^2)$]; (f)-(j) the corresponding power spectra $\tilde{e}(f)$ of $e(t)/e_0$. Pseudocolor plots, at a representative time, of (k)-(o) the  vorticity $\omega$ overlaid with the $\phi = 0$ contour (black lines) and (p)-(t) the energy spectra $\mathcal{E}(k_x, k_y)$ [see \cite{padhan2024interface} for details]; the capillary number $Ca$ increases from $0.01$ in the bottom row to $0.2$ in the top row.}
\end{figure}
\subsection{Low-Reynolds-Number Interface-induced turbulence in the CHNS system}
\label{subsec:interfaceturb}

Emergent turbulence-type states have been found at low Reynolds numbers in a variety of systems, for instance, in fluids with polymer additives [see, e.g., \cite{groisman2000elastic,majumdar2011universality,gupta2017melting,benzi2018polymers,steinberg2021elastic,singh2024intermittency}] and in dense bacterial suspensions [see, e.g., \cite{wensink2012meso,bratanov2015new,Dunkel_2013,linkmann2019phase,alert2021active,mukherjee2023intermittency,kiran2023irreversibility}]. In the former, this turbulence is driven by an increase in the Weissenberg number, whereas in the latter, it is obtained by increasing the activity of the bacterial suspension. Recently~\cite{padhan2024interface} have demonstrated that low-Re turbulence occurs in the 2D CHNS system if we increase the Weber number We or the Capillary number Ca [i.e., we decrease the interfacial tension [see Table~\ref{tab:Dimensionless}]] and hence enhance interfacial fluctuations.   

To obtain such interface-induced turbulence,  \cite{padhan2024interface} begin with a periodic arrangement of vortices and anti-vortices, which is referred to as a vortex crystal or a cellular flow. Such cellular flows, imposed by a spatially periodic forcing with an amplitude $f_0$ and wavenumber $k_f$, can be disordered via turbulence as shown for 2D fluid turbulence by \cite{perlekar2010turbulence} and for 2D fluids with polymer additives by \cite{gupta2017melting} and \cite{plan2017lyapunov}. For the 2D CHNS system, we follow the discussion of \cite{padhan2024interface} and examine the dependence of the statistically steady state of the system as we increase the Capillary number Ca. The natural length, time, and velocity scales are, respectively, $k_f^{-1}$, $\nu k_f/f_0$, and $U \equiv f_0/(\nu k_f^2)$.

In Figs.~\ref{fig:chns_melting}(a)-(e) we plot the scaled total kinetic energy $e(t)/e_0$ versus time $t$, with $e_0 = U^2$; Figs.~\ref{fig:chns_melting}(b)-(f) display the corresponding power spectra $|\tilde{e}(f)|$ of $e(t)/e_0$.  Figures~\ref{fig:chns_melting}(c)-(g) depict pseudocolor plots of the vorticity $\omega$ overlaid with the $\phi = 0$ contour (black lines) at a representative time. In Figs.~\ref{fig:chns_melting}(d)-(h) we present pseudocolor plots of the energy spectra $\mathcal{E}(k_x, k_y)$ at a representative time; these spectra are not averaged over wave-number shells because the underlying crystalline state is not isotropic. The first two columns of these figures help us to distinguish between states that are temporally periodic [Ca$=0.17$] from those that are chaotic [Ca$=0.19$]; the third and fourth columns aid the identification of the spatial order of the state. If the state is periodic in space, $\mathcal{E}(k_x, k_y)$ shows Bragg peaks in the reciprocal lattice of the vortex crystal [e.g., the strong red peaks in Fig.~\ref{fig:chns_melting} (h)], where we use the standard terminology of crystal physics. 
We find a rich variety of states: STPO, spatially and temporally periodic (i.e., a spatiotemporal crystal of the type discussed in ~\cite{perlekar2010turbulence,gupta2017melting}); STPOG, is like STPO, but with a grain boundary (G) separating two crystalline parts; STC denotes spatiotemporal chaos. Here, TPO denotes temporally periodic oscillations. For similar states in studies of turbulence-induced melting of a vortex crystals see~\cite{perlekar2010turbulence} for fluids and \cite{gupta2017melting} for fluids with polymer additives. In the latter case, the melting of the vortex crystal can be induced by elastic turbulence at low $Re$; this is akin to the low-$Re$ interface-fluctuation-induced melting we discuss here. In particular, the study of ~\cite{gupta2017melting} has noted that the boundary between spatially and temporally periodic states and ones that exhibit STC is non-monotonic, i.e., there is re-entrance from one state into another and back. Interface-induced turbulence in the 2D CHNS system also displays such re-entrance, as a function of $Ca$; 
\cite{padhan2024interface} have found the following re-entrant sequence of states: chaotic$\to$temporally periodic$\to$chaotic.
\section{Beyond Binary-fluid CHNS: Challenging Problems}
\label{sec:beyond}

\textcolor{black}{We now discuss some illustrative examples that go beyond the use of the binary-fluid CHNS PDEs that we have considered so far.
In particular, we consider three-fluid flows, at low and high Reynolds numbers, and the active-CHNS PDEs that have been used, \textit{inter alia}, to model scalar active turbulence. Subsection~\ref{subsec:chns3} outlines the CHNS framework for a ternary-fluid mixture; we give the Boussinesq approximation
for this case in Subsection~\ref{subsubsec:chns3boussinesq}. In Subsection~\ref{subsec:PhaseSepTernary} we describe how the CHNS
framework can be used to study phase separation, and its turbulence-induced suppression, in ternary-fluid mixtures.
Subsection~\ref{subsec:dropturb} contains an examination of the spatiotemporal evolution of droplets and compound droplets in turbulent flows. Subsection~\ref{subsec:bubblepass} discusses the passage of a bubble of one phase through
the interface between two other fluid phases. In Subsection~\ref{subsec:coalescence} we show how the CHNS framework allow us to study
the coalescence of liquid lenses and droplets quantitatively. Subsection~\ref{subsec:ActiveH} introduces the active-CHNS model (also called active Model H).
Subsections~\ref{subsec:activechnsturb} and \ref{subsec:dropprop} are devoted, respectively,
to turbulence in the active CHNS system and activity-induced droplet propulsion in the generalised active CHNS model~~\eqref{eq:phi}-\eqref{eq:Spsi}.}

\subsection{Ternary-fluid CHNS (CHNS3)}
\label{subsec:chns3}
\begin{figure}
  \centerline{\includegraphics[width=8cm]{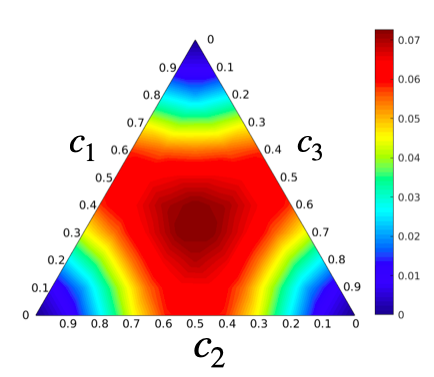}}
  \caption{Pseudocolor plot of $F(\{c_i\})$ projected onto a Gibbs triangle for the CHNS3 model~(\ref{eq:fe_ternary}). The three vertices yield the three minima of  $F(\{c_i\})$: the top vertex is $(c_1, c_2, c_3) = (1, 0, 0)$, the left vertex is $(c_1, c_2, c_3) = (0, 1, 0)$, and the right vertex is $(c_1, c_2, c_3) = (0, 0, 1)$.}
\label{fig:fe_ternary}
\end{figure}
The CHNS3 model for a ternary-fluid mixture uses the following variational free energy in the domain $\Omega$ [see, e.g., ~\cite{Boyer_2006,boyer2010cahn,kim2007phase,shek2022spontaneous,padhan2023unveiling}]:
    \begin{eqnarray}
    \mathcal{F}(\{c_i, \nabla c_i\}) = \mathlarger \int\displaylimits_{\Omega} d\Omega \left[\frac{12}{\epsilon}F(\{c_i\}) + \frac{3\epsilon}{8} \sum_{i=1}^{3}\gamma_i(\nabla c_{i})^2\right]\,,
    \label{eq:fe_ternary}
    \end{eqnarray}
where  the concentration fields $c_i (i = 1, 2, 3)$  are conserved order parameters that satisfy the constraint $\sum_{i=1}^{3} c_i = 1$, $\epsilon$ is the thickness of the interface, the variational bulk free energy
\begin{eqnarray}
    F(\{c_i\}) = \displaystyle\sum_{i=1}^{3} \gamma_i c_{i}^2(1-c_{i})^2\,, 
\end{eqnarray}
and the gradient terms give the surface-tension penalties for interfaces, with 
\begin{eqnarray}
    \sigma_{ij} = (\gamma_i + \gamma_j) / 2
\end{eqnarray}
 the bare surface (or interfacial) tension for the interface between the phases $i$ and $j$; the equilibrium values of $c_i$ follow from the global minima of $F(\{c_i\})$. In Fig.~\ref{fig:fe_ternary}, we show a pseudocolor plot of $F(\{c_i\})$ projected onto a Gibbs triangle [see, e.g., ~\cite{kim2007phase}]. The equilibrium chemical potential of fluid-$i$ is $\mu_i = \frac{\delta \mathcal F}{\delta c_i} + \beta (\{c_i\})$, with $\beta (\{c_i\})$ the Lagrange multiplier that ensures $\displaystyle\sum_{i=1}^3 c_i = 1$, whence we get 
\begin{eqnarray}
\mu_i &=& -\frac{3}{4}\epsilon \gamma_i \nabla^2 c_i + \frac{12}{\epsilon} [\gamma_i  c_i(1-c_i)(1-2c_i) \nonumber \\
&-& \frac{6\gamma_1 \gamma_2 \gamma_3 (c_1 c_2 c_3)}{\gamma_1 \gamma_2 + \gamma_1 \gamma_3 + \gamma_2 \gamma_3}]\;.
\label{eq:mu}
\end{eqnarray}

The incompressible CHNS3 equations can be written as follows [see, e.g., ~\cite{kim2005phase,Gyula_2016,boyer2010cahn,padhan2023unveiling}]:
\begin{eqnarray}
\rho(\{c_i\}) (\partial_t {\bm u} + (\bm{u} \cdot \nabla) \bm{u}) &=& -\nabla P + \nabla \cdot \left[\eta(\{c_i\}) (\nabla \bm u + \nabla \bm u^{T})\right] \nonumber \\
&-& \displaystyle\sum_{i=1}^{3}c_i \nabla \mu_i + \rho(\{c_i\}) \bm{g} - \alpha \bm u + \bm{f}^{ext}\,; \nonumber\\
\nabla \cdot \bm u &=& 0\,; \nonumber\\
 \partial_t{c_{i}} + (\bm u.{\nabla})c_{i} &=& \frac{M}{\gamma_{i}}{\nabla}^2 \mu_{i} , \;\; i = 1\; \rm{or}\; 2 \,; \nonumber\\
\eta(\{c_i\}) &=& \displaystyle\sum_{i=1}^{3}\eta_i c_i\,; \nonumber\\
\rho(\{c_i\}) &=& \displaystyle\sum_{i=1}^{3}\rho_i c_i\,.
\label{eq:3DCHNSC}
\end{eqnarray}
Here, $\bm{g}$, $\bm{f}^{ext}$, and $\alpha$ are, respectively, the acceleration because of gravity,
an external forcing, and the coefficient of friction. $\eta_i$ and $\rho_i$ are the viscosity and density of fluid $i$, respectively. The CHNS3 model becomes the binary-fluid CHNS model for $(c_1, c_2, c_3) = (c, 0, 0)$.

\subsubsection{Boussinesq Approximation} 
\label{subsubsec:chns3boussinesq}
If we use the Boussinesq approximation, the CHNS3 equations can be written as follows:
\begin{eqnarray}
\partial_t {\bm u} + (\bm{u} \cdot \bm{\nabla}) \bm{u} &=& - \frac{1}{\rho_0}\bm{\nabla} P + \nu \nabla^2 {\bm u}- \frac{1}{\rho_0}\sum_{i=1}^{3}(c_i \bm{\nabla} \mu_i) + \frac{[\rho(\{c_i\}) - {\rho}_0]}{{\rho}_0}\bf g - \alpha \bm{u},\\
\nabla \cdot \bm u &=& 0.
\label{eq:CHNS3A}
\end{eqnarray}
We write the density in the form:
\begin{eqnarray}
    \rho(\{c_i\}) &=& \sum\limits_{i=1}^{3} \rho_i c_i \nonumber\\
                  &=& \rho_1 c_1 + \rho_2 c_2 + \rho_3 (1-c_1-c_2) \nonumber\\
                  &=& \rho_3 + (\rho_1 - \rho_3) c_1 + (\rho_2 - \rho_3) c_2\,;
                  \label{eq:CHNS3B}
\end{eqnarray}
 we use $\rho_0 = \rho_3$, so
 \begin{eqnarray}
     \frac{[\rho(\{c_i\}) - {\rho}_0]}{{\rho}_0} &=& (\frac{\rho_1-\rho_3}{\rho_0}) c_1 + (\frac{\rho_2-\rho_3}{\rho_0}) c_2 \nonumber\\
     &=& \mathcal A_1 c_1 + \mathcal A_2 c_2\,,
     \label{eq:CHNS3C}
 \end{eqnarray}
where $\mathcal A_1$ and $\mathcal A_2$ are the Atwood numbers and $\mathcal A_1, A_2 \ll 1$.
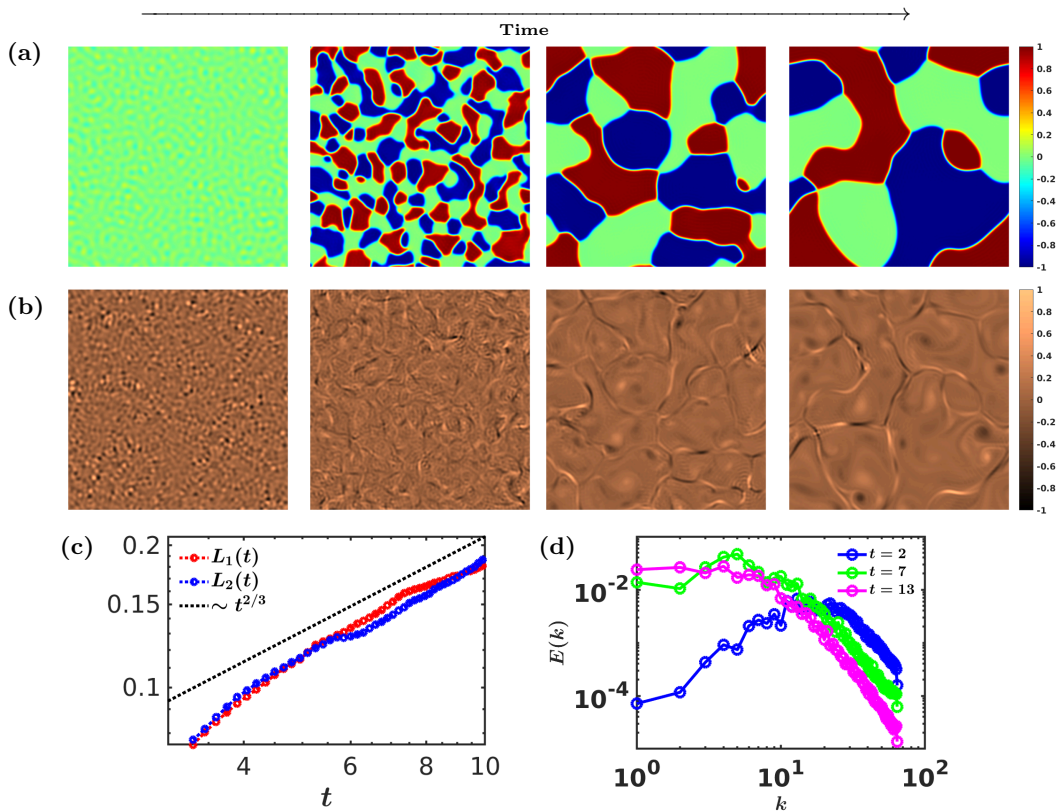
\begin{figure}
  \centerline{
  \includegraphics[width=\textwidth]{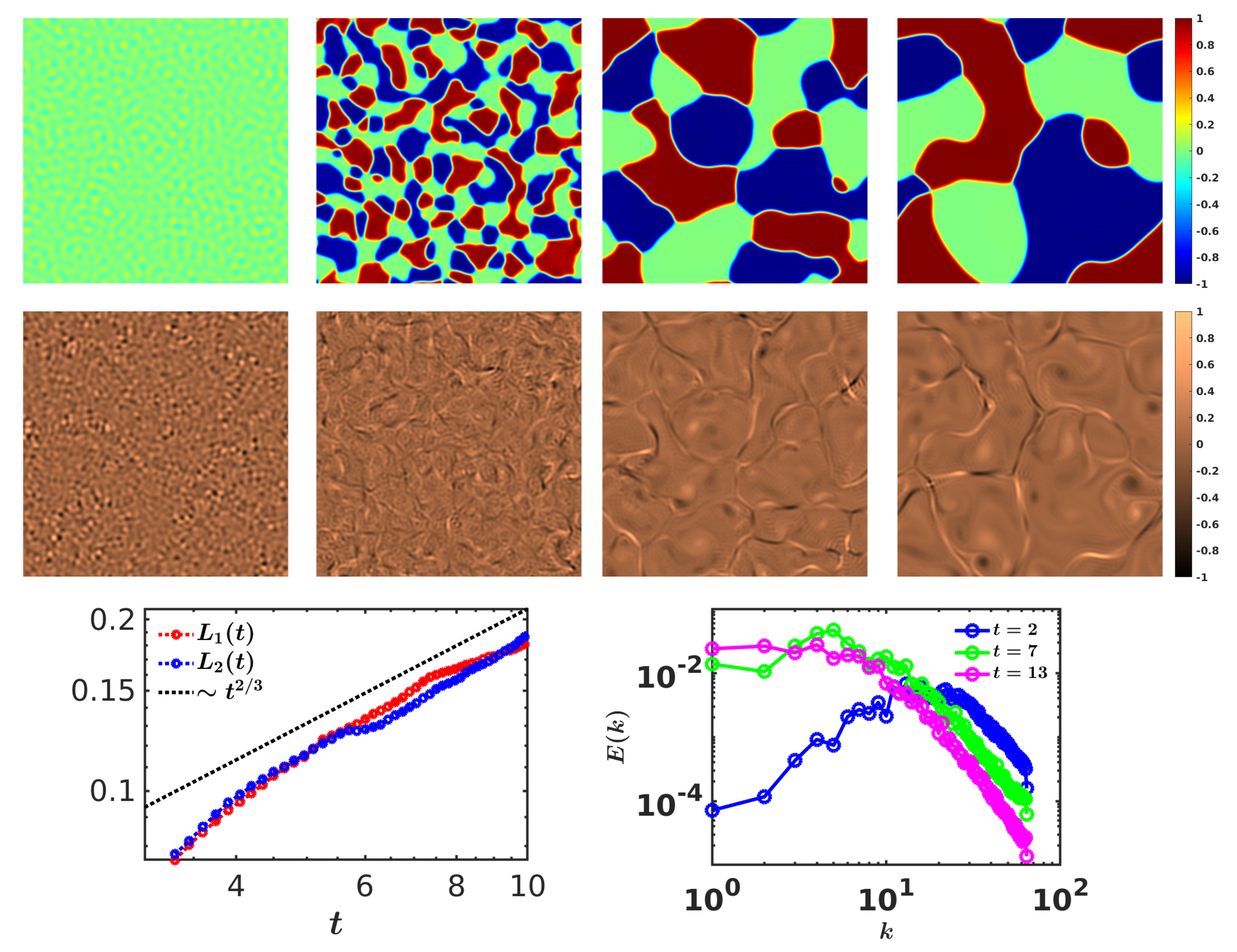}
  \put(-350,300){\rm {$\xrightarrow[\textit{\bf{Time}}]{\hspace*{10cm}}$}}
  \put(-400,285){\rm {\bf(a)}}
  \put(-400,190){\rm {\bf(b)}}
  \put(-380,100){\rm {\bf(c)}}
  \put(-200,100){\rm {\bf(d)}}
  }
  \caption{Coarsening in a ternary-fluid mixture from our DNS of the 2D CHNS3 equations. (a) Pseudocolour plots of the phase-fields $c_2 - c_1$ at simulation times $t = 0.8, 2, 7, 13$, increasing from left to right; (b) pseudo-colour plots of the corresponding vorticity fields $\omega(\bm x, t)$ (normalized by their absolute maximum for ease of visualization). (c) Log-log plots versus $t$ of the scaled lengths $L_1(t)/L_0$ and ${L}_2(t)/L_0$, with $L_0 = 2\pi$ the side of the simulation domain. (d) The energy spectrum $E(k,t)$ at simulation times $t = 2, 7, 13$. Simulation parameters: viscosity $\nu = 10^{-3}$, grid points $256 \times 256$, surface tension coefficients $(\sigma_{12}, \sigma_{23}, \sigma_{13}) = (1, 1, 1)$.}
\label{fig:spinodal_ternary}
\end{figure}
\subsection{Phase separation in the ternary-fluid CHNS3}
\label{subsec:PhaseSepTernary}
We turn now to the phase separation of ternary-fluid mixtures that occurs in a variety of settings [see, e.g., ~\cite{huang1995phase,C4SM02726D,wang2019progress,D2SM00413E}].
In Fig.~\ref{fig:spinodal_ternary}, we show illustrative results for coarsening in such a mixture, from our DNS of the 2D CHNS3 equations, which we visualise via 
pseudocolour plots of the difference of the phase-fields $(c_2 - c_1)$ [Fig.~\ref{fig:spinodal_ternary} (a)] and the corresponding vorticity fields $\omega(\bm x, t)$, normalized by their absolute maximum [Fig.~\ref{fig:spinodal_ternary} (b)], at simulation times $t = 0.8, 2, 7, 13$, which increase from left to right.  We define the following scaled lengths [cf. Eq.~\eqref{eq:LTST} for the two-fluid CHNS]:
\begin{equation}
    L_1(t) = 2\pi \tfrac{\displaystyle \sum_k S_1(k, t)}{\displaystyle \sum_k k S_1(k, t)}\,;\qquad 
    L_2(t) = 2\pi \tfrac{\displaystyle \sum_k S_2(k, t)}{\displaystyle \sum_k k S_2(k, t)}\,.
    \label{eq:L1L2}
\end{equation}
The scaled lengths $L_1(t)/L_0$ and ${L}_2(t)/L_0$, with $L_0 = 2\pi$, increase with time, in a manner that is consistent with $\sim t^{2/3}$, the power-law growth for the inertia-dominated regime [see the log-log plot in Fig.~\ref{fig:spinodal_ternary} (c)]. 
This power-law growth has also been seen in the molecular-dynamics simulations of \cite{C4SM02726D}. We expect that such ternary-fluid phase separation will also be suppressed by turbulence in the CHNS3 equatiions, just as turbulence leads to coarsening arrest in the binary-fluid case [see, e.g., \cite{perlekar2014spinodal} and \cite{perlekar2017two}],
but we must account for additional non-dimensional control parameters such as the ratios of the surface tensions between the three fluid phases that coexist in equilibrium. To the best of our knowledge, a complete study of such turbulence-induced coarsening arrest in three-phase fluid mixtures has not been carried out so far. 

\subsection{Spatiotemporal evolution of droplets in turbulent flows}
\label{subsec:dropturb}

We follow and generalise, to the case of compound droplets, the investigation of the spatiotemporal evolution of droplets in binary-fluid turbulent flows by \cite{pal2016binary}, who studied the advection of a droplet, initially circular with a diameter $d_0$, by a turbulent binary-fluid flow for which they used the CHNS system in 2D. The droplet was active, insofar as it affected the flow, and was, in turn, deformed by the flow.  \cite{pal2016binary} obtained the acceleration components of the droplet center of mass and showed that their probability distribution function (PDF) had wide, non-Gaussian tails. They uncovered multifractal fluctuations in the time series of the scaled perimeter (see below) of the droplet. Finally they showed that the droplet fluctuations led to an enhancement of the energy spectrum $E(k)$ for large wave numbers $k$, and thence to dissipation reduction, as in fluid turbulence with polymer additives [see, e.g., \cite{perlekar2006manifestations}, \cite{perlekar2010direct}, and \cite{gupta2015two}]. We refer the reader to \cite{pal2016binary} for the details of their DNS.
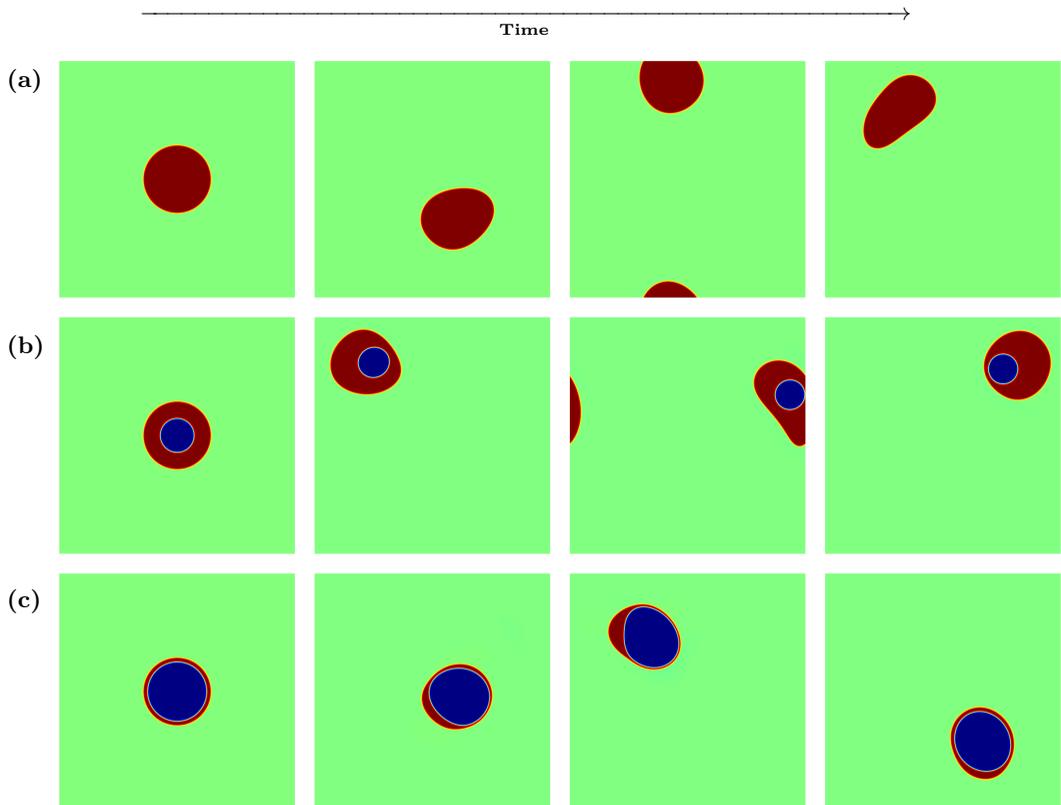
\begin{figure}
  \centerline{
  \includegraphics[width=\textwidth]{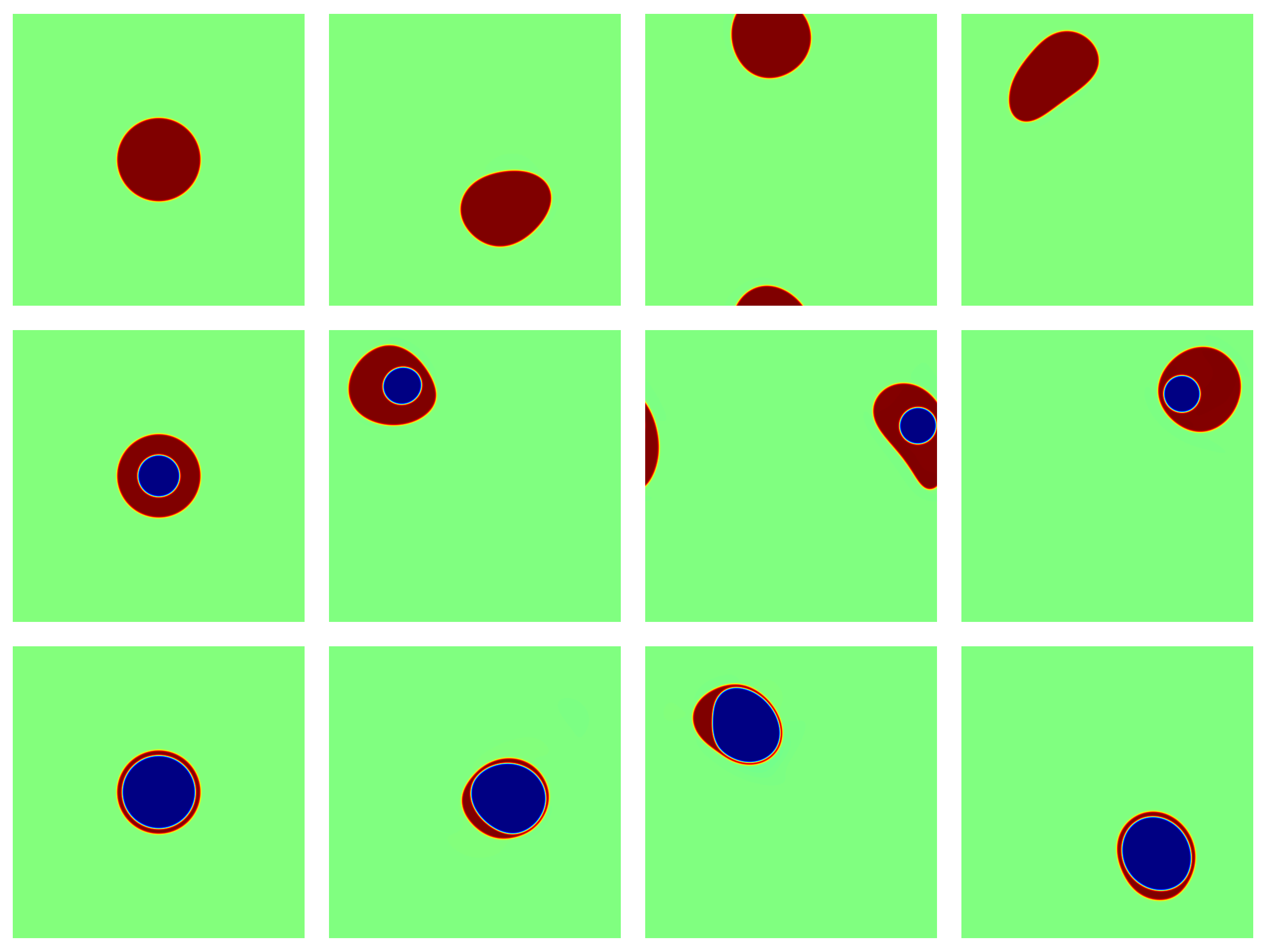}
  \put(-350,300){\rm {$\xrightarrow[\textit{\bf{Time}}]{\hspace*{10cm}}$}}
   \put(-400, 275){\rm {\bf(a)}}
   \put(-400, 175){\rm {\bf(b)}}
   \put(-400, 80){\rm {\bf(c)}}
  }
  \caption{The pseudocolor plots of (a) the phase field $c_2$ for a simple droplet, and the phase field $c_2 - c_1$ for the compound droplets with radius ratios (b) $R_{in}/R_{out} = 0.5$ and (c) $R_{in}/R_{out} = 0.9$. }
\label{fig:compound_droplet_pcolor}
\end{figure}
Our investigations of a compound droplet are motivated by studies of compound droplets in external electric fields [see ~\cite{PhysRevFluids.5.063602}], the examination of the break-up of double-emulsion droplets in linear flows [see ~\cite{Stone_Leal_1990}], the 
numerical and theoretical investigations of the dynamics of a compound vesicle (a lipid bilayer membrane enclosing a fluid with a suspended particle) in a shear flow [see ~\cite{PhysRevLett.106.158103}] and a compound vesicle in shear flow [see ~\cite{sinha2019theoretical}]. Experiments have also suggested that a compound droplet can be used as a model for a white-blood cell (WBC) cell [see ~\cite{levant2014complex}]. Furthermore, the deformation and breakup of a compound droplet  has been studied in a 3D oscillatory shear flow [see ~\cite{liu2021deformation}] and in a channel flow [see ~\cite{lanjewar2024dynamics}].

We now give illustrative results for the turbulent advection of a simple droplet, initially circular with a radius $R_0$, and for a compound droplet, 
initially concentric circles with inner and outer radii $R_{in}$ and $R_{out}$; for the former we employ the binary-fluid CHNS in 2D; and for the latter we use the 2D ternary-fluid CHNS; in both cases we impose periodic boundary conditions, obtain statistically steady turbulence via a Kolmogorov-type forcing 
with amplitude $F_{0}$, as in \cite{pal2016binary}, and a forcing wavenumber $k_f$ that yields an energy spectrum in which the forward-cascade regime is
dominant. The pseudocolor plots in Fig.~\ref{fig:compound_droplet_pcolor} (a) show the phase field $\phi$, for a simple droplet, 
at different representative times, which increase from left to right. For the compound droplet we present pseudocolor plots of 
$(c_2 - c_1)$ with the radius ratios (b) $R_{in}/R_{out} = 0.5$ [Fig.~\ref{fig:compound_droplet_pcolor} (b)] and (c) $R_{in}/R_{out} = 0.9$ [Fig.~\ref{fig:compound_droplet_pcolor} (c)]. 

\begin{figure}
  \centerline{
  \includegraphics[width=\textwidth]{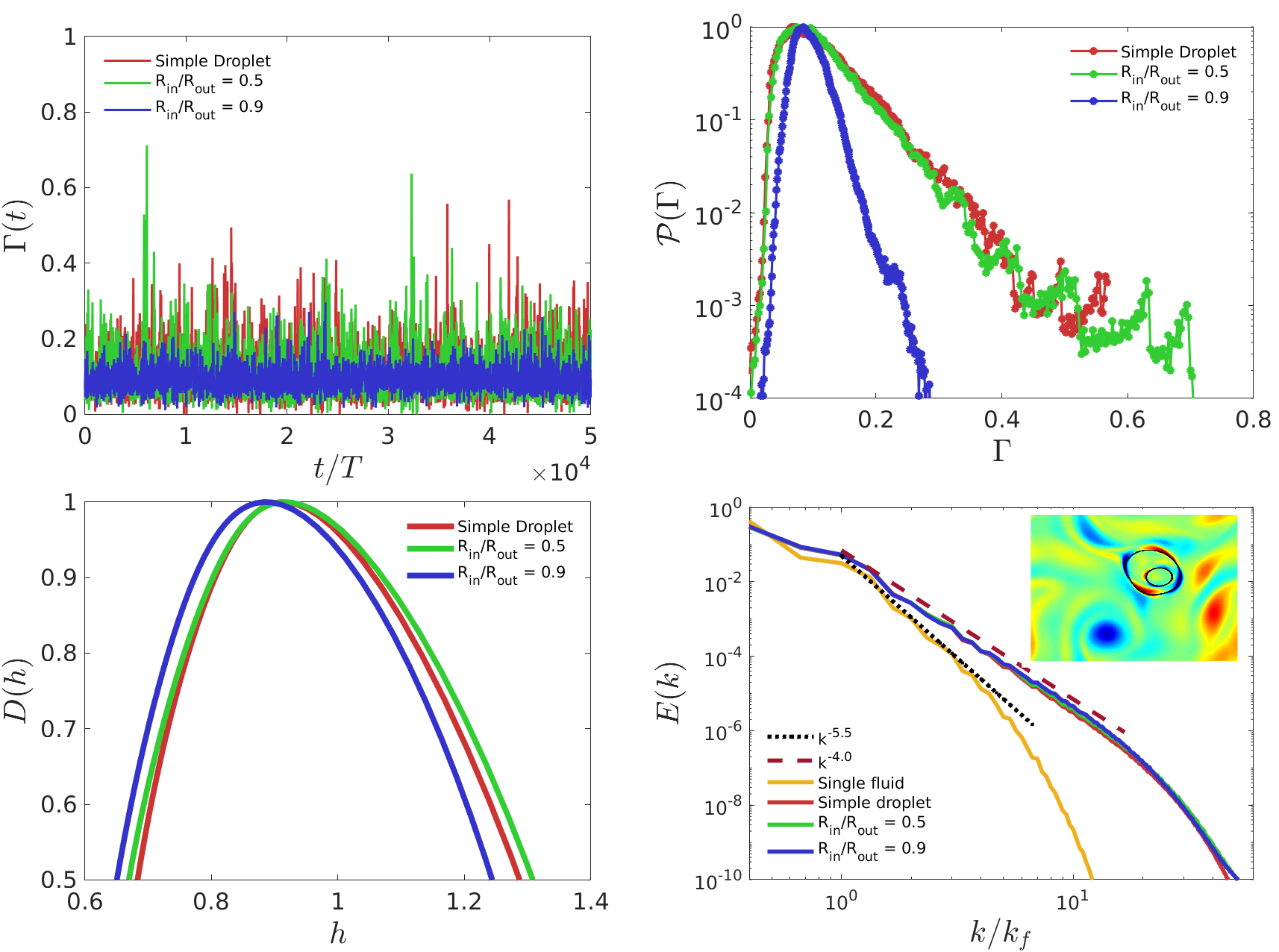}
   \put(-400, 270){\rm {\bf(a)}}
   \put(-195, 270){\rm {\bf(b)}}
   \put(-400, 120){\rm {\bf(c)}}
   \put(-195, 120){\rm {\bf(d)}}
  }
  \caption{(a) Plot versus time of the deformation parameter $\Gamma(t)$ for a simple droplet (red line) and compound droplets with radius ratios $R_{in}/R_{out} = 0.5$ (green line) and $R_{in}/R_{out} = 0.9$ (blue line); the horizontal axis is scaled by the forcing time scale $T = \nu k_f/f_0$. The corresponding (b) PDFs of $\Gamma$ (semi-log plots) and (c) the multifractal spectra $D(h)$ of $\Gamma$. (d) Log-log plots of the energy spectrum $E(k)$ for a single-phase turbulent fluid, a binary fluid with a simple droplet, and a ternary fluid with a compound droplet [with radius ratios $R_{in}/R_{out} = 0.5$ and $R_{in}/R_{out} = 0.9$]; the inset shows a pseudocolor plot of the vorticity in the presence of a compound droplet.}
\label{fig:compound_droplets_graphs}
\end{figure}
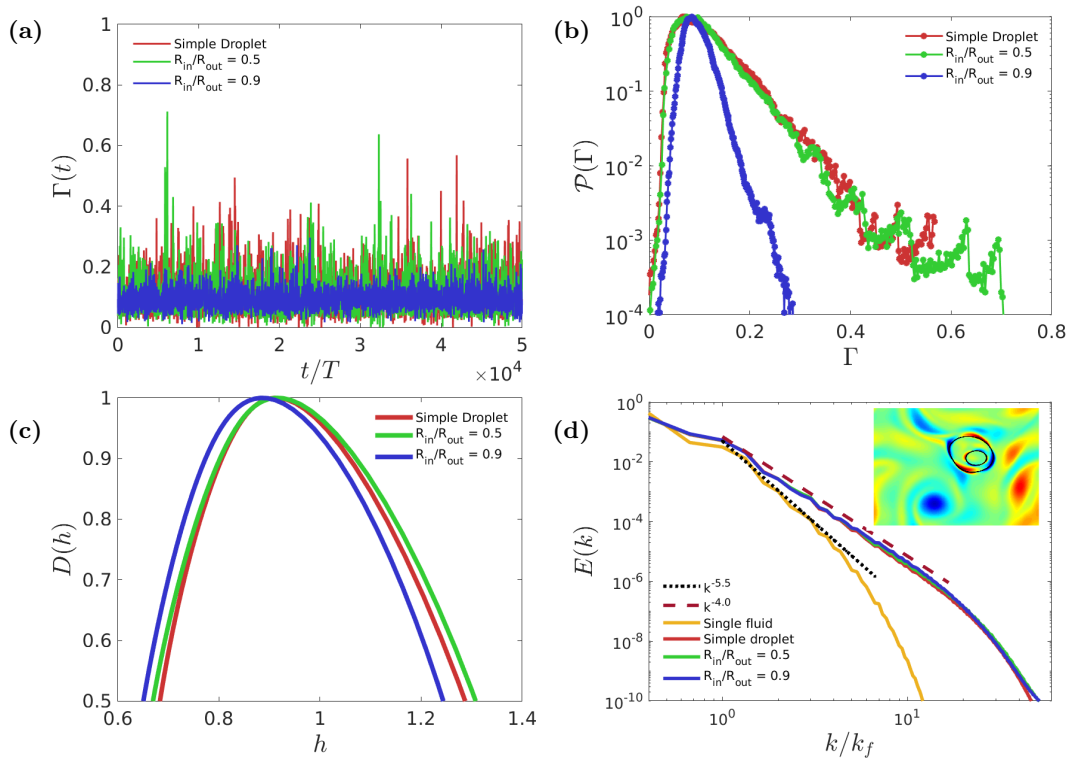

To characterise droplet fluctuations, we use the time dependence of the deformation parameter defined in ~\cite{pal2016binary}:
\begin{equation}
\Gamma(t)=\frac{{\mathcal {S}}(t)}{{\mathcal {S}}_{0}(t)}-1,
\label{eq:peri}
\end{equation}
with ${\mathcal {S}}(t)$, the perimeter of the droplet, and  ${\mathcal {S}}_0(t)$, the perimeter of an un-deformed
droplet. In the binary-fluid case we track the perimeter via the $\phi=0$ contour; in the ternary-fluid case we use the contours
$c_1 = 0.5$ and $c_2 = 0.5$ for the perimeters of the inner and outer droplets, respectively. The strength of the non-dimensionalised forcing is given by the Grashof number $Gr=L^{4}F_{0}/\nu^{2}$; and the forcing-scale Weber number $We\equiv \rho
k_f^{-3}F_0/\sigma$ measures the non-dimensionalised inverse surface tension; these parameters are chosen such that the droplets are not torn asunder 
during our DNSs.  In Fig.~\ref{fig:compound_droplets_graphs}(a) we plot $\Gamma(t)$ versus time [scaled by the forcing time scale $T = \nu k_f/F_0$] for a simple droplet (red line) and compound droplets with radius ratios $R_{in}/R_{out} = 0.5$ (green line) and $R_{in}/R_{out} = 0.9$ (blue line). 
In Figs.~\ref{fig:compound_droplets_graphs} (b) and (c) we plot, respectively, the PDF  of $\Gamma$ and the multifractal spectrum $D(h)$
of the time series $\Gamma(t)$. From Figs.~\ref{fig:compound_droplets_graphs} (a)-(c) we clearly see that the temporal fluctuations $\Gamma$, its PDF, and its multifractal spectrum $D(h)$ are similar for simple droplet interface and outer droplet interface in the compound droplet with a radius ratio $R_{in}/R_{out} = 0.5$. By contrast, all these measures are reduced for the perimeter of the outer droplet interface in a compound droplet with a radius ratio $R_{in}/R_{out} = 0.9$ because the fluctuations of the outer interface are constrained by the presence of the inner droplet.

The energy spectra $E(k)$ for a single-phase turbulent fluid, a binary fluid with a simple droplet, and a ternary fluid with a compound droplet [with radius ratios $R_{in}/R_{out} = 0.5$ and $R_{in}/R_{out} = 0.9$] are given in Fig.~\ref{fig:compound_droplets_graphs} (d). These spectra show that droplet fluctuations, of both single and compound droplets, enhance $E(k)$ at large $k$, relative to its counterpart for single-fluid turbulence. This enhancement
is reminiscent of a similar enhancement of $E(k)$ in fluid turbulence by polymer additives, where it is associated with dissipation reduction can be understood as a $k$-dependent correction to the viscosity [see ~\cite{perlekar2010direct,gupta2015two}]. Such a scale-dependent correction to the viscosity  also occurs for a single droplet in a turbulent flow as has been discussed by \cite{pal2016binary}.

\subsection{Bubble passing through an interface between two fluids}
\label{subsec:bubblepass}
\begin{figure}
  \centerline{
  \includegraphics[width=0.825\textwidth]{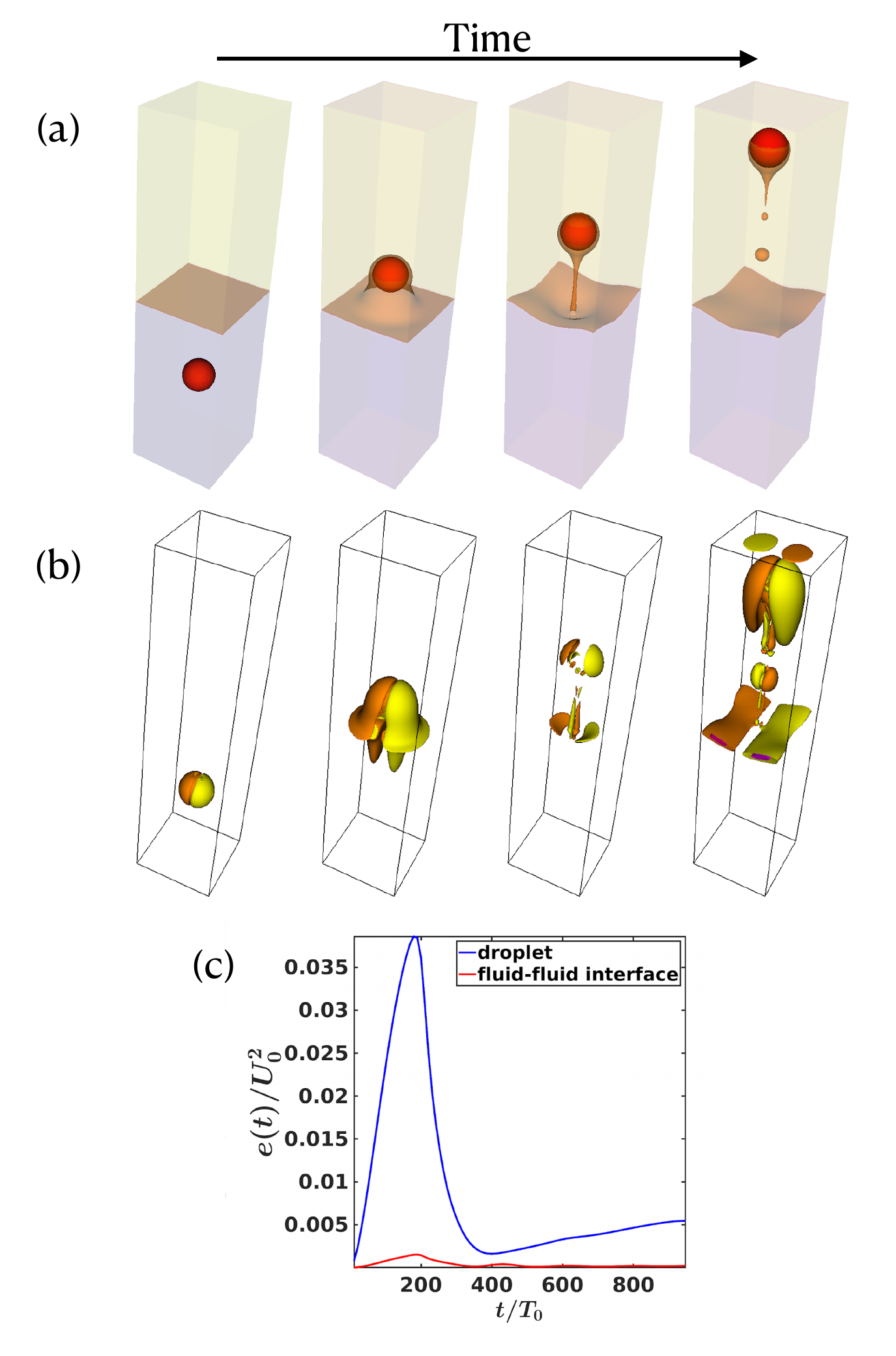}
  }
  \caption{A bubble passing through a fluid-fluid interface. (a) The isosurface plot of the $(c_1, c_2)$ fields. (b) The isocontour plot of the $z$ component of the vorticity field. (c) The kinetic energy time series of the droplet (blue line), where $e(t) = \langle |\bm u(\bm x)|^2\rangle_{\bm x\in (c_1 \ge 0.5)}$. The red line shows the kinetic energy time series of the fluid-fluid interface defined as $e(t) = \langle |\bm u(\bm x)|^2\rangle_{\bm x \in (0.1 \le c_2\le 0.9)}$. We use the characteristic velocity and time scales as $U_0 = g\epsilon^2/\nu$ and $T_0 = \frac{\nu}{g\epsilon}$.} 
\label{fig:bubble_ternary}
\end{figure}
How do bubbles or droplets pass through an interface between two fluids? This problem has attracted considerable attention [see, e.g.,~\cite{manga1995low,dietrich2008passage,bonhomme2012inertial,natsui2014multiphase,li2014satellite,singh2015passage,feng2016dynamics,prosperetti2017vapor,emery2018flow,emery2019modeling,kumar2019passage,choi2021interfacial,chowdhury2022wettability,rabbani2024interaction}] in the fluid-dynamics, chemical-engineering, and microfluidics literature. Modern experiments that use high-speed cameras to track the passage of bubbles through fluid-fluid interfaces have led to theoretical and numerical investigations of this problem. 

In Fig. \ref{fig:bubble_ternary} we present our results from an illustrative DNS of a bubble passing through a fluid-fluid interface; for this we employ the three-component 3D CHNS3 equations \eqref{eq:CHNS3A}-\eqref{eq:CHNS3C} and choose parameters
such that the bubble does go through the interface and does not get trapped there. For such studies it is natural to use the characteristic velocity and time scales $U_0 = g\epsilon^2/\nu$ and $T_0 = \frac{\nu}{g\epsilon}$, an elongated simulation domain [$(L_x, L_y, L_z) = (4\pi, \pi, \pi)$, with $512 \times 128 \times 128$ grid points] with periodic 
boundary conditions in the directions normal to gravity, and volume penalization in the direction of gravity to incorporate solid boundaries. We use $6$ grid points at both top and bottom boundaries in our volume-penalization scheme and the following simulation parameters: $\nu = 3.5\times 10^{-3}$, $g=1$, $\mathcal A_1 = 0.132$, $\mathcal A_2 = 0.132$, $\rho_1 = 1.132, \rho_2 = 0.868, \rho_3 = 1$  [with $\rho_3$ as the reference density], and $\sigma_{12} =  \sigma_{13}= 0.5$, and $\sigma_{23} = 0.01$. Figures \ref{fig:bubble_ternary} (a) and (b) show, respectively, isosurface plots of the $c_1$ and $c_2$ fields and isocontour plots of the $z$ component of the vorticity field, which illustrate how a $c_1$ droplet of the first fluid (in red), with concentration field $c_1$, passes through the $c_2-c_3$ interface (light brown) between the second and third fluids. As it passes through the interface, this droplet entrains some of the heavy fluid, which forms a slender neck that collapses eventually to yield droplets of the the heavy fluid that fall back onto the interface
[see, e.g., \cite{singh2015passage} and \cite{emery2018flow}].
In Fig.~\ref{fig:bubble_ternary}(c) we show that we can track the passage of this bubble through the interface by monitoring the temporal evolution of the droplet's kinetic energy [$e_d(t) \equiv \langle |\bm u(\bm x)|^2\rangle_{\bm x\in (c_1 \ge 0.5)}$ (blue line)] and the energy of the $c_2-c_3$ interface [$e_I(t) = \langle |\bm u(\bm x)|^2\rangle_{\bm x \in (0.1 \le c_2\le 0.9)}$ (red line)]; both these quantities display maxima when the bubble passes through the interface.

\subsection{The coalescence of liquid lenses and droplets}
\label{subsec:coalescence}

\begin{figure}
  \centerline{
  \includegraphics[width=\textwidth]{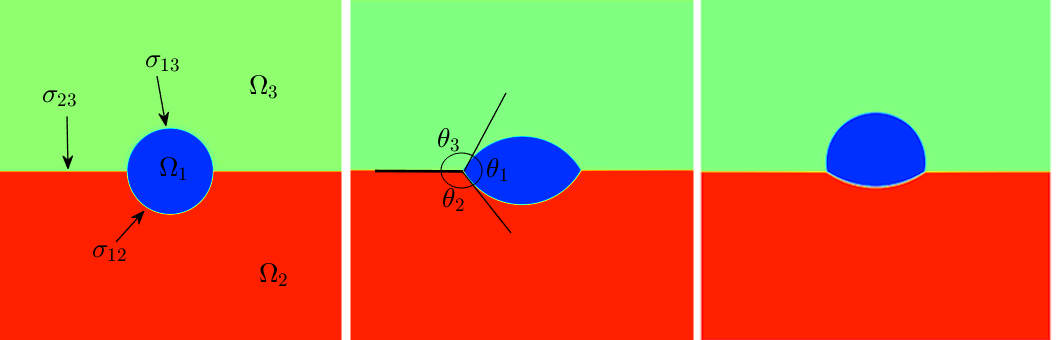}
  \put(-380,115){\rm {\bf(a)}}
  \put(-250,115){\rm {\bf(b)}}
  \put(-125,115){\rm {\bf(c)}}
  }
  \caption{Plots showing the three coexisting phases 1 (blue), 2 (red), and 3 (green) in regions $\Omega_1$, $\Omega_2$, and
  $\Omega_3$, respectively: (a) The intial profile with a circular droplet of radius $R_0/L = 0.15$ of fluid 1 at the interface between fluids 2 and 3. [$L = 2\pi$ is the side length of the simulation domain.]  (b) The equilibrium profile of the droplet is a symmetrical lens, because we choose $(\sigma_{12}, \sigma_{13}, \sigma_{23}) = (1, 1, 1)$; the three contact angles are $\theta_1$, $\theta_2$, and $\theta_3$. (c) The equilibrium profile of the droplet is an asymmetrical lens if we choose $(\sigma_{12}, \sigma_{13}, \sigma_{23}) = (1.4, 0.8, 1)$.}
\label{fig:init_lens}
\end{figure}

The coalescence of liquid droplets and lenses is a problem of fundamental importance in fluid mechanics and statistical mechanics [see, e.g., ~\cite{paulsen2014coalescence,palphdthesis,heinen2022droplet,padhan2023unveiling,scheel2023viscous}]. Our DNS for the CHNS3 [~\cite{padhan2023unveiling}] model can be used to examine the development of liquid-lens mergers in phase-separated ternary-fluid systems as we summarise below. The coexistence of three immiscible fluids leads to three distinct interfaces with three interfacial tensions: $\sigma_{ij}$ is the surface-tension coefficient for the ${ij}$ interface, where the integers $i$ and $j$ ($= 1, 2,$ or $3$) label the coexisting phases. We prepare neutrally buoyant, symmetrical or asymmetrical, lenses in 2D by starting our DNS with the following configuration for a single circular droplet of fluid 1, with radius $R_0$ and centre at $(\pi, \pi)$, placed at the interface between fluids 2 and 3 (see Fig.~\ref{fig:init_lens}(a)):
\begin{eqnarray}
    c_1(x,y,0) &=& \frac{1}{2}\left[1-\tanh \left(\frac{\sqrt{(x-\pi)^2+(y-\pi)^2}-R_{0}}{2 \sqrt{2} \epsilon}\right)\right]\,;\nonumber\\
    c_2(x, y, 0) &=& \frac{1}{2}\left[1-\tanh \left(\frac{y-\pi}{2 \sqrt{2} \epsilon}\right)\right] - c_1(x,y,0)\,;\\
    \omega(x, y, 0) &=& 0\,. \nonumber
    \label{eq:init}
\end{eqnarray}
The initial and equilibrium configurations in 3D are a sphere and a lenticular biconvex lens, respectively.
As time evolves in our (2D) DNS, the initial circular droplet relaxes to its equilibrium-lens shape as shown in Figs.~\ref{fig:init_lens}(b) and (c) for $(\sigma_{12},\sigma_{13},\sigma_{23}) = (1, 1, 1)$ and $(\sigma_{12},\sigma_{13},\sigma_{23}) = (1.4, 0.8, 1)$, respectively. 
Now we verify the Young relations for liquid lenses at equilibrium [see, e.g., ~\cite{mchale2022liquid, Boyer_2006}]. The theoretical distance $d^{th}$ between the two triple-phase junctions in Figs.~\ref{fig:init_lens} (b) and (c) are given by the following Young relations:
\begin{eqnarray}
    d^{th} &=& (l_1 + l_2)^{-{\frac{1}{2}}}\,;\nonumber \\
    l_1 &=& \frac{2(\pi-\theta_3) - \sin(2(\pi-\theta_3))}{8A\sin(\pi-\theta_3)}\,;\\
    l_2 &=& \frac{2(\pi-\theta_1) - \sin(2(\pi-\theta_1))}{8A\sin(\pi-\theta_1)}\,,\nonumber
    \label{eq:Young}
\end{eqnarray}
The contact angles are related as follows:
\begin{eqnarray}
    \frac{\sin\theta_1}{\sigma_{23}} &=& \frac{\sin\theta_2}{\sigma_{13}} = \frac{\sin\theta_3}{\sigma_{12}}\,.
    \label{eq:angle}
\end{eqnarray}
We calculate the distance between the triple-phase junctions $d^{sim}$ from our simulations [see Figs.~\ref{fig:init_lens}(b) and (c)] and compare them with $d^{th}$ in Table~\ref{tab:junction}. The agreement between these values is good.
\begin{table}
\centering
\begin{tabular}{c c c c c c c c} 
 \hline
 \hspace{0.3cm}{Runs} \hspace{0.5cm}&  \hspace{0.5cm}$(\sigma_{12},\sigma_{13},\sigma_{23})$  \hspace{0.5cm} & \hspace{0.75cm}$d^{th}$ \hspace{0.75cm} & \hspace{0.75cm}$d^{sim}$\hspace{0.75cm} & Relative error (\%)\\ [0.5ex] 
 \hline\hline
 $\mathcal{R}1$ & (1, 1, 1) & 2.15 & 2.18 & 1.4\\ 
 \hline
 $\mathcal{R}2$ & (1.4, 0.8, 1) & 1.3 & 1.35 & 3\\
 \hline
\end{tabular}
 \caption {The distance between two triple-phase junctions calculated from theory $d^{th}$ [see Eq.~\eqref{eq:Young}] and numerical simulations $d^{sim}$ for the symmetrical lens (Fig.~\ref{fig:init_lens}(b)) and the asymmentrical lens (Fig.~\ref{fig:init_lens}(c)).}
 \label{tab:junction}
\end{table} 
\begin{figure}
\includegraphics[width=\textwidth,height=0.5\textwidth]{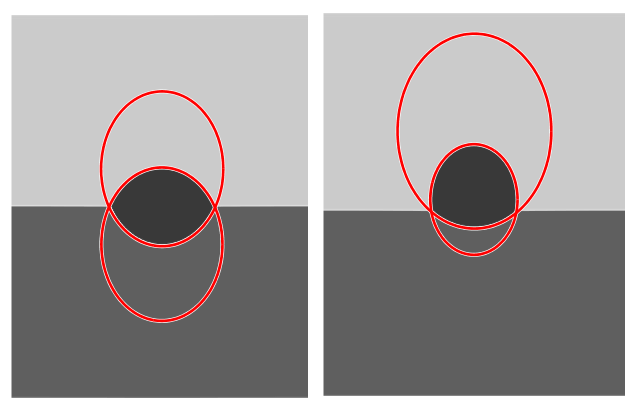}
    \put(-365,170) {\bf(a)} 
    \put(-175,170) {\bf(b)}
    \caption{Illustrations of the circle Hough transform (see text) that we use to fit circles for the lens interfaces for (a) $(\sigma_{12}, \sigma_{13}, \sigma_{23}) = (1, 1, 1)$ (run $\mathcal{R}1$) and (b) $(\sigma_{12}, \sigma_{13}, \sigma_{23}) = (1.4, 0.8, 1)$ (run $\mathcal{R}2$).}
    \label{fig:Hough}
\end{figure}
\begin{table}
\caption{\label{tab:Laplace}
Illustrative comparisons of theoretical and our DNS results for Laplace-pressure jumps for different lens shapes. ${\Delta P} \equiv P_1 - P_2 = P_1 - P_3$. There is good agreement between these results. While evaluating $\Delta P$, we calculate the values of $P_1$, $P_2$, and $P_3$ at points that are far from the interface.}
\begin{tabular}{c c c c c c}
\hline
 \hspace{0.3cm} {Runs} \hspace{0.5cm}& \hspace{0.5cm}$(\sigma_{12}, \sigma_{23}, \sigma_{13})$ \hspace{0.5cm}&\hspace{0.5cm} $R_0/L$ \hspace{0.5cm}&
  \hspace{0.5cm}Theory \hspace{0.5cm}& \hspace{0.5cm}Simulation \hspace{0.5cm}& Relative error(\%)\\
 & & & $\Delta P$& $ \Delta P$& \\
\hline
RN1& (1, 1, 1) & 0.2 & 0.989  & 0.994 & 0.5\\
\hline
RN2& (1, 0.8, 1) & 0.2 & 1.148  & 1.132 & 1.3\\
\hline
RN3& (0.6, 1, 0.8) & 0.2 & 0.465  & 0.460 & 1\\
\hline
RN4& (1.4, 1, 0.6) & 0.2 & 0.889 & 0.891 & 0.2\\
\hline
RN-3D& (1, 1, 1) & 0.13 & 1.54 & 1.56 & 1.2\\
\hline
\end{tabular}
\end{table}
\begin{figure}
\includegraphics[width=0.475\textwidth]{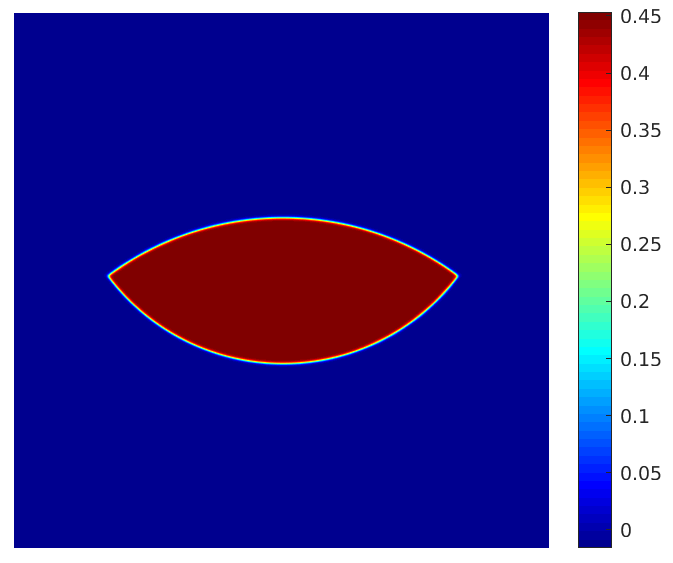}
\includegraphics[width=0.475\textwidth]{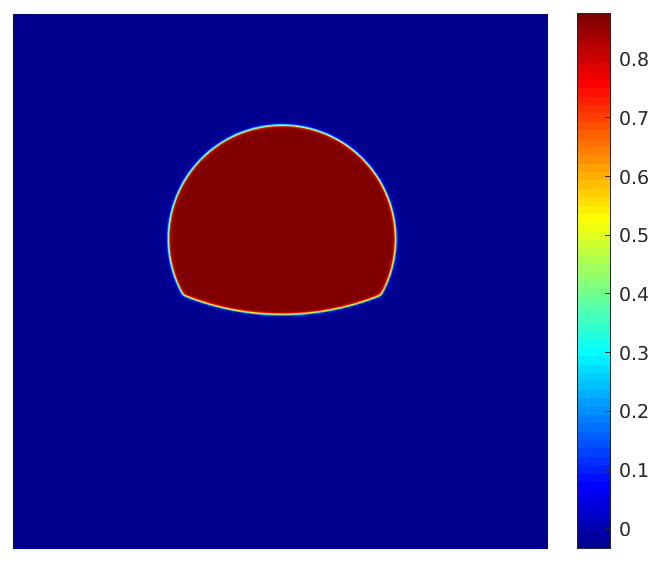}
\put(-350,130) {\bf\color{magenta}(a)} 
\put(-160,130) {\bf\color{magenta}(b)}
\caption{\label{fig:pr}Pseudocolor plot of illustrative equilibrium-pressure profiles for (a) $(\sigma_{12}, \sigma_{23}, \sigma_{13}) = (0.6, 1, 0.8)$ (run RN3) and (b) $(\sigma_{12}, \sigma_{23}, \sigma_{13}) = (1.4, 1, 0.6)$ (run RN4).}
\end{figure}
Furthermore, we can compare the results of our DNS with the prediction of the Laplace law for pressure jumps, at equilibrium; these jumps are defined as folllows:
\begin{eqnarray}
    \frac{\sigma_{13}}{R_{13}} &=& P_1 - P_3 = P_1 - P_2 = \frac{\sigma_{12}}{R_{12}}\,;\nonumber\\
    P_2 - P_3 &=& 0\,;
\end{eqnarray}
$R_{12}$ and $R_{13}$ are the radii of curvatures of the interfaces between fluids 1 and 2 and between fluids 1 and 3, respectively. To calculate the radii of curvatures, we use the circle Hough transform (CHT) in MATLAB [see ~\cite{atherton1999size}]. In Fig.~\ref{fig:Hough}, we illustrate the CHT of the images in Figs.~\ref{fig:init_lens}(b) and (c); red circles are the best-fit circles to the curves. The CHT gives the coordinates of the centres and the radii of the red circles. We calculate the theoretical values of the pressure jumps from these radii of curvatures. To compare our DNS results with these theoretical values, we evaluate the pressure jumps from the following pressure-Poisson equation:
\begin{eqnarray}
    \nabla^2 P = \bm{\nabla} \cdot \left(-{\sum_{i=1}^{3}c_{i} \bm{\nabla} \mu_{i}}\right)\,,
    \label{eq:Lap_press}
\end{eqnarray}
which we portray via pseudocolor plots of illustrative equilibrium-pressure profiles for runs RN1 and RN2 in Fig.~\ref{fig:pr}. We present our DNS results in TABLE.~\ref{tab:Laplace} for various runs. These results are in good agreement with their theoretical counterparts. 
\begin{figure}
    \includegraphics[width=\textwidth]{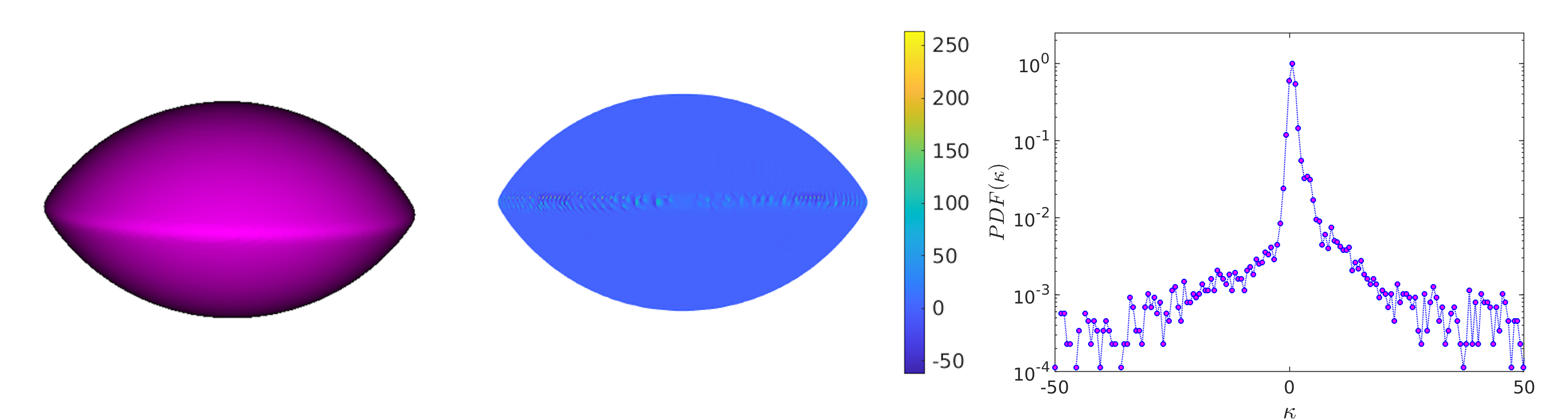}
    \put(-340,90) {\bf(a)} 
    \put(-220,90) {\bf(b)}
    \put(-110,80) {\bf(c)}
    \caption{(a) The isosurface plot of $c_1 = 0.5$, at equilibrium, illustrating a lens in 3D. (b) The isosurface plot of the corresponding Gaussian curvature $\kappa$. (c) The PDF of the Gaussian curvature. }
    \label{fig:GC_lens}
\end{figure}
We obtain similar results from 3D DNSs of the CHNS3 model: In Fig.~\ref{fig:GC_lens}(a) we give an isosurface plot, with $c_1 = 0.5$, for a 3D lenticular biconvex lens. We then calculate its Gaussian curvature $\kappa$ by implementing, in MATLAB, the algorithm described in ~\cite{meyer2003discrete}. The isosurface plot of $\kappa$ is shown in Fig.~\ref{fig:GC_lens}(b); clearly, $\kappa$ is constant throughout the lens surface, except at the edges. So, in Fig.~\ref{fig:GC_lens}(c), we present the probability distribution function (PDF) of $\kappa$ to find out the most probable value of $\kappa$. We consider the values of  $\kappa$ with the highest probability, namely, $0.59, -0.04, 1.23$; the average value is $\kappa \simeq 0.593$. 
The Gaussian curvature $\kappa = 1/R_{G}^2$ for a sphere of radius $R_G$ [see, e.g., ~\cite{nothard1996gaussian}]. The symmetric 3D lens in Fig.~\ref{fig:GC_lens}(a) is a combination of two surfaces that are parts of spheres of equal radii. Then we follow the Laplace law in 3D to evaluate the pressure jumps:
\begin{eqnarray}
    \frac{2\sigma_{13}}{R_{13}} &=& P_1 - P_3 = P_1 - P_2 = \frac{2\sigma_{12}}{R_{12}}\,;\nonumber\\
    P_2 - P_3 &=& 0\,;
\end{eqnarray}
here $R_{13} = R_{12} = R_G = 1/\sqrt{\kappa}$. The theoretical values of the pressure jumps are given in Table~\ref{tab:Laplace} (see run RN-3D); we solve the pressure-Poisson equation~\eqref{eq:Lap_press} numerically; we find good agreement between the theoretical and numerical values of these jumps .
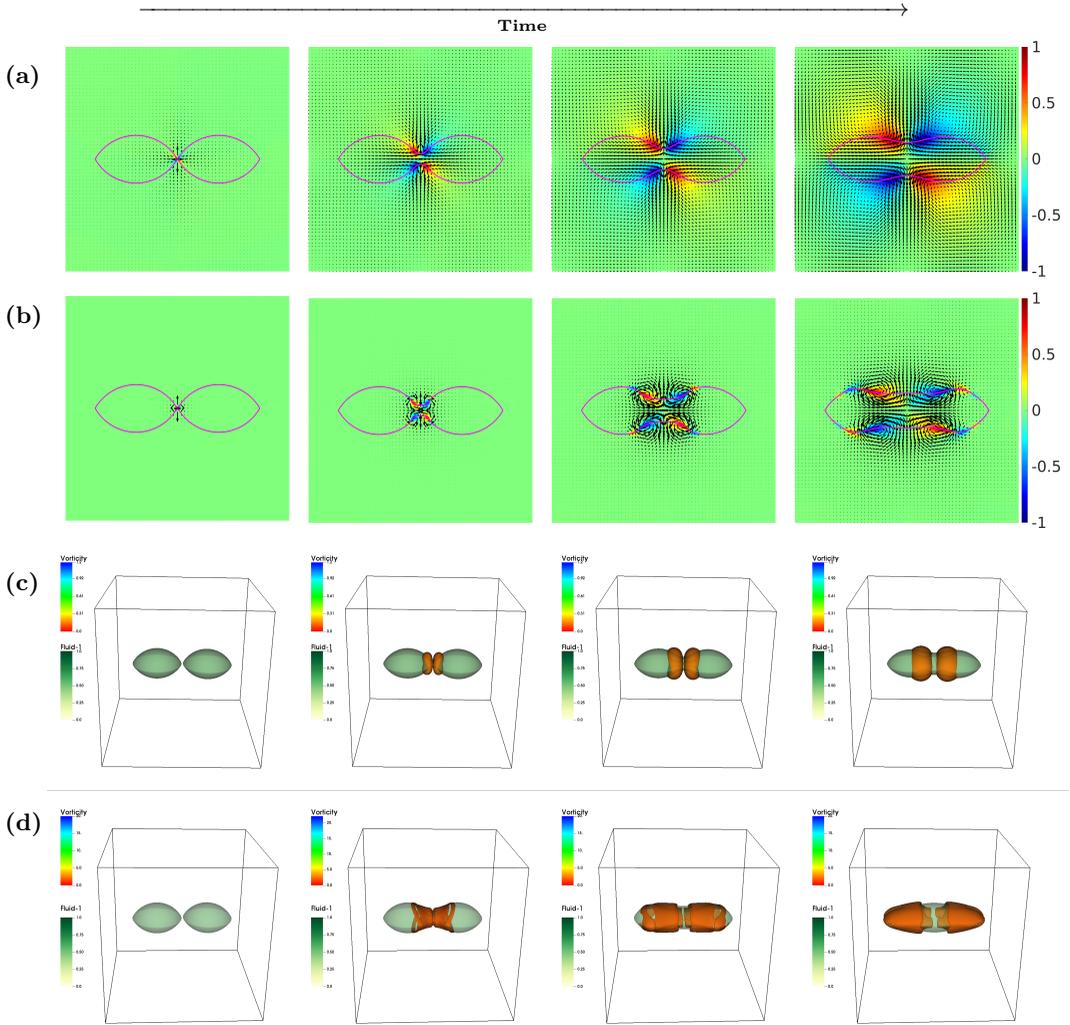
\begin{figure}
  \centerline{
  \includegraphics[width=\textwidth]{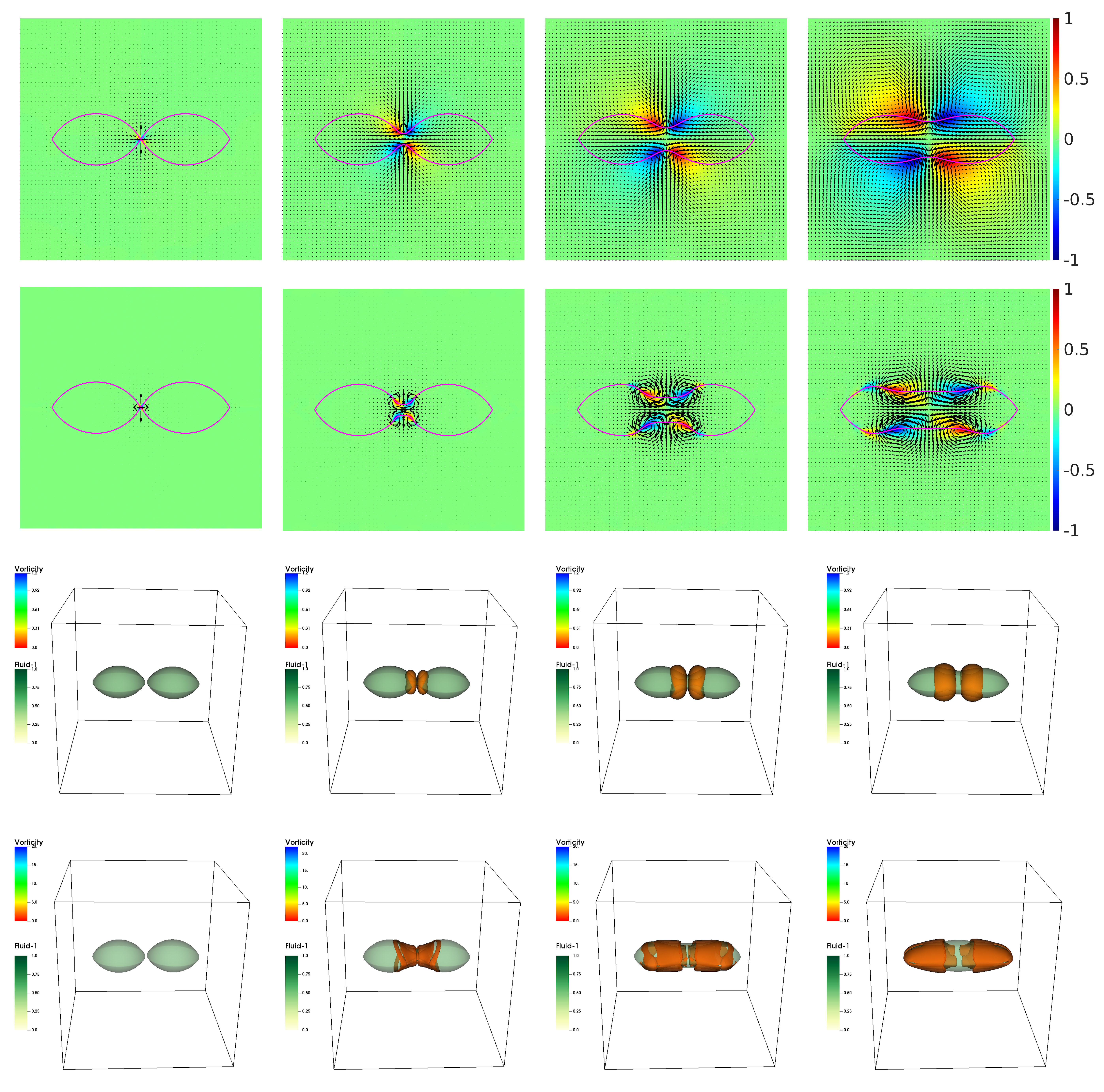}
  \put(-350,385){\rm {$\xrightarrow[\textit{\bf{Time}}]{\hspace*{10cm}}$}}
  \put(-400,360){\rm {\bf(a)}}
  \put(-400,270){\rm {\bf(b)}}
  \put(-400,170){\rm {\bf(c)}}
  \put(-400,80){\rm {\bf(d)}}
  }
  \caption{Illustrative results from our DNSs of liquid-lens mergers in the CHNS3 model in 2D (top two rows) and 3D (bottom two rows): Pseudocolor plots of $\omega$, with overlaid velocity vectors in (a) the viscous regime and (b) the inertial regime; the $c_1 = 0.5$ contour (magenta line) indicates the lens interface; $\omega$ is normalized by its maximal absolute value for ease of visualization. Isosurface plots of $c_1$ (green) and $|\omega|$ (brown)
for (c) the viscous regime and (d) the inertial regime.}
\label{fig:lens_pcolor}
\end{figure}

\begin{figure}
  \centerline{
  \includegraphics[width=\textwidth]{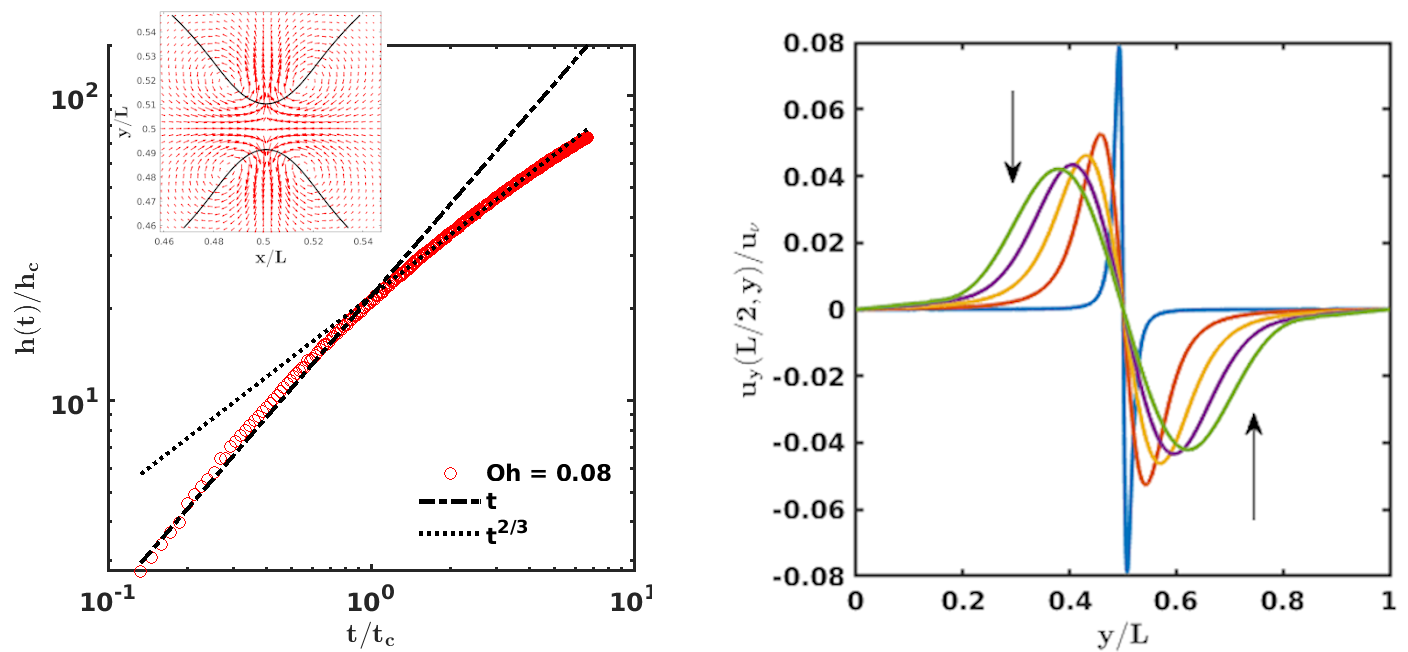}
  \put(-380,165){\rm {\bf(a)}}
  \put(-190,165){\rm {\bf(b)}}
  }
  \caption{(a) Log-log plot of the neck height $h$ versus time $t$. The axes are scaled by their respective viscous length scales for $Oh = 0.08$. The plot shows the crossover in the neck growth from the viscous regime with $h(t) \sim t$ to the inertial regime with $h(t) \sim t^{2/3}$. The inset shows the velocity vectors near the neck region at a representative time. (b) The profile of the $y$ component of the velocity field $u_y(L/2, y)$ along the vertical direction; the arrows show the direction of the evolution of the profiles.}
\label{fig:lens_height}
\end{figure}

We turn now to an overview of our recent study [\cite{padhan2023unveiling}] that has shown how to use DNSs of the CHNS3 system \eqref{eq:CHNS3A}-\eqref{eq:CHNS3C} and their 2D counterparts to study the spatiotemporal evolution of the merger of liquid lenses in both 2D and 3D. In Fig.~\ref{fig:lens_pcolor} we present illustrative results from these DNSs in 2D (top two rows) and 3D (bottom two rows). In 2D we give pseudocolor plots of $\omega$, with overlaid velocity vectors in (a) the viscous regime and (b) the inertial regime; the $c_1 = 0.5$ contour (magenta line) indicates the lens interface; $\omega$ is normalized by its maximal absolute value for ease of visualization. Isosurface plots of $c_1$ (green) and $|\omega|$ (brown) for (c) the viscous regime and (d) the inertial regime. In the viscous regime, i.e., at large values of the Ohnesorge number $Oh$, a vortex quadrupole dominates the flow in the region of
the neck in both 2D and 3D lens mergers, shown in Figs.~\ref{fig:lens_pcolor} (a) and (c), respectively. In the small-$Oh$ inertial regime [Figs.~\ref{fig:lens_pcolor} (b) and (d) for 2D and 3D, respectively] this quadrupole moves away from the region of the neck with the passage of time.  
We quantify the growth of the height $h(t)$ of the neck, in the low-$Oh$ case, in Fig.~\ref{fig:lens_height} that contains a log-log plot of $h$ versus 
the time $t$. This plot shows clearly the crossover from the viscous regime with $h(t) \sim t$, at early times, to $h(t) \sim t^{2/3}$, at late times. The exponent $2/3$ is typical of inertial-regime neck growth; the early-time
growth is similar to that found in the viscous case (because the early-time quadrupolar configuration is similar to that in the viscous case). The inset shows the velocity vectors near the neck region at a representative time. In Fig.~\ref{fig:lens_height} (b) we show the profile of the $y$-component of the velocity field $u_y(L/2, y)$, along the vertical direction; the arrows show how the profiles flatten as $t$ increases.

\subsection{Active CHNS model}
\label{subsec:ActiveH}

We consider the following incompressible CHNS equations (also called active model H) to study active turbulence in systems of contractile swimmers [see, e.g., ~\cite{tiribocchi2015active,padhan2023activity,D4SM00163J,padhan2025suppression,cates2024active}] in two spatial dimensions (2D):
\begin{eqnarray}
    \partial_t \phi + (\bm u \cdot \nabla) \phi &=& M \nabla^2\left(\frac{\delta \mathcal F}{\delta \phi}\right)\, ;\label{eq:ch}\\ 
    \partial_t \omega + (\bm u \cdot \nabla) \omega &=& \nu \nabla^2\omega + \frac{3}{2}\epsilon \nabla \times (\nabla \cdot \bm \Sigma^A) - \alpha \omega\, ;\;\label{eq:ns}\\
    \nabla \cdot \bm u &=& 0\,; \label{eq:incomp} 
\end{eqnarray}
$\omega$ is the vorticity field; $\nu$, $\alpha$, and $M$ are the kinematic viscosity, bottom friction, and mobility, respectively. 
$\mathcal F$ is the Landau-Ginzburg variational free-energy functional
\begin{eqnarray}
     \mathcal F[\phi, \nabla \phi] = \int_{\Omega} \left[\frac{3}{16} \frac{\sigma}{\epsilon}(\phi^2-1)^2 + \frac{3}{4} \sigma \epsilon |\nabla \phi|^2\right]\,,\label{eq:functional}
\end{eqnarray}
in which the first term is a double-well potential with minima at $\phi = \pm 1$. The scalar order parameter $\phi$ is positive (negative) in regions where the microswimmer density is high (low); in the interfaces between these regions, $\phi$ varies smoothly, over a width $\epsilon$. The free-energy penalty for an interface is given by the bare surface tension $\sigma$. In the inherently nonequilibrium active model H all terms in the stress tensor do not follow from $\mathcal F$. In particular, we must include the stress tensor $\bm \Sigma^A$, which has the form of a nonlinear Burnett term and has the components [see, e.g., ~\cite{padhan2023activity,D4SM00163J,tiribocchi2015active,bhattacharjee2022activity,das2020transition}]
\begin{eqnarray}
    \Sigma^{A}_{ij} = -\zeta \left[\partial_i \phi \partial_j \phi - \frac{\delta_{ij}}{2} |\nabla \phi|^2\right]\,,
    \label{eq:tensor}
\end{eqnarray}
where $\zeta$, the activity coefficient, can take both positive and negative values: $\zeta < 0$ ($\zeta > 0$) for contractile (extensile) swimmers. \textcolor{black}{We emphasize that the free-energy functional used in this active model is a mathematical construct without a direct physical origin. The activity is introduced phenomenologically. For instance, an active-stress term with an effective negative surface tension coefficient has been incorporated into the model to capture the coarsening-arrest mechanism in contractile systems [see, e.g., \cite{tiribocchi2015active,cates2024active}].}

\subsection{Generalised Active CHNS model for an active self-propelling droplet}
\label{subsec:genactivechns}

To study active, self-propelling  droplets, we follow \cite{padhan2023activity} and use two scalar fields $\phi$ and $\psi$, with $\psi$ an active scalar, in the active-matter sense [see, e.g., \cite{marchetti2013hydrodynamics}]\footnote{\textcolor{black}{The terminology used in active matter and conventional fluid dynamics differs slightly. In fluid dynamics, both $\phi$ and $\psi$ are considered active scalars because they influence the velocity field $\bm{u}$. However, in active matter, only $\psi$ is regarded as active, whereas $\phi$ is not.}}. We employ the following free-energy functional:
\begin{eqnarray}
\mathcal{F}[\phi, \nabla \phi, \psi, \nabla \psi] &=& \int_{\Omega} \frac{3}{16}\left(\frac{\sigma_1}{\epsilon_1}(\phi^2-1)^2 + \frac{\sigma_2}{\epsilon_2} (\psi^2-1)^2 \right) - \beta \phi \psi  \nonumber \\
&+& \frac{3}{4}\left( \sigma_1 \epsilon_1 |\nabla \phi|^2 + \sigma_2 \epsilon_2 |\nabla \psi|^2\right) d\Omega\,,
\label{eq:fren}
\end{eqnarray}
where $\Omega$ is the region we consider. This model allows for interfaces of $\phi$ and $\psi$, with (bare) positive interfacial tensions  $\sigma_1$ and $\sigma_2$ and widths $\epsilon_1$ and $\epsilon_2$, respectively; the coupling constant $\beta > 0$, so there is an attractive coupling between $\phi$ and $\psi$. Experiments on active droplets 
are carried out confined planar domains, so we use the following generalisation of the 2D \textit{active} incompressible CHNS equations given in Subsection~\ref{subsec:ActiveH}:
\begin{eqnarray}
\partial_t \phi + (\bm u \cdot \nabla) \phi &=& M_1 \nabla^2 \left( \frac{\delta \mathcal F}{\delta \phi}\right)\,; \label{eq:phi}\\
\partial_t \psi + (\bm u \cdot \nabla) \psi &=& M_2 \nabla^2 \left( \frac{\delta \mathcal F}{\delta \psi}\right)\,;\label{eq:psi}\\
\partial_t \omega + (\bm u \cdot \nabla) \omega &=& \nu \nabla^2 \omega -\alpha \omega +[\nabla \times (\mathfrak{S}^{\phi} +  \mathfrak{S}^{\psi})]\,;\label{eq:omega}\\
\nabla \cdot \bm u &=& 0\,; \quad \omega = (\nabla \times \bm u)\,;\label{eq:incom}\\
\mathfrak{S}^{\phi} &=& -(3/2)\sigma_1 \epsilon_1 \nabla^2 \phi \nabla \phi \,;\label{eq:Sphi}\\
 \mathfrak{S}^{\psi} &=&    -(3/2) \tilde\sigma_2 \epsilon_2 \nabla^2 \psi \nabla \psi\,; \label{eq:Spsi}
\end{eqnarray} 
here, the vorticity, kinematic viscosity, and the bottom friction are, respectively, $\omega$, $\nu$, and  $\alpha$ and we set the constant fluid density $\rho=1$.
We use the constant mobilities $M_1$ and $M_2$ for $\phi$ and $\psi$, respectively;  the $\phi$ interfacial stress $\mathfrak{S}^{\phi}$ [Eq.~(\ref{eq:Sphi})] follows from $\mathcal F$; in contrast, for the \textit{active stress} $\mathfrak{S}^{\psi}$ [Eq.~(\ref{eq:Spsi})] from $\psi$, we use the active-model-H formulation [see, e.g., ~\cite{wittkowski2014scalar,tiribocchi2015active,shaebani2020computational,padhan2023activity}]. It is important to note that
(a) both $\omega$ and $[\nabla \times (\mathfrak{S}^{\phi} +  \mathfrak{S}^{\psi})]$ are orthogonal to the 2D plane and (b)
the mechanical surface tension $\tilde \sigma_2 \neq \sigma_2$, which can be either negative or positive values, unlike $\sigma_1$ and $\sigma_2$
that are positive. For contractile swimmers $\tilde \sigma_2 < 0$ and for extensile swimmers $\tilde \sigma_2 > 0$; the former show arrested phase separation, whereas the latter display complete phase separation [see, e.g., ~\cite{tiribocchi2015active} and \cite{padhan2023activity}]. 
We will show in Subsection~\ref{subsec:dropprop} that the \textit{activity}  
\begin{equation}
A= |\tilde \sigma_2|/\sigma_2\,,
\label{eq:activityA}
\end{equation}
 is the most important control parameter here.
\begin{figure}
  \centerline{
  {\includegraphics[width=\textwidth]{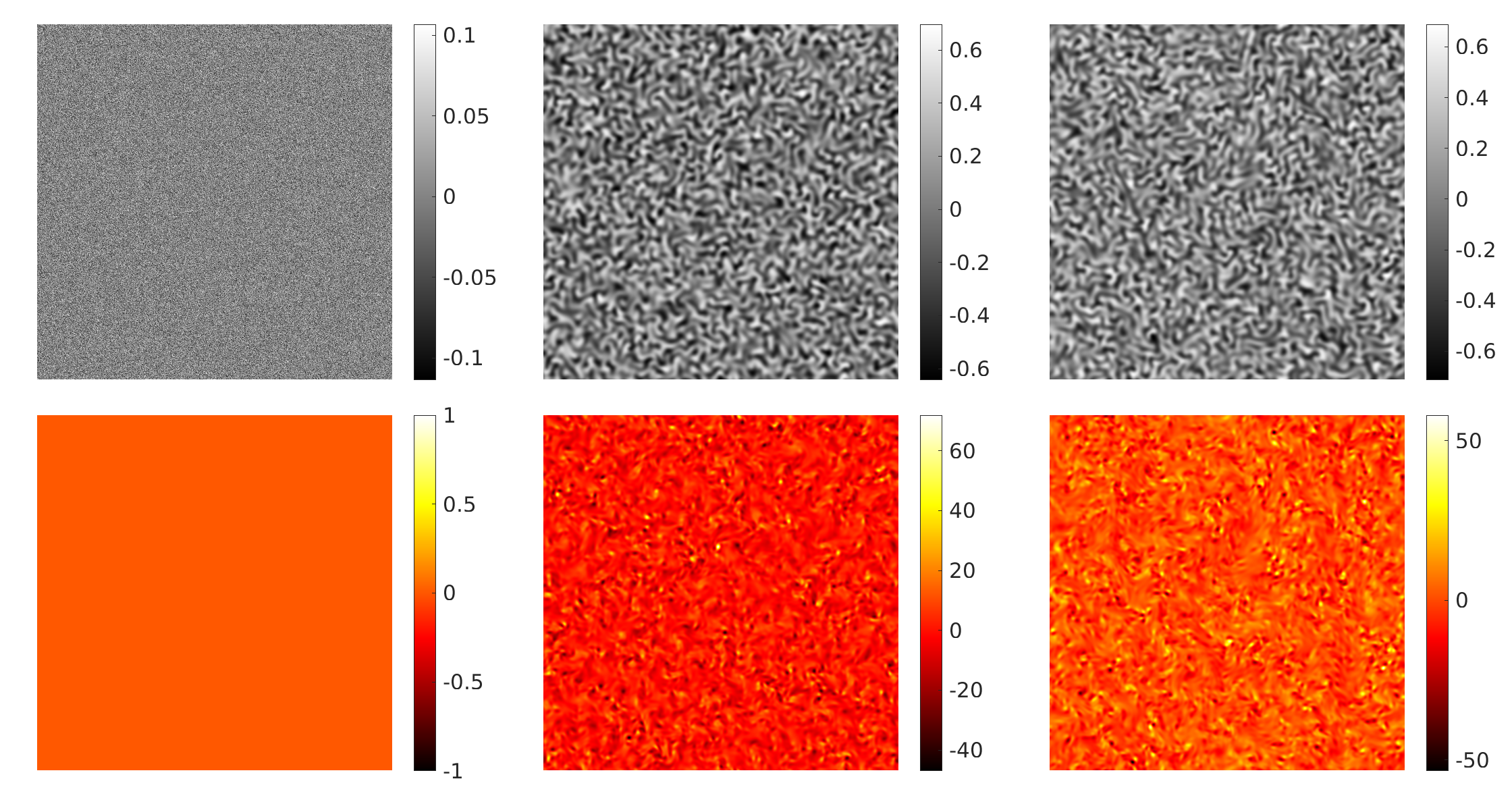}
   \put(-330,220){\rm {$\xrightarrow[\textit{\bf{Time}}]{\hspace*{10cm}}$}}
    \put(-390,190){\rm {\bf(a)}}
    \put(-260,190){\rm {\bf(b)}}
    \put(-134,190){\rm {\bf(c)}}
    \put(-390,90){\rm {\bf(d)}}
    \put(-260,90){\rm {\bf(e)}}
    \put(-135,90){\rm {\bf(f)}}
  }
  }
\caption{(a)-(c) Pseudocolor plots of the active-scalar field $\phi$, at three representative times, which increase from left to right, for the activity parameter $\zeta = 0.1$ [see Eqs.~\eqref{eq:ch}-~\eqref{eq:tensor}]; (d)-(f)  pseudocolor plots of the vorticity $\omega$ corresponding, respectively, to the pseudocolor plots $\phi$ in (a)-(c).}
\label{fig:activechns}
\end{figure}
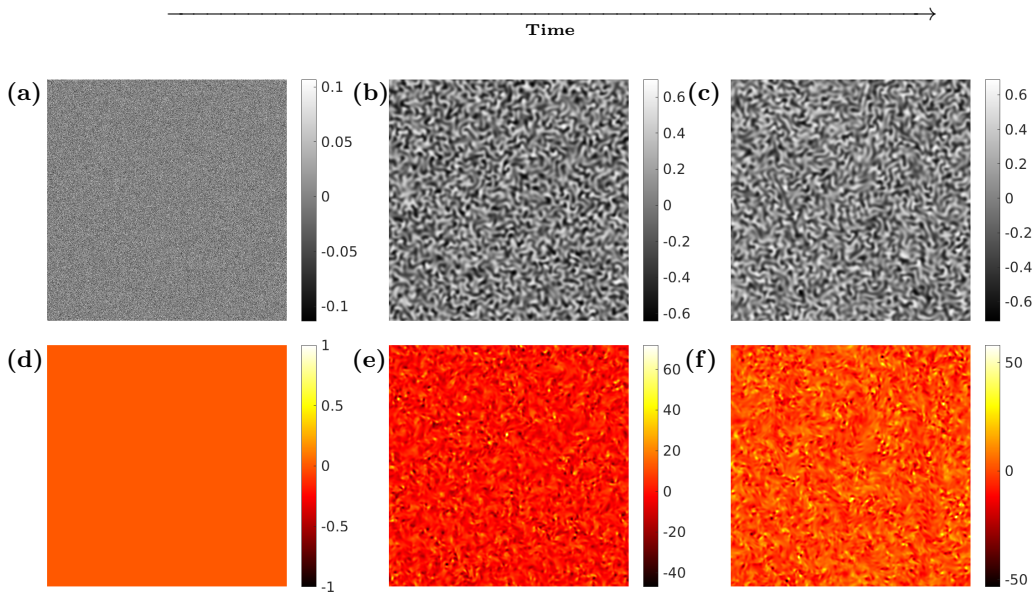
\subsection{Active CHNS Turbulence}
\label{subsec:activechnsturb}
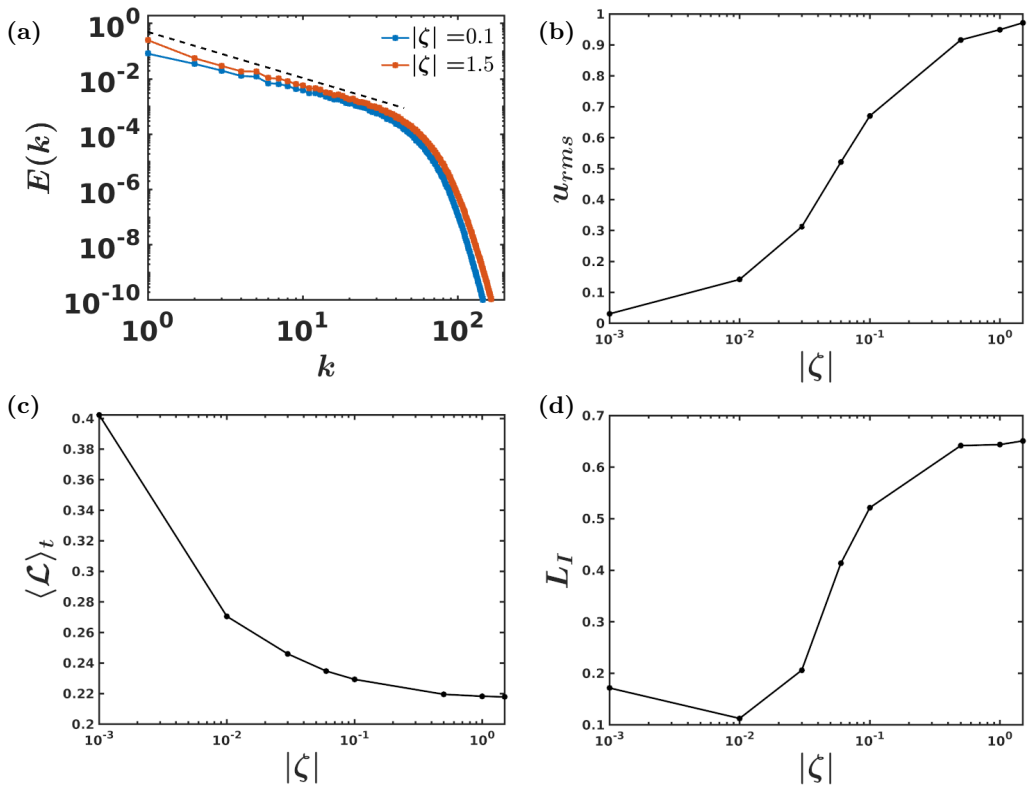
\begin{figure}
  \centerline{
  {\includegraphics[width=\textwidth]{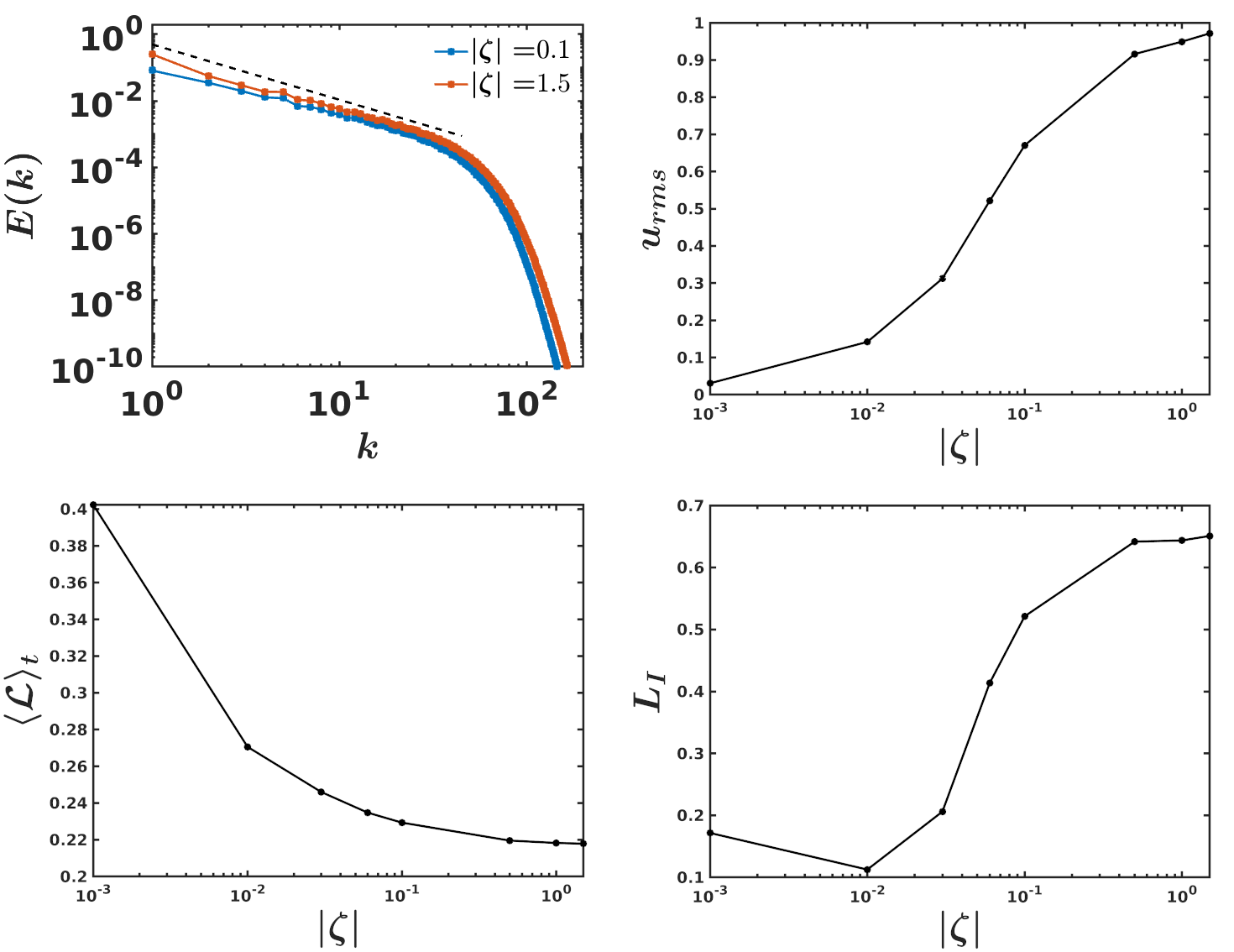}
    \put(-390,280){\rm {\bf(a)}}
    \put(-190,280){\rm {\bf(b)}}
    \put(-390,140){\rm {\bf(c)}}
    \put(-190,140){\rm {\bf(d)}}
  }
  }
\caption{(a) Log-log plots of the energy spectra $E(k)$  versus the wavenumber $k$, for the activities $|\zeta| = 0.1$ and $|\zeta| = 1.5$ in Eqs.~\eqref{eq:ch}-~\eqref{eq:tensor}; power-law regimes in these spectra are consistent with the dashed line $E(k) \sim k^{-5/3}$; (b) plot of the root-mean-squared velocity $u_{rms}$ versus $|\zeta|$; (c) log-linear plots versus $|\zeta|$ of (c) the mean coarsening length scale $L_c \equiv <\mathcal L(t)_t>$ and (d) the integral length scale $L_I$.}
\label{fig:turb_active}
\end{figure}
Turbulence in active fluids, which include dense bacterial suspensions, has garnered considerable attention over the past decade [see, e.g., \cite{wensink2012meso}, \cite{Dunkel_2013}, \cite{bratanov2015new}, \cite{alert2021active}]. Many models of active fluids consider systems of polar active swimmers [e.g., \cite{rana2024defect,jain2024inertia}] or Toner-Tu type systems and their generalisations [see, e.g., \cite{toner1998flocks}, \cite{toner2005hydrodynamics}, \cite{rana2020coarsening}, \cite{alert2021active}, \cite{mukherjee2021anomalous}, \cite{gibbon2023analytical}, \cite{kiran2024onset}, and \cite{kiran2023irreversibility}].  In a recent paper \cite{D4SM00163J} have demonstrated that a new type of active-scalar turbulence occurs in active-model H [see Eqs.~\eqref{eq:ch}-\eqref{eq:tensor}], whose stochastic version has been studied in the context of motility-induced phase separation (MIPS) that has been discussed at very-low Reynolds numbers by \cite{tiribocchi2015active} and \cite{cates2015motility}. We give an overview of the work of  \cite{D4SM00163J} in Figs.~\ref{fig:activechns} and \ref{fig:turb_active}, which examines activity-induced turbulence in Eqs.~\eqref{eq:ch}-\eqref{eq:tensor} by increasing
$\zeta$ in this active model H; positive values of $\zeta$ are used for contractile swimmers, whereas negative values of $\zeta$ are appropriate for extensile swimmers. \cite{D4SM00163J} concentrate on $\zeta < 0$, which yields activity-induced turbulence that suppresses phase separation. It has been suggested in  \cite{D4SM00163J} that this model, with $\zeta < 0$,  might be applicable to a dense suspension of \textit{Chlamydomonas reinhardtii}. 

 In Figs.~\ref{fig:activechns} (a)-(c) we show gray-scale plots of the active-scalar field $\phi$, at three representative times, which increase from left to right, for the activity parameter $|\zeta| = 0.1$ [see Eqs.~\eqref{eq:ch}-~\eqref{eq:tensor}]; Figs.~\ref{fig:activechns} (d)-(f) contain pseudocolor plots of the vorticity $\omega$ corresponding, respectively, to the pseudocolor plots $\phi$ in Figs.~\ref{fig:activechns} (a)-(c). These plots indicate that, as time increases, the activity induces spatiotemporal chaos and turbulence; eventually the systems reaches a nonequilibrium statistically steady state in which coarsening is arrested by active turbulence [much as it is arrested by conventional fluid turbulence as we have discussed in Subsection~\ref{subsec:PhaseSepBinary}]. We can characterise the statistical properties of this turbulence using the spectra and lengths that 
 we have defined in Eq.~\eqref{eq:LTST} for phase separation in the binary-fluid case. In Fig.~\ref{fig:turb_active} (a) we present log-log plots of the energy spectra $E(k)$  versus the wave number $k$, for the activities $|\zeta| = 0.1$ and $|\zeta| = 1.5$ in Eqs.~\eqref{eq:ch}-~\eqref{eq:tensor}.
 Clearly, the energy is spread out over a large range of $k$ as it is in fluid turbulence; furthermore, the power-law regimes in these spectra are consistent with $E(k) \sim k^{-5/3}$ (indicated by the dashed line). The root-mean-squared velocity $u_{rms}$ grows with $|\zeta|$ [Fig.~\ref{fig:turb_active} (b)]. In Figs.~\ref{fig:turb_active} (c) and (d) we present log-linear plots versus $|\zeta|$ of the mean coarsening length scale $L_c \equiv \langle\mathcal L(t)\rangle_t$ and the integral length scale $L_I$, respectively. Figures ~\ref{fig:turb_active} (b) and (d) quantify the enhancement of turbulence in Eqs.~\eqref{eq:ch}-\eqref{eq:tensor} with increasing $|\zeta|$; and Fig.~\ref{fig:turb_active} (c) characterises coarsening
 arrest by this form of active turbulence.
\subsection{Activity-induced droplet propulsion}
\label{subsec:dropprop}
\begin{figure}
  \centerline{
  \includegraphics[width=\textwidth]{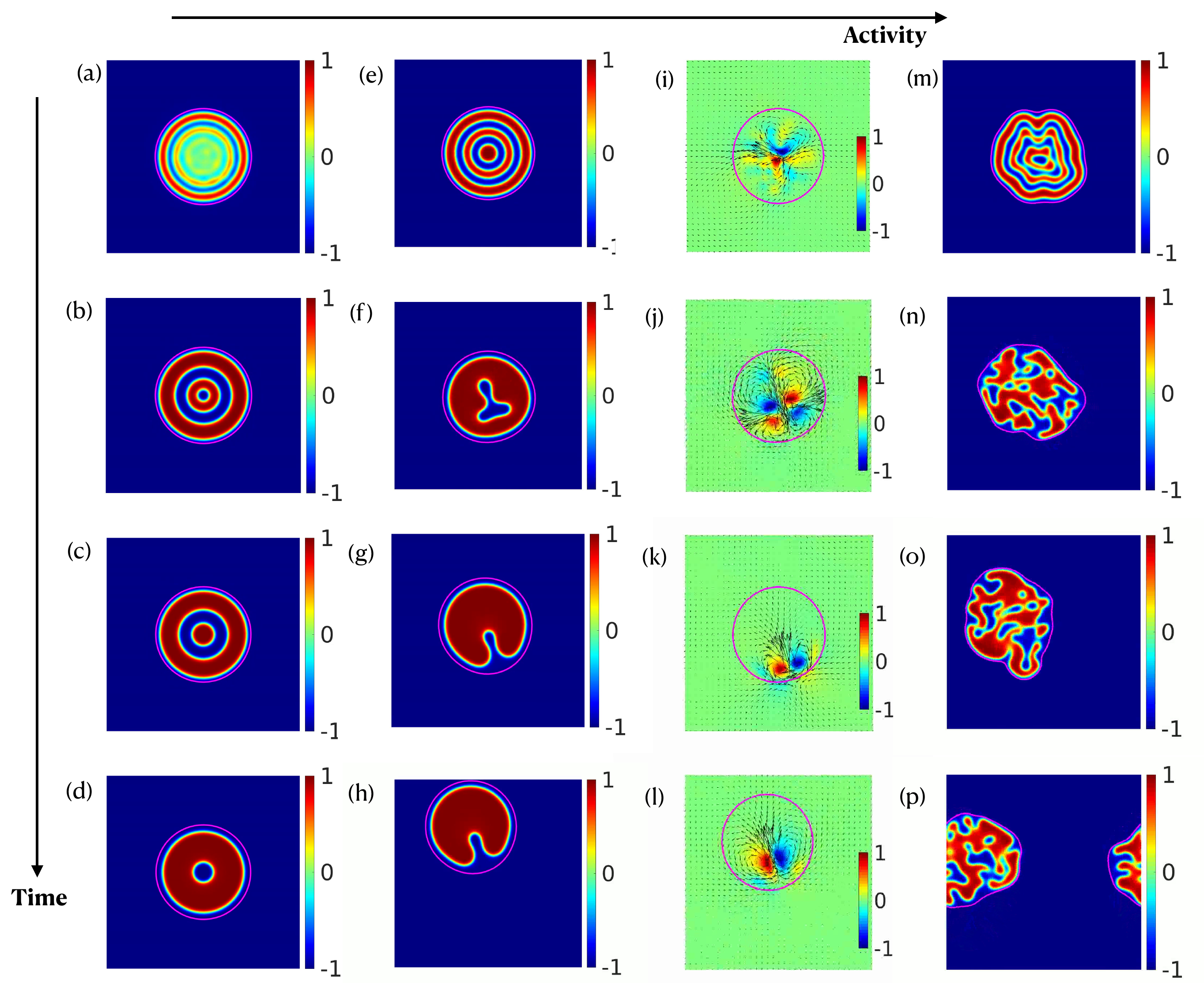}
  }
  \caption{Activity-induced droplet propulsion: Pseudocolor plots of $\psi$ [the  magenta contour shows $\phi = 0$], at various times $t$ and activities $A$ ($t$ increases from the top row to the bottom row): (a)-(d) $A = 0$ (complete phase separation in $\psi$ and no droplet propulsion); (e)-(h) $A = 0.15$ (rectilinear droplet propulsion); and (m)-(p) $A = 1$ (chaotic droplet propulsion). (i)-(l) For $A = 0.15$: vector plots of $\bm u$, with the overlaid $\phi = 0$ contour line (magenta) and the pseudocolor plot of $\omega$, normalised by its maximal value [velocity vectors have lengths proportional to $|\bm u|$].}
\label{fig:active_droplet}
\end{figure}
We now use Eqs.~(\ref{eq:fren})-(\ref{eq:Spsi}) [see Subsection~\ref{subsec:genactivechns}] to demonstrate activity-induced droplet propulsion in
a model that has been studied in detail by~\cite{padhan2023activity}. This model employs two scalar fields $\phi$ and $\psi$; both affect the velocity field $\bm{u}$ by which they are advected; but only $\psi$ is active in the parlance of active matter. Negative and positive values of $\phi$ and $\psi$ lead, respectively, to low and high densities of these scalars. We begin with the following initial data: a circular droplet, of radius $R_0$ and centre at $(x_{0,1}, x_{0,2}) = (\pi, \pi)$:
\begin{eqnarray}
\bm{u}(\bm{x},t=0)&=&0\,;\nonumber \\
\phi(\bm x, t=0) &=& \tanh{\left(\frac{R_0 - \sqrt{(x_1-x_{0,1})^2 + (x_2-x_{0,2})^2}}{\epsilon_1}\right)}\,;\nonumber \\
    \psi(\bm x, t=0) &=& \begin{cases}
    \psi_0(\bm x) &\text{for}\quad |\bm x|\leq R_0 \,;\\
    -1 &\text{for}\quad |\bm x| > R_0 \,;
    \label{eq:init}
    \end{cases}
\end{eqnarray}
 $\psi_0 (\bm x)$ is distributed uniformly and randomly on the interval $[-0.1,0.1]$. We then monitor the spatiotemporal evolution of $\phi$, $\psi$, and the normalised $\omega$, which we depict in Fig.~\ref{fig:active_droplet} via pseudocolor plots with an overlaid $\phi=0$ contour; the pseudocolor plots of $\omega$ also have superimposed vector plots of the velocity field $\bm u$. The non-dimensional Weber numbers $\text{We}_1 = R_0 U_0^2 /\sigma_1$ and $\text{We}_2 = R_0 U_0^2/\sigma_2$, Cahn numbers $\text{Cn}_1 = \epsilon_1/R_0$ and $\text{Cn}_2 = \epsilon_2 / R_0$,  Peclet numbers $\text{Pe}_1 = R_0 U_0 \epsilon_1/(M_1 \sigma_1)$ and $\text{Pe}_2 = R_0 U_0 \epsilon_2/(M_2 \sigma_2)$, Reynolds number $\text{Re} = R_0U_0/\nu$, where $U_0 = {\left< U_{CM}(t) \right>_t}$, with $U_{CM}$, the speed of the droplet's centre of mass ($CM$), the order-parameter couplings $\beta_1^{\prime} = \beta \epsilon_1/\sigma_1$ and $\beta_2^{\prime} = \beta \epsilon_2 / \sigma_2$, and the friction $\alpha^{\prime} = \alpha R_0 / U_0$, all affect the detailed dynamics of this initial droplet. However, most important of all these control parameters is the activity $A$ [Eq.~\eqref{eq:activityA}].

\begin{figure}
  \centerline{
  \includegraphics[width=\textwidth, height=6cm]{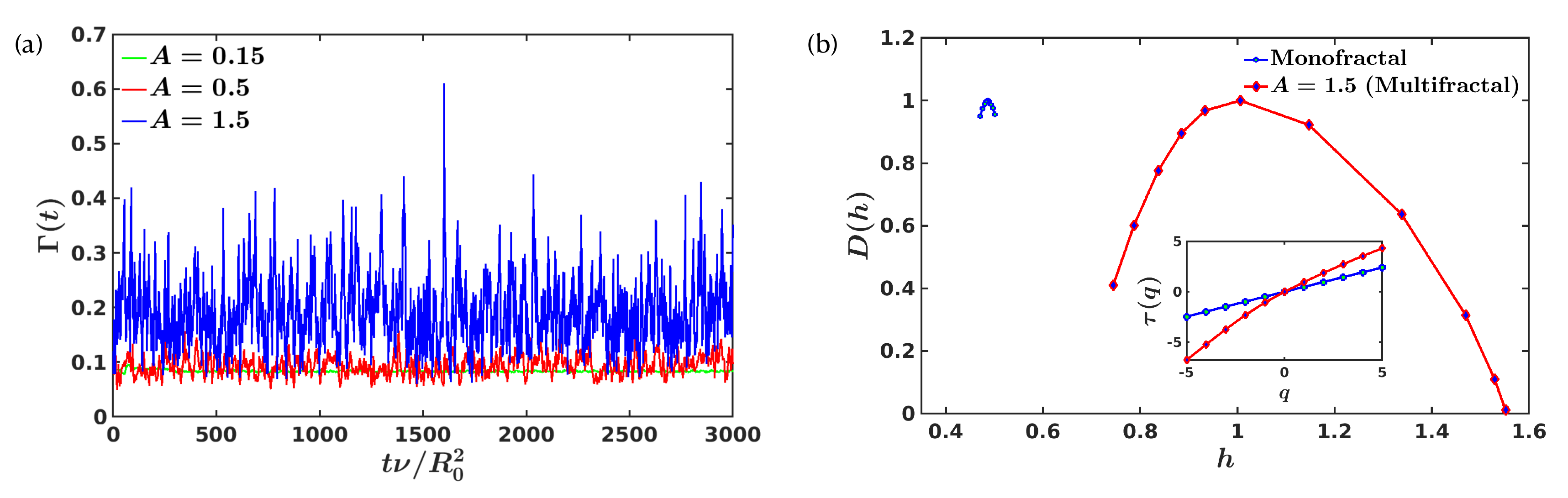}
  }
  \caption{(a) The multifractal time series for the scaled perimeter-deformation parameter $\Gamma(t)$ for various values of the activity. (b) The multifractal spectrum for representative activity $A = 1.5$. We present the spectrum  for a monofractal time series (given in blue) to show the robustness of the multifractal spectrum. The inset shows the plot of generalized exponent $\tau(q)$ as a function of the order $q$ for the representative value $A=1.5$; the deviation from the linearity suggests the multifractality of $\Gamma(t)$.}
\label{fig:multifractality}
\end{figure}

In Fig.~\ref{fig:active_droplet} we exhibit activity-induced droplet propulsion in this model by illustrative pseudocolor plots of $\psi$, with the $\phi = 0$ contour shown in magenta; we show such plots at different representative times, which increase from top to bottom, and three values of $A$, 
which move from low to high values, as we move from left to right. In Figs.~\ref{fig:active_droplet}(a)-(d) we show the spatiotemporal of this droplet 
for the case of vanishing activity $A = 0$; there is no droplet propulsion and the system proceeds towards complete phase separation of $\psi$, inside the $\phi = 0$ contour, by the formation of alternating annuli of regions with $\psi > 0$ and $\psi < 0$, which is reminiscent of the phase separation of oil and water in a microfluidic droplet [see ~\cite{moerman2018emulsion}]. Figures~\ref{fig:active_droplet} (e)-(h) show that, when $A = 0.15$, the system displays rectilinear propulsion of an active droplet; this is driven by the formation of an oscillating dipole that is visible clearly in Figs.~\ref{fig:active_droplet} (i)-(l), where we show, for $A = 0.15$, vector plots of the velocity field $\bm u$, with the $\phi = 0$ contour line (magenta), overlaid on a pseudocolor plot of the vorticity $\omega$ normalised by its maximal value. Finally, we show in Figs.~\ref{fig:active_droplet} (m)-(p), where $A = 1$, chaotic droplet propulsion, which is characterised by significant fluctuations inside the droplet and on its boundary; the former suppress phase separation within the droplet and lead to diffusive or super-diffusive meandering of the centre of mass of the droplet [see ~\cite{padhan2023activity} for details]. The fluctuations of the boundary can be quantified by using the scaled droplet perimeter $\Gamma(t)$ [see Eq.~\eqref{eq:peri} and Fig.\ref{fig:compound_droplets_graphs} in Subsection~\ref{subsec:dropturb}]. In Fig.~\ref{fig:multifractality} (a) we show that $\Gamma(t)$ has multifractal time series for various values of the activity $A$; we characterise this
in  Fig.~\ref{fig:multifractality} (b) by showing the multifractal spectrum for the representative value $A = 1.5$; for comparison we present this spectrum  for a monofractal time series in blue; the inset shows a plot of the generalized exponent $\tau(q)$ as a function of the order $q$; the deviation of the red curve from linearity quantifies the multifractality of $\Gamma(t)$.
\section{Conclusions and Perspective}
\label{sec:Conclusions}

We have demonstrated that the Cahn-Hilliard-Navier-Stokes (CHNS) framework offers an excellent theoretical foundation for probing diverse aspects of multi-phase fluid flows in binary and ternary systems and in active fluids.  We have given an introduction to the statistical mechanics of systems in which two or more coexisting phases, distinguished from each other by one or more scalar order parameters, are separated by an interface. Our discussion 
of systems with non-conserved and conserved order parameters leads, respectively, to the time-dependent Ginzburg-Landau (TDGL) and Cahn-Hilliard (CH) PDEs.  We have then considered models in which the coexisting phases are fluids; in particular, we have shown that two immiscible fluids require that we use the Cahn-Hilliard-Navier-Stokes (CHNS) equations. We have given generalisations of the CHNS equations for (a) coexisting phases with different viscosities, (b) CHNS with gravity, (c) three-component fluids (CHNS3), and (d) CHNS for active fluids. We have provided brief discussions of the methods we use for our DNSs of these CHNS systems and, in the antibubble case, we have contrasted the CHNS diffuse-interface approach with the volume-of-fluid scheme that tracks the spatiotemporal evolution of sharp fluid-fluid interfaces. Furthermore, we have discussed mathematical issues of the regularity of solutions of the CHNS PDEs. Then we have provided a survey of the rich variety of results that have been obtained by numerical studies of CHNS-type PDEs for diverse systems, including droplets in turbulent flows, antibubbles, droplet and liquid-lens mergers, turbulence in the active-CHNS model, and its generalisation that can lead to a self-propelled droplet. We hope that our overall perspective of this field will lead to more studies of multiphase flows in which interfaces and their fluctuations play important roles.

There are several other exciting areas in which the CHNS system can, or has already, played an important role. We have not been able to cover all these areas here. We give an illustrative list of such areas along with representative references:
\begin{itemize}
\item We do not cover quasi-compressible CHNS models; for these we refer the reader to ~\cite{lowengrub1998quasi} and ~\cite{abels2023global}; and for high-order CHNS PDEs readers should consult ~\cite{pan2020uniform}, \cite{dlotko2022navier}, and references therein.
\item There are intriguing links between the 2D CHNS system and 2D magnetohydrodynamics (MHD); these have been explored in, e.g., \cite{fan2016cascades,fan2017formation,fan2018chns,ramirez2024staircase}.
\item For simplicity we have considered coexisting phases with equal viscosities and densities. This constraint can be relaxed easily by using the CHNS equations~\eqref{eq:chnsphi}; and then this system can be used to study a variety of laboratory experiments, such as the droplet coalescence considered in 
\cite{Paulsen_2011,paulsen2014coalescence}.
\item There has been considerable interest in the study of the Cahn-Hilliard-type PDEs on curved surfaces; we refer the reader to \cite{voigt2002asymptotic} and \cite{ratz2006pde}.
\item There has been a lot of recent work on non-reciprocal Cahn-Hilliard systems [see, e.g., \cite{you2020nonreciprocity}, \cite{saha2020scalar}, \cite{frohoff2023nonreciprocal}, \cite{suchanek2023entropy}, and \cite{brauns2024nonreciprocal}]; we expect that these models will be coupled to the Navier-Stokes PDEs in future studies [for a recent study see~\cite{pisegna2025can}]. 
\item Studies of the statistics of Lagrangian tracers or heavy inertial particles in CHNS systems are in their infancy [see, e.g., \cite{padhan2024interfaces}]; we expect that such investigations will increase in the coming years. 
\end{itemize}

\section*{Declaration of Interests}
The authors report no conflict of interest.
\section*{Acknowledgements}
We thank J.K. Alageshan, J.D. Gibbon, A. Gupta, K.V. Kiran, B. Maji, D. Mitra, N. Pal, P. Perlekar, D. Vincenzi, and M. Wortis for discussions on different aspects of the CHNS 
system. We thank the Anusandhan National Research Foundation (ANRF), the Science and Engineering Research Board (SERB), and the National Supercomputing Mission (NSM), India for support, and the Supercomputer Education and Research Centre (IISc) for computational resources.

\bibliographystyle{jfm}

\bibliography{main}

\end{document}